\documentclass[11pt]{article}
\pdfoutput=1
\usepackage[margin=1.0in]{geometry}
\usepackage{amsmath,amssymb,graphicx}
\usepackage{hyperref}
\usepackage{slashed}
\usepackage{wasysym}
\usepackage{hhline}
\usepackage{subcaption}
\usepackage{graphicx}
\usepackage{multirow}
\usepackage{color}
\usepackage{booktabs}
\usepackage{xifthen}
\usepackage{xspace}
\usepackage{multirow}
\usepackage{placeins}
\usepackage[title]{appendix}
\usepackage[normalem]{ulem}

\usepackage[utf8]{inputenc}

\usepackage{cite}

\numberwithin{equation}{section}

\newcommand\w[1]{_{\mathrm{#1}}}

\newcommand\neut  [1][\relax]{{\tilde\chi^0_{#1}}}

\newcommand\charPM[1][\relax]{{\tilde\chi^\pm_{#1}}}

\newcommand\smuL{\tilde\mu\w L}
\newcommand\smuR{\tilde\mu\w R}
\newcommand\snumu{\tilde\nu_\mu}
\newcommand\slepL{\tilde l\w L}

\newcommand{\WHL}[1][]{\textbf{WHL}\ifthenelse{\equal{#1}{}}{}{$_{#1}$}\xspace}
\newcommand{\BHL}[1][]{\textbf{BHL}\ifthenelse{\equal{#1}{}}{}{$_{#1}$}\xspace}
\newcommand{\BHR}[1][]{\textbf{BHR}\ifthenelse{\equal{#1}{}}{}{$_{#1}$}\xspace}
\newcommand{\BLR}[1][]{\textbf{BLR}\ifthenelse{\equal{#1}{}}{}{$_{#1}$}\xspace}

\newcommand\mL{\tilde m_{l\w L}}
\newcommand\mR{\tilde m_{l\w R}}

\newcommand\vev[1]{\langle#1\rangle}
\DeclareMathOperator{\Order}{\mathcal O}

\newcommand\unit[1]{\,\mathrm{#1}}
\newcommand\GeV{\unit{GeV}}
\newcommand\TeV{\unit{TeV}}

\newcommand\pT{p\w T}
\newcommand{\met}{E\w T^{\rm miss}}
\newcommand\amu[1][\relax]{\ifx#1\relax{a_\mu}\else{a_\mu^{\text{#1}}}\fi}

\begin{document}


\begin{center}


~
\vskip 1cm

{\LARGE \bf
Supersymmetric explanation of the muon $g-2$ anomaly \\[0.3cm] 
with and without stable neutralino
}

\vskip 1.5cm

{\large 
Manimala Chakraborti$^{(a)}$,
Sho Iwamoto$^{(b)}$,
Jong Soo Kim$^{(c)}$,\\[3mm]
Rafa\l{} Mase\l{}ek$^{(d)}$
and
Kazuki Sakurai$^{(d)}$
}

\vskip 1cm

$^{(a)}${\em
Astrocent, Nicolaus Copernicus Astronomical Center of the Polish Academy of Sciences,\\
ul. Rektorska 4, 00-614 Warsaw, Poland\\[0.1cm]
}

$^{(b)}${\em
Institute for Theoretical Physics, ELTE E\"otv\"os Lor\'and University, \\
P\'azm\'any P\'eter s\'et\'any 1/A, H-1117 Budapest, Hungary \\[2mm]
}

$^{(c)}${\em
School of Physics, University of the Witwatersrand, Johannesburg, South Africa\\[2mm]
}

$^{(d)}${\em
Institute of Theoretical Physics, Faculty of Physics,\\
University of Warsaw, ul.~Pasteura 5, PL-02-093 Warsaw, Poland\\[0.1cm]
}

\vskip 0.3cm

\end{center}

\vskip 1.5cm

\begin{abstract}

In this paper we explore the possibility of explaining 
the muon $g-2$ anomaly in various types of supersymmetric extensions of the Standard Model.
In particular, we investigate and compare the phenomenological constraints 
in the MSSM with stable neutralino 
and the other types of scenarios where the neutralino is unstable.
For the latter case 
we study the Gauge Mediated SUSY Breaking (GMSB) scenario with very light gravitino
and the $UDD$-type R-Parity Violating (RPV) scenario.
In the MSSM with stable neutralino, 
the parameter region favoured by the $(g-2)_\mu$
is strongly constrained by the neutralino relic abundance
and the dark matter direct detection experiments, as well as by the
LHC searches in the lepton plus missing transverse energy channel.
On the other hand, the scenarios without stable neutralino
are free from the dark matter constraints,
while the LHC constraints depends strongly on the decay of the neutralino.
We find that in GMSB the entire parameter region favoured by the muon $g-2$
is already excluded if the Next Lightest SUSY Particle (NLSP) is the neutralino.
In the GMSB scenario with a stau NSLP
and in the RPV scenario,
LHC constraints are 
weaker than the stable neutralino case
and a larger region of parameter space is available to fit the  $(g-2)_\mu$ anomaly.

\end{abstract}


\vskip 1.5cm

\section{Introduction}
\label{sec:intro}

The $4\sigma$-level anomaly on the muon anomalous magnetic moment (muon $g-2$), established by the measurements at the Brookhaven National Laboratory~\cite{Muong-2:2002wip,Muong-2:2004fok,Muong-2:2006rrc} and more recently at the Fermilab~\cite{Muong-2:2021ojo},
is arguably one of the most interesting hints for physics beyond the Standard Model (BSM).
The anomaly, expressed in terms of $a_\mu = (g_\mu-2)/2$, amounts to
\begin{equation}
  \Delta \amu = \amu[BNL+FNAL] - \amu[SM] = \left( 25.1\pm5.9 \right) \times 10^{-10},
  \label{eq:gmin2}
\end{equation}
where we use the averaged value of the two measurements as the experimental value and the expected value based on the Standard Model (SM) is taken from the white paper~\cite{Aoyama:2020ynm, Aoyama:2012wk,Aoyama:2019ryr,Czarnecki:2002nt,Gnendiger:2013pva,Davier:2017zfy,Keshavarzi:2018mgv,Colangelo:2018mtw,Hoferichter:2019mqg,Davier:2019can,Keshavarzi:2019abf,Kurz:2014wya,Melnikov:2003xd,Masjuan:2017tvw,Colangelo:2017fiz,Hoferichter:2018kwz,Gerardin:2019vio,Bijnens:2019ghy,Colangelo:2019uex,Blum:2019ugy,Colangelo:2014qya
}.\footnote{
 A new calculation of the hadronic light-by-light contribution is reported in Ref.~\cite{Chao:2021tvp}, which is consistent with the white paper value.
On the other hand, a new lattice-based result of the hadronic vacuum-polarization contribution~\cite{Borsanyi:2020mff} disagrees with the white paper value.
Although the $(g-2)_\mu$ anomaly
is ameliorated when the new lattice value is adopted,
the tension arises elsewhere 
\cite{Crivellin:2020zul,Keshavarzi:2020bfy,Colangelo:2020lcg}.
}

Taking the $4\sigma$-level anomaly as a face value, it may be attributed to the effect of physics 
beyond the Standard Model.
There is a wealth of literature pursuing this possibility.
The list of BSM scenarios that can explain the $(g-2)_\mu$ anomaly
includes, for instance, models with leptoquarks \cite{Das:2016vkr,Chakraverty:2001yg,ColuccioLeskow:2016dox,Bigaran:2020jil}, vector-like leptons \cite{Dermisek:2013gta,Chun:2020uzw,Crivellin:2021rbq},
axion-like particles \cite{Bauer:2019gfk}, and dark matter (DM) \cite{Arcadi:2021cwg}.
Among these scenarios, supersymmetry (SUSY) \cite{Nilles:1983ge}, in particular the Minimal Supersymmetric extension
of the Standard Model (MSSM) \cite{Haber:1984rc}, is especially attractive 
since it also offers other benefits,
such as providing candidates for DM \cite{Goldberg:1983nd,Ellis:1983ew}, gauge coupling unification \cite{Ellis:1990wk},
and radiative breaking of the electroweak symmetry \cite{Ibanez:1982fr}.   

The SUSY explanation of the $(g-2)_\mu$ anomaly
has been widely studied in the literature \cite{Lopez:1993vi,Chattopadhyay:1995ae,Moroi:1995yh,
Martin:2001st,Everett:2001tq,Baltz:2001ts,Feng:2001tr,Chattopadhyay:2001vx}.\footnote{For simultaneous SUSY explanations for the electron and muon $g-2$, see e.g.~\cite{Dutta:2018fge,Endo:2019bcj, Badziak:2019gaf,Ali:2021kxa} }
The majority of these analyses assumes that the lightest supersymmetric particle
(LSP) is the lightest neutralino, $\tilde \chi_1^0$, and it is stable due to the
conservation of a discrete ${\mathbb Z}_2$ symmetry called the R-parity.
Moreover, many of the studies try to identify the lightest neutralino as the DM particle \cite{Cox:2018vsv,Cox:2018qyi,Abdughani:2019wai,Chakraborti:2021kkr,Chakraborti:2020vjp}, selecting the parameter region in which the relic abundance of the neutralino, $\Omega_{\tilde \chi_1^0}$, agrees with the observed density of the DM in the present Universe, $\Omega_{\rm DM}$.  
In this setup, the parameter region favoured by the $(g-2)_\mu$ anomaly
is severely constrained by the requirement 
$\Omega_{\tilde \chi_1^0} \simeq \Omega_{\rm DM}$ 
and the upper bounds on the spin-independent 
DM-nucleon scattering cross-section 
set by the DM direct detection experiments \cite{Abdughani:2019wai,Chakraborti:2021kkr,Chakraborti:2020vjp}. 
Even if one 
gives up the possibility to identify the neutralino 
as the DM candidate, the requirement that the relic neutralinos do not overclose
the Universe, $\Omega_{\tilde \chi_1^0} < \Omega_{\rm DM}$,
and the constraint from the DM direct detection experiments apply
and these exclude a large part of the parameter region preferred 
by the $(g-2)_\mu$ anomaly. 
The tension between the SUSY explanation of the $(g-2)_\mu$ anomaly
and the DM constraint 
is not a coincidence.
As we will discuss in the next section, the requirement on
the SUSY parameters to give large contributions to $(g-2)_\mu$ 
is directly related to the enhancement of the spin-independent 
DM-nucleon cross-section and 
suppression of the $\tilde \chi_1^0$-$\tilde \chi_1^0$ annihilation cross-section, leading to 
the overproduction of the relic neutralinos.

Another important constraint 
comes from the direct BSM searches by the ATLAS and CMS
experiments at the LHC.
The SUSY contribution to $(g-2)_\mu$ is large enough only
when the masses of sleptons and electroweak (EW) gauginos are below ${\cal O}(500)$ GeV.
Such light sleptons and electroweakinos (EWinos) would be produced at the LHC
and detected in search channels looking for, e.g., multiple leptons and a sizable missing transverse energy ($\met$).
Provided $\neut[1]$ is the stable LSP,
the slepton and electroweakino mass spectrum must be compressed in order to evade the LHC exclusion,
so that decays of the produced SUSY particles do not produce high-$\pT$ leptons.

In the first part of this paper,
we review the viable MSSM $(g-2)_\mu$ parameter region assuming $\tilde \chi_1^0$ to be the stable LSP
and confront it with the latest phenomenological constraints. 
Several studies have explored this possibility in the recent 
times~\cite{deVries:2015hva,Bagnaschi:2017tru,Chakraborti:2021kkr,Chakraborti:2020vjp,Chakraborti:2021dli,Chakraborti:2021mbr,Iwamoto:2021aaf,VanBeekveld:2021tgn,Cox:2021gqq,Baum:2021qzx,Athron:2021iuf,Endo:2021zal,Gomez:2022qrb}.
Refs.~\cite{deVries:2015hva,Bagnaschi:2017tru,Chakraborti:2021kkr,Chakraborti:2020vjp,Chakraborti:2021dli,Chakraborti:2021mbr}, e.g.
have performed inclusive random scans
over all the relevant MSSM parameters to provide upper and lower limits on
the superpartner masses satisfying $(g-2)_\mu$, DM and LHC data.
The MSSM scenarios considered in these analyses are categorized according 
to the DM relic density generation mechanism.
The authors of Ref.~\cite{Athron:2021iuf} have analyzed various MSSM scenarios depending
on the mass hierarchy among charginos and sleptons. They have also
allowed for the possibility of non-universality between the input stau masses 
and the masses of the first two generations of sleptons.
A similar study is performed in Ref.~\cite{Cox:2021gqq} focusing on
bino DM scenario.
In our analysis, we investigate
minimal MSSM scenarios where only one of the 1-loop contributions to $(g-2)_\mu$
is dominant at a time. This helps us to systematically scan over the relevant
MSSM parameters and identify the appropriate LHC constraints.\footnote{Similar analyses have been
performed in Refs.~\cite{Endo:2013bba,Endo:2017zrj,Endo:2020mqz,Iwamoto:2021aaf,Endo:2021zal}.
In particular, Refs.~\cite{Endo:2017zrj,Endo:2021zal} have found MSSM model points that explain both the $(g-2)_\mu$ anomaly
and the dark matter abundance,
$\Omega\w{DM}\simeq\Omega\w{\neut[1]}$, and are consistent with various phenomenological constraints, such as LHC.
In those analyses, the SUSY contributions are dominated by the BHL and BLR diagrams defined below.}
In this way, we gain an overview of the existing tension between 
the SUSY explanation of the $(g-2)_\mu$ anomaly and 
the other phenomenological constraints in the stable neutralino case.

The analysis of the stable neutralino case leads us to the following question:
{\it How does the phenomenologically viable SUSY $(g-2)_\mu$ parameter region 
change if the lightest neutralino is unstable?}
Clearly, if $\tilde \chi_1^0$ is unstable, it can no longer serve as a
candidate for DM and thus can easily evade the stringent constraints
coming from DM relic density measurements and DM direct detection experiments.
The LHC constraints would also change since the missing transverse energy 
signature produced by the stable $\tilde \chi_1^0$s in LHC events 
will be replaced by the signatures responsible for the particles produced 
from the $\tilde \chi_1^0$ decays. 
In this study we focus on two concrete scenarios of unstable neutralino:
(1) a scenario with an almost massless gravitino LSP and
(2) a scenario of R-parity violation, in particular that by a $UDD$ operator. 
We recast various direct BSM searches performed by the ATLAS and CMS collaborations
and estimate the impact of those constraints on the  $(g-2)_\mu$ favoured parameter region.

The rest of the paper is organised as follows.  
In the next section we discuss the SUSY contribution to the $(g-2)_\mu$
and find four 1-loop diagrams defined in the mass-insertion approximation.
Motivated by these distinctive contributions,
we define several 2-dimensional parameter planes in which
we study the $(g-2)_\mu$ and phenomenological constraints.
In section \ref{sec:constrant}
we discuss the relevant phenomenological constraints in our analyses.
We lay down the concrete procedure of our numerical analysis in section \ref{sec:procedure}.
The results of our numerical analyses 
for the stable neutralino,
R-parity violation
and
gravitino LSP scnarios 
are shown and discussed in sections \ref{sec:stable},
\ref{sec:rpv} and \ref{sec:grav}, respectively.
Section \ref{sec:concl} is devoted 
to the conclusion.

\section[SUSY contribution to muon g-2]{SUSY contribution to $(g-2)_\mu$}
\label{sec:1-loop-diagram}

The SUSY contribution to $(g-2)_\mu$ at the one-loop level
depends on the following six parameters:\footnote{Strictly speaking, the trilinear 
SUSY breaking coupling, $A_\mu$, also contributes to the $(g-2)_\mu$.  However, this contribution is suppressed by $1/\tan\beta$
and negligible in the large or moderate $\tan\beta$ region we are interested in this paper.}
\begin{equation}
    M_1,~M_2,~\mu,~ \tilde m_{l_L},~ \tilde m_{l_R},~\tan\beta\,,
    \label{eq:param}
\end{equation}
where
$\mu$ is the Higgsino mass parameter,
$\tan \beta \equiv \vev{H_u^0} / \vev{H_d^0}$ is the ratio of the vacuum expectation values of the two Higgs doublets,
and
$M_{1}$, $M_2$, $\tilde m_{l_L}$ and $\tilde m_{l_R}$ are the soft SUSY breaking masses
for the Bino ($\mathrm{U}(1)_Y$ gaugino), the Wino ($\mathrm{SU}(2)_L$ gaugino),
the left-handed slepton doublet and the right-handed slepton singlet, respectively.
{To reduce the number of free parameters we adopt the universal slepton mass assumption:
$\tilde m_{ l_1}= \tilde m_{ l_2}= \tilde m_{ l_3} \equiv \tilde m_{l_L}$
and
$\tilde m_{ e_R} = \tilde m_{ \mu_R}= \tilde m_{ \tau_R} \equiv \tilde m_{l_R}$.
With this assumption, 
the model can evade 
the strong constraints from
lepton-flavour violating (LFV) processes such as $\mu \to e \gamma$.
Strictly speaking, the universal slepton mass assumption is not necessary to avoid the LFV constraints.
It is sufficient to assume that the charged slepton mass matrices, ${\mathbf m}_{l_L}$ and ${\mathbf m}_{l_R}$, are diagonal in the same basis in which the charged lepton mass matrix is diagonal.
{However, in many known scenarios of SUSY breaking mediation,
flavour universal sfermion masses are realised at the mediation scale.
Although some of those scenarios predict
particular gaugino mass relations,
these ratios are relatively unconstrained when there are multiple sources of SUSY breaking mediation.\footnote{ 
For example, in the minimal gauge mediation, the messenger fields are a vector-like pair of $SU(5)$, (${\bf 5}$, $\bf \bar 5$), and the gaugino masses 
are proportional to the square of the corresponding gauge couplings, $M_i \propto g^2_i$. 
However, this relation can be easily modified by allowing mass splitting between the triplet and doublet parts of 
(${\bf 5}$, $\bf \bar 5$),
while keeping the flavour universality of sfermion masses intact.
The mixed modulus and anomaly mediation (or the mirage mediation) \cite{Choi:2004sx, Choi:2005ge, Endo:2005uy, Choi:2005uz, Falkowski:2005ck}
and a non-singlet F-term breaking 
scenario (see e.g.\ \cite{Martin:2009ad})
are other examples to achieve 
non-standard gaugino mass relations.
}
Our analysis is also applicable to such situations.}
We also assume {for simplicity} that the SUSY-breaking parameters and $\mu$ are real 
and do not contribute to the CP violating observables.\footnote{Again, such an assumption is not required once we allow the possibility of cancellation among the imaginary parts of various parameters.  However, we do not consider such a contrived case and simply adopt real parameters.}

\begin{figure}[t!]
\centering
      \includegraphics[width=0.99\textwidth]{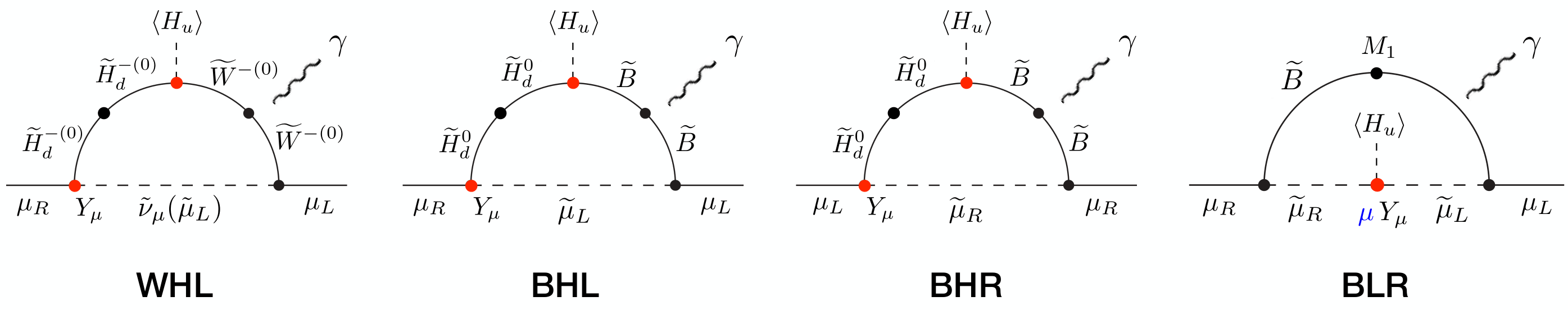}
\caption{\label{fig:diagrams}  \small
  One-loop diagrams representing SUSY contribution to $(g-2)_\mu$. Each diagram is labelled by the name corresponding to sparticles taking part in it, e.g. \textit{BHL} is a diagram involving Bino, Higgsino and left-handed slepton.
  The red dots represent the interactions responsible
  for the $\tan \beta$ enhancement.
}
\end{figure}

For large or moderate values of $\tan \beta$ ($5 \lesssim \tan \beta \lesssim 50$), 
the one-loop SUSY contribution is approximated by~\cite{Moroi:1995yh}
\begin{equation}
\amu[SUSY] \,\simeq~ \amu[WHL] + \amu[BHL] + \amu[BHR] + \amu[BLR]\,.
\end{equation}
{In the mass-insertion approximation,
each term on the right-hand side is represented by the respective diagram in Fig.~\ref{fig:diagrams}, which is given by}
\begin{subequations}
\label{eq:aMI} 
\begin{align}
    a_\mu^{\rm WHL}(M_2, \mu, \mL) &=
    \frac{\alpha_2}{8 \pi} \frac{m_\mu^2}{M_2 \mu}
    \tan \beta \cdot 
    \left[ 2 f_C \left( \frac{M_2^2}{m^2_{\tilde \nu_\mu}}, \frac{\mu^2}{m^2_{\tilde \nu_\mu}} \right)
    - 
    f_N \left( \frac{M_2^2}{m^2_{\tilde \mu_L}}, \frac{\mu^2}{m^2_{\smuL}} \right)
    \right]\,,
    \label{eq:aWHL}
\\
    a_\mu^{\rm BHL}(M_1, \mu, \mL) &=
    \frac{\alpha_Y}{8 \pi} \frac{m_\mu^2}{M_1 \mu}
    \tan \beta \cdot 
    f_N \left( \frac{M_1^2}{m^2_{\tilde \mu_L}}, \frac{\mu^2}{m^2_{\smuL}} \right)\,,
    \label{eq:aBHL}
\\
    a_\mu^{\rm BHR}(M_1, \mu, \mR) &=
    - \frac{\alpha_Y}{8 \pi} \frac{m_\mu^2}{M_1 \mu}
    \tan \beta \cdot 
    f_N \left( \frac{M_1^2}{m^2_{\tilde \mu_R}}, \frac{\mu^2}{m^2_{\smuR}} \right)\,,
    \label{eq:aBHR}
\\
    a_\mu^{\rm BLR}(M_1, \mL, \mR; \mu) &=
    \frac{\alpha_Y}{4 \pi} \frac{m_\mu^2 M_1 \mu}{ m^2_{\smuL} m^2_{\smuR}}
    \tan \beta \cdot 
    f_N \left( \frac{m^2_{\smuL}}{M_1^2}, \frac{m^2_{\smuR}}{M_1^2} \right)\,.
    \label{eq:aBLR}    
\end{align}
\end{subequations}
The functions $f_C$ and $f_N$ are given by
\begin{subequations} 
 \begin{align}
    f_C(x,y) &= xy \left[ \frac{5 - 3(x+y) + xy}{(x-1)^2 (y-1)^2} 
    - \frac{2 \ln x}{(x-y)(x-1)^3} + \frac{2 \ln y}{(x-y)(y-1)^3}
    \right]\,,
    \\
    f_N(x,y) &= xy \left[ \frac{-3 +x+y + xy}{(x-1)^2 (y-1)^2} 
    + \frac{2 x \ln x}{(x-y)(x-1)^3} - \frac{2 y \ln y}{(x-y)(y-1)^3}
    \right]\,,
 \end{align}
\end{subequations}
and
{
$m_{\smuL} = (\tilde m_{l_L}^2 + m^2_\mu + \Delta_L)^{\frac{1}{2}}$ and $m_{\smuR} =
(\tilde m^2_{l_R} + m^2_\mu + \Delta_R)^{\frac{1}{2}}$
with $\Delta_L \equiv m_Z^2 (-\frac{1}{2} + \sin^2 \theta_W) \cos 2 \beta $ and
$\Delta_R \equiv m_Z^2 \sin^2 \theta_W \cos 2 \beta$ 
are the diagonal (LL and RR) entries 
of the smuon mass matrix,
and $m_{\snumu}$, $m_\mu$ and $\theta_W$  are the masses of the sneutrinos and the muon and the weak mixing angle
respectively.}

If all SUSY particles have the same mass,
the $a_\mu^{\rm WHL}$ is an order of magnitude larger than the other contributions.
{Although we define our parameter planes 
based on the four 1-loop diagrams
described above,
we do not use the mass insertion approximation in our numerical study.
To compute $a_\mu^{\rm SUSY}$
we use
{\tt GM2Calc} \cite{Athron:2015rva} instead,
which includes the full 1-loop contribution as well as the leading 2-loop effect.}


\subsection{Parameter planes}
\label{sec:planes}

Since there are six parameters in the problem,
exploration and visualization of the whole parameter space are challenging.
In this paper, we instead define eight two-dimensional parameter planes motivated by the above one-loop contributions
and systematically study the interplay between $\amu[SUSY]$ and the other phenomenological constraints.
As $\amu[SUSY]$ at the one-loop level is proportional to $\tan \beta$,
we take $\tan \beta = 50$ as the default value in order to maximise the SUSY contributions and specify three of the remaining five parameters to define two-dimensional parameter spaces.

We set the masses of gluinos, squarks, and heavy Higgs bosons arbitrary large in our analysis
because
$\amu[SUSY]$ at the one-loop level is independent of their mass and
recent SUSY searches at the LHC have placed stringent lower bounds on the masses of the coloured SUSY particles \cite{ATLAS:2020syg,ATLAS:2021twp,CMS:2019zmd}.

\begin{table}[t!]\small
\centering
 \begin{tabular}[t]{|c|c|c|r@{~~}l|c|}\hline
 name     & axes & range [TeV] & \multicolumn{2}{|c|}{other parameters} & $\tan\beta$
\\\hline
\WHL[\mu] & $(M_2, \mu)$ & ([0.2, 4], [0.2, 4]) & $\mL = \min(M_2,\mu)+20\GeV$, & $M_1 = \mR = 10\TeV$& 50 \\\hline
\WHL[L]   & $(M_2, \mL)$ & ([0.2, 4], [0.2, 2]) & $\mu = \min(M_2,\mL)-20\GeV$, & $M_1 = \mR = 10\TeV$& 50 \\\hline
\BHL[\mu] & $(M_1, \mu)$ & ([0.12, 0.6], [0.12, 0.35]) & $\mL = \min(M_1,\mu)+20\GeV$, & $M_2 = \mR = 10\TeV$& 50 \\\hline
\BHL[L]   & $(M_1, \mL)$ & ([0.12, 0.8], [0.14, 0.22]) & $\mu = \min(M_1,\mL)-20\GeV$, & $M_2 = \mR = 10\TeV$& 50 \\\hline
\BHR[\mu] & $(M_1, |\mu|)$ & ([0.12, 0.7], [0.12, 0.7]) & $\mR = \min(M_1,|\mu|)+20\GeV$, & $M_2 = \mL = 10\TeV$& 50 \\\hline
\BHR[L]   & $(M_1, \mR)$   & ([0.12, 0.8], [0.14, 0.25]) & $-\mu = \min(M_1,\mR)-20\GeV$,  & $M_2 = \mL = 10\TeV$& 50 \\\hline
\BLR[50]  & $(\mL, \mR)$ & ([0.15, 0.6], [0.12, 1.2]) & \multicolumn{2}{|c|}{$M_1 = m_{\tilde\tau_1} - 20\GeV$,~~$\mu=\mu\w{max}$,~~$M_2=10\TeV$} & 50 \\\hline
\BLR[10]  & $(\mL, \mR)$ & ([0.15, 0.6], [0.12, 1.2]) & \multicolumn{2}{|c|}{$M_1 = m_{\tilde\tau_1} - 20\GeV$,~~$\mu=\mu\w{max}$,~~$M_2=10\TeV$} & 10 \\\hline
 \end{tabular}
\caption{\small The parameter planes
and choices of the other parameters.
{$\mu_{\rm max}$ is defined as the maximum value allowed 
by the vacuum stability constraint.}
}
\label{tab:planeRPC}
\end{table}

The eight parameter planes are summarized in Table~\ref{tab:planeRPC}.
In each plane, three mass parameters among the five (cf.~Eq.~\eqref{eq:param}) are set as light as $\Order(100)\GeV$ and the remaining two are set heavier, so that one of the four contributions in Eq.~\eqref{eq:aMI} dominates.
We prepare two planes, which result in different collider phenomenology, for each of the four scenarios.
In these parameter planes 
the LSP 
among the MSSM spectrum is always given by the lightest neutralino, $\tilde \chi_1^0$.
Beyond MSSM, the gravitino can be lighter than 
$\tilde \chi_1^0$, which we study in detail 
section \ref{sec:grav}.

\subsubsection*{WHL planes}

The WHL planes are characterized by
\begin{equation*}
|M_2|,\, |\mu|,\, \mL \text{~at~} \Order(100)\GeV,
\qquad
|M_1|,\,\mR \gg 1\TeV,
\qquad
\mu \cdot M_2 > 0.
\end{equation*}
Thanks to the first requirement, the WHL contribution $\amu[WHL]$ yielded by the WHL diagram becomes large enough to explain the muon $g-2$ anomaly, where the last requirement guarantees $\amu[WHL]$ is positive (cf.~Eq.~\eqref{eq:aWHL} and Fig.~\ref{fig:diagrams})\footnote{%
  Only the relative sign between $\mu$ and $M_i$ ($i=1,2,3$) is physical.
  We take $M_i>0$ throughout the paper.
}.
On the other hand, the other three contributions, i.e., $\amu[BHL]$, $\amu[BHR]$, and $\amu[BLR]$, are smaller because of the second requirement.
The SUSY contribution $\amu[SUSY]$ is thus dominated by $\amu[WHL]$.

As discussed in Sec.~\ref{sec:intro}, if the LSP is $\neut[1]$ and stable,
the parameter space under these requirements receives stringent constraints from
SUSY searches at the LHC, such as searches for multi-lepton plus large missing transverse momentum ($\met$) signature.
In order to avoid emergence of high-$\pT$ leptons,
we require two of the three small parameters are close to each other and
define the following two planes:
\begin{itemize}
      \item {\bf WHL$_\mu$} ($M_2$ vs $\mu$) plane: \\
      $\tilde m_{l_L} = \min(M_2, \mu) + 20$\,GeV, 
      $\tan \beta = 50$, $(M_1, \tilde m_{l_R}) > $ a few TeV 
      \item {\bf WHL$_L$} ($M_2$ vs $\tilde m_{l_L}$) plane: \\
      $\mu = \min(M_2, \tilde m_{l_L}) - 20$\,GeV, $\tan \beta = 50$, $(M_1, \tilde m_{l_R}) > $ a few TeV 
\end{itemize}
In both planes the MSSM-LSP is $\neut[1]$. 
Both $\neut[1]$ and $\charPM[1]$ are composed either of the Wino ($M_2 < \mu$) or the Higgsino ($\mu < M_2$).
The $\neut[1]$ and $\charPM[1]$
are almost mass degenerate 
and the scale of the mass splitting is given by $v_{\rm EW}^2/m_{\neut[1]}$.

On the {\bf WHL$_\mu$} plane, $\mL$ is set close to $\neut[1]$.
By this assumption, the decay $\slepL\to l+\neut[1]$, which may occur in the cascade-decay chain from the SUSY particles produced at the LHC, does not yield a high-$\pT$ lepton.
The models are therefore elusive at the LHC, even though the electroweakino pair-productions have sizable cross sections.
{At the same time, if $\neut[1]$ is stable and mostly Wino-like, $\charPM[1]$ might be long-lived enough so that $\charPM[1]$ can be detected by the disappearing track searches at the LHC.}

On the {\bf WHL$_L$} plane, $\neut[1]$ is close in mass either to the other neutralinos $\neut[2,3]$ and chargino $\charPM[2]$ ($M_2 < \tilde m_{l_L}$), or to $\slepL$ ($\tilde m_{l_L} < M_2$).
The latter case is similar to the ${\bf WHL}_\mu$ plane and thus elusive at the LHC.
{In the former case with $M_2 < \tilde m_{l_L}$, high $p_T$ leptons may be produced from the slepton decays in the $pp\to\slepL\slepL^*$ process. 
However, the cross-section of the slepton pair production is much smaller compared to the electroweakino pair production at the same mass and the LHC constraint in this region can be relatively easily evaded.}

\subsubsection*{BHL planes}

The BHL diagram is obtained from the WHL diagram 
by the replacement
$(\widetilde W^\pm, \widetilde H^\pm, \tilde \nu_\mu) \to (\widetilde B, \widetilde H^0, \tilde \mu_L)$, and thus $\amu[BHL]$ dominates over the other three contributions if
\begin{equation*}
|M_1|,\, |\mu|,\, \mL \text{~at~} \Order(100)\GeV,
\qquad
|M_2|,\,\mR \gg 1\TeV,
\qquad
\mu \cdot M_1 > 0,
\end{equation*}
where the last condition guarantees $\amu[BHL]>0$.
Similarly to the WHL scenarios,
we define two planes by
\begin{itemize}
      \item {\bf BHL$_\mu$} ($M_1$ vs $\mu$) plane: \\
      $\tilde m_{l_L} = \min(M_1, \mu) + 20$\,GeV, $\tan \beta = 50$, $(M_2, \tilde m_{l_R}) > $ a few TeV 
      \item {\bf BHL$_L$} ($M_1$ vs $\tilde m_{l_L}$) plane: \\
      $\mu = \min(M_1, \tilde m_{l_L}) - 20$\,GeV, $\tan \beta = 50$, $(M_2, \tilde m_{l_R}) > $ a few TeV 
\end{itemize}
Unlike the WHL planes, $\charPM[1]$ is always Higgsino-like
in the BHL planes
and short-lived enough
not to be detected by the disappearing track searches at the LHC.
Also, $\neut[1]$ is composed of Bino or neutral Higgsino.
Because the pure-Bino cross-section is vanishingly small when squarks are decoupled, SUSY-particle production is only through the Higgsino pair-production or slepton pair-production.
Their cross sections are smaller than the Wino pair-production (compared at the same masses) and the parameter space tends to be less constrained by the current LHC searches.

\subsubsection*{BHR planes}

The BHR diagram can be obtained from the BHL diagram 
simply by flipping the chirality of the muon and smuon.
Hence $\amu[BHR]$ becomes dominant if
\begin{equation*}
|M_1|,\, |\mu|,\, \mR \text{~at~} \Order(100)\GeV,
\qquad
|M_2|,\,\mL \gg 1\TeV,
\qquad
\mu \cdot M_1 < 0,
\end{equation*}
where, as can be seen in Eq.~\eqref{eq:aBHR}, we take $\mu<0$ and $M_1>0$ to have $\amu[BHR]>0$,
which fits the observed $(g-2)_\mu$ anomaly.

The BHR planes are defined as follows:
\begin{itemize}
      \item {\bf BHR$_\mu$} ($M_1$ vs $|\mu|$) plane: \\
      $\tilde m_{l_R} = \min(M_1, \mu) + 20$\,GeV, $\tan \beta = 50$, $(M_2, \tilde m_{l_L}) > $ a few TeV 
      \item {\bf BHR$_L$} ($M_1$ vs $\tilde m_{l_R}$) plane: \\
      $|\mu| = \min(M_1, \tilde m_{l_R}) - 20$\,GeV, $\tan \beta = 50$, $(M_2, \tilde m_{l_L}) > $ a few TeV 
\end{itemize}
The nature of MSSM-LSP and the LHC phenomenology is similar to those in the corresponding BHL planes, but sneutrinos are always decoupled and do not participate in LHC signatures.

\subsubsection*{BLR planes}

The BLR diagram in Fig.~\ref{fig:diagrams} is special because, unlike the other diagrams, the chirality flip occurs 
in the smuon sector.
The size of the smuon chirality flip is given by the off-diagonal element of the smuon mass-squared matrix and thus proportional to the $\mu$ parameter (cf.~Eq.~\eqref{eq:staumassmatrix}).
The BLR contribution is therefore proportional to $\mu$ as well as to $\tan\beta$, whereas
all the other one-loop contributions are vanishing in the limit of $|\mu|\to\infty$.
Therefore, $\amu[BLR]$ becomes dominant if smuons and Bino are as light as $\Order(100)\GeV$ and $\mu\tan\beta$ is large enough.

Under these conditions, we have to pay attention to two constraints from the stau sector, i.e.,
the LEP limit on the stau mass
and
the vacuum stability.
These constraints are critical because
the SUSY breaking mass parameters 
are common for all generations.
The mass-squared matrices for the left- and right-handed charged sleptons are given by
\begin{equation}
\label{eq:staumassmatrix}
\begin{pmatrix}
\mL^2 + m_l^2 + (\sin^2\theta_W-1/2)m_Z^2\cos2\beta & m_l(A_l^* - \mu^* \tan \beta) \\
m_l(A_l-\mu \tan \beta) & \mR^2 + m_l^2 - {m_Z^2\sin^2\theta_W\cos2\beta}
\end{pmatrix}
\end{equation}
at the tree-level, where $l=e,\mu,\tau$.
Here, $m_l$ is the charged lepton mass, $\theta_W$ is the weak mixing angle, and
$A_l$ is the SUSY-breaking trilinear coupling of the slepton, which we set to be zero.
Because $\mL$ and $\mR$ are common for all generations and the off-diagonal element is proportional to the lepton mass,
the left-right mixing of staus is larger than that of smuons and the lighter stau (heavier stau) becomes lighter (heavier) than the lighter smuon (heavier smuon).
The LEP bounds on the stau mass, $m_{\tilde \tau_1} > 81.9\GeV$ \cite{Zyla:2020zbs}, is therefore critical in this scenario.
In addition, with the large left-right mixing of the staus, the effective potential may develop a global minimum with non-zero vacuum expectation values of the stau fields.
If the vacuum transition rate from our EW vacuum to this charge-breaking global minimum
is significantly faster than the age of the Universe, the parameter point is excluded.
The vacuum stability thus provides an upper bound on $|\mu\tan\beta|$~\cite{Kitahara:2013lfa,Endo:2013lva}.

Accordingly, the BLR planes are characterized by
\begin{align*}
&|M_1|,\, \mL,\, \mR \text{~at~} \Order(100)\GeV,
&&100\GeV\ll |\mu| \le \frac{(\mu\tan\beta)\w{max}}{\tan\beta},
&&|M_2| \gg 1\TeV,
&&\mu \cdot M_1 > 0,
\end{align*}
where $(\mu\tan\beta)\w{max}$ is the maximal value of $|\mu\tan\beta|$ under the vacuum-stability condition given in Refs.~\cite{Kitahara:2013lfa,Endo:2013lva}, which is dependent on $\mL$ and $\mR$.
The last requirement guarantees $\amu[BLR]>0$ to explain the muon $g-2$ anomaly and $\amu[WHL]$ is subdominant because of the third condition, while $\amu[BLR]$ is enhanced by the first and second ones.
However, one should note that $\amu[BHL]$ or $\amu[BHR]$ may not be negligible because $|\mu|$ has an upper bound
$\mu\w{max}\equiv(\mu\tan\beta)\w{max}/\tan\beta$.

We define two BLR planes, in which $\neut[1]$ is the MSSM-LSP, as follows:
\begin{itemize}
      \item {\bf BLR$_{50}$} ($\tilde m_{l_L}$ vs $\tilde m_{l_R}$) plane: \\
      $M_1 = m_{\tilde \tau_1} - 20$\,GeV, $\mu = \mu_{\rm max}$, $\tan \beta = 50$, $M_2 > $ a few TeV 
      \item {\bf BLR$_{10}$} ($\tilde m_{l_L}$ vs $\tilde m_{l_R}$) plane: \\
      $M_1 = m_{\tilde \tau_1} - 20$\,GeV, $\mu = \mu_{\rm max}$, $\tan \beta = 10$, $M_2 > $ a few TeV 
\end{itemize}
In both planes, we fix $\mu = \mu_{\rm max}$ so that BLR contribution is maximised 
without violating the vacuum stability condition.
SUSY-particle production at the LHC is through the slepton pair-production or the Higgsino pair-production, but the latter is available only if $\mu$ is at $\Order(100)\GeV$ due to the vacuum stability condition.
On the \BLR[10] plane, the condition is relaxed due to the non-default value of $\tan \beta$, and thus 
the LHC constraints originated from the Higgsino production is relaxed.

\section{Phenomenological constraints}
\label{sec:constrant}

\subsection{The neutralino relic density}
\label{sec:relic}
In the cases with the lightest neutralino $\neut[1]$ being stable, one must consider the phenomenological constraint 
that the relic neutrinos do not overclose the Universe.
In the early Universe, neutralinos are coupled to the thermal bath and their number density 
is fixed by the temperature.
While the Universe expands and the temperature drops, their number density decreases 
and at some point the lightest neutralino freezes out as the dark matter.
Those relic neutralinos contribute to the energy density of the dark matter in the present Universe.
Clearly, the energy density of the relic neutralino, $\Omega_{\tilde \chi_1^0}$,
cannot exceed the measured DM density, $\Omega_{\rm DM}$.
Namely, we require $\Omega_{\tilde \chi_1^0} \le \Omega_{\rm DM}$ in our analysis 
for the case of stable neutralino.
If the equality is satisfied in the above condition, 
the thermal relic neutralino explains the observed 
density of the dark matter,
while if the inequality is unsaturated, 
one needs extra components (e.g.~axions) to explain the dark matter abundance.

It is well known that if the neutralino is Higgsino- or Wino-like,
they can annihilate rapidly into a pair of EW bosons.
Due to the relatively large annihilation rate, they can stay in the thermal bath 
rather longer and the number density at the time of thermal decoupling 
tends to be sufficiently low.
For the pure-Higgsino neutralino, 
the bound $\Omega_{\tilde \chi_1^0} \le \Omega_{\rm DM}$
is translated to $m_{\tilde \chi_1^0} \sim |\mu| \lesssim 1$ TeV \cite{Profumo:2004at}.
For the case of pure-Wino, this condition corresponds to 
$m_{\tilde \chi_1^0} \sim |M_2| \lesssim 3$ TeV \cite{Profumo:2004at}.
These mass scales are significantly higher than those required to fit the $(g-2)_\mu$ anomaly.
Therefore, the neutralino thermal relic abundance does not give strong constraints 
on the SUSY $(g-2)_\mu$ solution when the neutralino is Higgsino- or Wino-like

Unlike Higgsino- and Wino-like neutralinos, the annihilation cross-section is suppressed 
for the Bino-like neutralino, since Bino is gauge-singlet and does not directly couple to EW bosons.
The leading annihilation process comes from a $t$-channel sfermion exchange diagram, 
but the Majorana nature of Bino forces the $s$-wave cross-section 
to be chirally suppressed.  
In order to bring the Bino relic abundance into the acceptable level,
one needs to resort the coannihilation mechanism \cite{Griest:1990kh},
in which the Bino-number changing process, $\xi_{\rm SM} \widetilde B \leftrightarrow \xi^\prime_{\rm SM} \widetilde \eta$,
is operative, where $\xi_{\rm SM}$
and $\xi^{\prime}_{\rm SM}$ are SM particles
and $\widetilde \eta$ is a coannihilation partner.
For this mechanism to work, the coannihilation partner must have
a large annihilation rate and the mass gap between $\widetilde \eta$ and $\widetilde B$ 
must be smaller than $5-10\,\%$ of
the Bino mass.

The Bino-like neutralino appears on the BLR plane and the BHL and BHR planes with $M_1<|\mu|$.
These regions are tightly constrained by the neutralino relic abundance condition.
In the BLR planes, the Bino mass is placed 20 GeV below $m_{\tilde \tau_1}$,
such that the Bino may efficiently coannihilate with $\tilde \tau_1$.

\subsection{The dark matter direct detection}
\label{sec:dmdd}
In the cases with the lightest neutralino $\neut[1]$ being stable,
the thermal relic neutralinos in the present Universe may be detected at the dark matter direct detection (DMDD) experiments.\footnote{See e.g.~\cite{Billard:2021uyg} 
for the current status and future prospects.}
These experiments are located in deep underground and look for 
a delicate signature from nucleons
recoiled by dark matter particles. 
The DMDD experiments have not found
any convincing evidence for such events so far,
which in turn set a strong upper limit (UL) 
on the DM-nucleon scattering cross-section
as a function of the DM mass.
For the neutralino DM, the most stringent 
constraints are placed on the 
spin-independent (SI) cross-section
and we have $\sigma_{\rm SI}^{\tilde \chi_1^0}(m) < \sigma_{\rm SI}^{\rm UL}(m)$,
where $m$ is the neutralino mass.
Even if the neutralino is not the main component of the dark matter, $\Omega_{\tilde \chi_1^0} < \Omega_{\rm DM}$, we still have the constraint 
\begin{equation}
    \sigma_{\rm SI}^{\tilde \chi_1^0}(m) \frac{\Omega_{\tilde \chi_1^0}(m) }{\Omega_{\rm DM}} < \sigma_{\rm SI}^{\rm UL}(m)\,.
    \label{eq:SI_const}
\end{equation}
We generally adopt this condition throughout our analysis for stable neutralinos.

The spin-independent $\tilde \chi_1^0$-nucleon scattering is obtained 
by the $s$-channel squark exchange 
or the $t$-channel Higgs ($h, H, A$) exchange diagrams. 
Since our analysis adopts decoupled squarks and heavy Higgses, the contribution comes entirely from the $h$ exchange process,
which requires the 
$\tilde \chi_1^0$-$\tilde \chi_1^0$-$h$
3-point interaction.
In the MSSM, however,
the pure-Higgsino and pure-gaugino 
do not have this interaction, 
i.e.~the coupling for
the
$\widetilde H$-$\widetilde H$-$h$
and
$\widetilde \lambda$-$\widetilde \lambda$-$h$
are vanishing in the gauge eigenbasis, 
where $\widetilde \lambda$ is the EW gauginos ($\widetilde B$, $\widetilde W^0$),
while the couplings for
$\widetilde H$-$\widetilde \lambda$-$h$
are non-vanishing.
This implies that the SI cross-section is
enhanced when both Higgsino and one of the EW gauginos are light and largely mixed in $\tilde \chi_1^0$.
It happens that this is precisely what is required to have large WHL, BHL and BHR contributions to $(g-2)_\mu$ as discussed 
in the previous section.
As we will see in the next section,
the current constraint from the DMDD experiments 
excludes large parts of 
the WHL, BHL and BHR planes
if the neutralino is stable.

\subsection{The LHC constraints}
\label{sec:lhc}

As discussed in the previous section,
both sleptons and electroweakinos need to be as light as $\Order(100)\GeV$.
More precisely, at least one of $\smuL$ and $\smuR$ needs to be light in any scenarios.
Higgsinos and one of the EW gauginos must be light in the WHL, BHL, and BHR scenarios, whereas only the Bino needs to be light in the BLR scenario.
Since these light particles should be produced at the LHC, the models are constrained by the direct BSM searches by the ATLAS and CMS collaborations at the LHC.

If the LSP is the lightest neutralino $\neut[1]$ and is stable,
production of these SUSY particles
contributes to the lepton plus $\met$ channel.
There are ATLAS and CMS analyses
targeting exactly the same kinematics 
and final states,
and they set the upper limit 
on the cross-section 
for their simplified model
as a function of the masses 
relevant to the kinematics.
Whenever this is the case,
we assess the exclusion of
our scenario at a particular mass point
using the 
methodology outlined in \cite{Kraml:2013mwa, Papucci:2014rja}.
Namely, we confront 
the simplified model cross-section limit 
published by an experimental collaboration
with the production cross-section times the branching ratios to the same final state 
of our model.
This comparison must be performed 
with consistent mass assumptions 
between the
simplified model and our scenario.
The procedure is described 
in detail in section \ref{sec:LHCimpl}.

If $\neut[1]$ is not the LSP, or if it is the LSP but unstable, the LHC signature drastically changes.
In the baryonic RPV scenario studied in section \ref{sec:rpv},
$\neut[1]$ decays into three light-flavour (anti-)quarks,
$\tilde \chi_1^0 \to q q q$ or $\bar q \bar q \bar q$.
Events at the LHC then lose their characteristic $\met$ signature, but can still be searched for in the multi-jet with small $\met$ channel if $\met$ is yielded by neutrinos produced in the upper stream of the SUSY-particle cascade decays.
In the gravitino ($\widetilde G$) LSP scenarios, the signature largely depends 
on the type of the next-to-the-lightest SUSY particle (NLSP).
If the gravitino is almost massless and NLSP is $\tilde \tau_1$,
the large $\met$ in the stable neutralino signature 
is replaced by the moderately large $\met$
and energetic taus produced from the NLSP decays $\tilde \tau_1 \to \tau \widetilde G$.

These unconventional signatures cannot be analysed 
systematically in the simplified model framework 
by ATLAS and CMS, and the methodology mentioned above
cannot be used to assess the exclusion. 
In order to confront these scenarios with the LHC data,
we resort the Monte Carlo simulation.
At each mass point, we generate inclusive signal events with {\tt Pythia 8} \cite{pythia, pythia-susy} for the 8 and 13 TeV LHC.
These event samples are then passed to {\tt CheckMATE 2} \cite{checkmate1,checkmate2,analysismanager},
which emulates various ATLAS and CMS analyses of direct BSM searches 
and computes the expected number of signal events $S$ for all signal regions and compares to the published experimental results. We perform the comparison in the standard CheckMATE approach, the number $S$ is tested against a pre-calculated model independent 95$\%$ CL limit S95.
The detector response is internally simulated with {\tt Delphes 3} \cite{delphes3}
within CheckMATE.
The list of the analyses included 
in our Monte Carlo based study is given in Appendix \ref{app:checkmate}.

\section{Analysis procedure}
\label{sec:procedure}

We use {\tt GM2Calc} \cite{Athron:2015rva} for calculation of $\amu[SUSY]$, which includes the leading two-loop contributions.
For consistency, we take 
physical masses and mixing matrices of non-colored SUSY particles from the {\tt GM2Calc} output. 
These observables are calculated at tree-level
but the corrections to the lepton Yukawa couplings that can be sizable for large $\mu\tan\beta$ \cite{Marchetti:2008hw,Endo:2013lva} are included.
We also include the EW radiative correction 
to the chargino mass, $m_{\tilde \chi_1^\pm}$, using the fitting function provided in \cite{Ibe:2012sx}.
This correction is very small, $m_{\tilde \chi_1^\pm} \sim {\cal O}(100)$ MeV, and 
barely affects $a_\mu^{\rm SUSY}$, whilst 
it may give significant effect 
on the chargino lifetime 
through the mass splitting, $\Delta m_{\pm,0} = m_{\tilde \chi_1^\pm} - m_{\tilde \chi_1^0}$,
when $\tilde \chi_1^0$ is Wino-like.
The decay widths and branching ratios of sparticles are
calculated with {\tt SDecay}\cite{sdecay}
within the {\tt SUSY-HIT} package \cite{susyhit}.
In the compressed mass region, 
$\Delta m_{\pm,0} < 1$ GeV,
we calculate and include the $\tilde \chi_1^+ \to \pi^+ \tilde \chi_1^0$ 
mode since {\tt SDecay} does not support this decay mode.
For the scenarios where $\neut[1]$ is stable, the neutralino relic density and $\sigma^{\neut[1]}\w{SI}$ are calculated
with {\tt MicrOmegas 5.2} \cite{Belanger:2001fz,Belanger:2004yn}.

\section{Stable neutralino}
\label{sec:stable}

In this section, we discuss the eight parameter planes defined in section \ref{sec:planes} (cf.~Table~\ref{tab:planeRPC}) 
assuming $\neut[1]$ is LSP and stable
due to the R-parity conservation.
The constraints on the neutralino relic density (section \ref{sec:relic}) and the DMDD (section \ref{sec:dmdd}) are of importance, while the LHC experiments benefit from the large $\met$ signature discussed in section \ref{sec:lhc}.

\subsection{Implementation of LHC constraints}\label{sec:LHCimpl}


\begin{table}[t!]
\begin{center}
\begin{tabular}{ |l|c|c|c|c| } 
\hline
Analysis & $E/{\rm TeV}$ & ${\cal L}/{\rm fb}^{-1}$ & Simplified Model & Colour \\
\hline
\multirow{2}{7em}{CMS $\ell^+ \ell^-$ \cite{CMS:2020bfa}} & \multirow{2}{1em}{13} & \multirow{2}{2em}{137} &
$\tilde \ell^+ \tilde \ell^- \to (\ell^+ \tilde \chi_1^0) (\ell^- \tilde \chi_1^0)$ & Red \\ 
\cline{4-5}
& & & $\tilde \chi_2^0 \tilde \chi_1^\pm \to 
(Z \tilde \chi_1^0) (W^\pm \tilde \chi_1^0)$ & -- \\ 
\hline
CMS 1$\ell$+2$b$ \cite{CMS:2021lzg} & 13 & 137~\, & $\tilde \chi_2^0 \tilde \chi_1^\pm \to 
(h \tilde \chi_1^0) (W^\pm \tilde \chi_1^0)$ & -- \\
\hline
\multirow{2}{9em}{ATLAS soft-$\ell$ \cite{ATLAS:2019lng} } & \multirow{2}{1em}{13} & \multirow{2}{2em}{139} &
$\tilde \ell^+ \tilde \ell^- \to (\ell^+ \tilde \chi_1^0) (\ell^- \tilde \chi_1^0)$ & Blue \\ 
\cline{4-5}
& & & $\tilde \chi_2^0 \tilde \chi_1^\pm \to 
(Z^* \tilde \chi_1^0) (W^{\pm*} \tilde \chi_1^0)$ & Purple \\ 
\hline
ATLAS DT \cite{ATLAS:2022rme} & 13 & 136~\, & $\chi_1^\pm \to X_{\rm soft} + \tilde \chi_1^0$ (disappearing tracks) & Orange \\
\hline
\end{tabular}
\caption{\label{tb:stable}
\small
The experimental analyses and simplified models used in 
the inspection of a stable neutralino scenario.
The colour indicates to the
region where the simplified model limit
provides 95\% CL exclusion in Figs.~\ref{fig:MSSM_WHL},
\ref{fig:MSSM_BHLR} and \ref{fig:MSSM_BLR}.
The simplified models without designated colours did not give excluded regions
in our parameter planes.
}
\end{center}
\end{table}

In the stable neutralino case, the main LHC constraint comes from the lepton plus $\met$ channel.
In order to constrain our parameter planes 
we use the simplified model limits on the cross-section and the limit on the mass vs lifetime plane from the disappearing track searches for the Wino-like LSP scenario
listed in Table \ref{tb:stable}.

CMS $\ell^+ \ell^-$ (search with
an opposite charge same flavour lepton pair plus $\met$) \cite{CMS:2020bfa}
interpreted their data for the 
$\tilde \ell^+ \tilde \ell^- \to (\ell^+ \tilde \chi_1^0) (\ell^- \tilde \chi_1^0)$ topology 
and provided the cross-section limit 
on the ($m_{\tilde \ell}$ vs $m_{\tilde \chi_1^0}$) plane, assuming a 100\% branching ratio.
We impose this constraint not only on the cross-section of 
$\tilde \ell^+ \tilde \ell^- \to (\ell^+ \tilde \chi_1^0) (\ell^- \tilde \chi_1^0)$ process,
but also on the cross-section times branching ratios\footnote{i.e.
$\sigma(pp \to \tilde \xi \tilde \xi^\prime) \cdot 
{\rm Br(\xi \to \ell^+ \tilde \eta)}\cdot {\rm Br(\xi^\prime \to \ell^- \tilde \eta^\prime)}$ 
} of
$\tilde \xi \tilde \xi^\prime \to (\ell^+ \tilde \eta) (\ell^- \tilde \eta^\prime)$
if $m_{\tilde \eta^{(\prime)}} - m_{\tilde \chi_1^0} \le 20$ GeV,
where $\tilde \xi^{(\prime)} = (\tilde \ell \,\,{\rm or}\,\, \tilde \nu)$
and $\tilde \eta^{(\prime)} = (\tilde \chi^\pm_1 \,\,{\rm or}\,\, \tilde \chi^0_{1/2})$ in our model point.
This is because with this condition the decay products of
$\eta^{(\prime)} \to \tilde \chi_1^0$ are very soft and do not alter
the signal efficiency significantly.
{When $\tilde \chi_1^0$ is Wino- or Higgsino-like,
$\chi^\pm_1$ (as well as $\chi^0_{2}$ in the Higgsino-like case)
is quasi mass degenerate with $\chi_1^0$
and the above condition is generally satisfied.}
The corresponding cross-section limit is read from
the original ($m_{\tilde \ell}$ vs $m_{\tilde \chi_1^0}$) plane
by mapping
$(m_{\tilde \xi}, m_{\tilde \eta}) \to (m_{\tilde \ell}, m_{\tilde \chi_1^0})$.

In the same philosophy, we also use this simplified model limit 
to constrain the cross-section times branching ratios of 
$\tilde \eta \tilde \eta^\prime \to (\ell^+ \tilde \xi) (\ell^- \tilde \xi^\prime)$ process
when $m_{\tilde \xi^{(\prime)}} - m_{\tilde \chi_1^0} \le 20$ GeV,
where $\tilde \xi^{(\prime)} = (\tilde \ell \,\,{\rm or}\,\, \tilde \nu)$
and $\tilde \eta^{(\prime)} = (\tilde \chi^\pm_{1/2} \,\,{\rm or}\,\, \tilde \chi^0_{2/3})$.
The mass degeneracy condition here is often satisfied 
since the slepton mass is placed 20 GeV above the LSP mass parameter in some of our parameter planes.

The CMS $\ell^+ \ell^-$ analysis \cite{CMS:2020bfa}
also interpreted their result 
for the
$\tilde \chi_2^0 \tilde \chi_1^\pm \to 
(Z \tilde \chi_1^0) (W^\pm \tilde \chi_1^0)$
topology and published the signal cross-section limit 
on the $(m_{\tilde \chi_2^0}, m_{\tilde \chi_1^0})$ plane 
assuming $m_{\tilde \chi_1^\pm} = m_{\tilde \chi_2^0}$.
Similarly, 
CMS 1$\ell$+2$b$ \cite{CMS:2021lzg}
provided the signal cross-section limit 
on the same mass plane 
for the
$\tilde \chi_2^0 \tilde \chi_1^\pm \to 
(h \tilde \chi_1^0) (W^\pm \tilde \chi_1^0)$
topology.
We use these simplified model limits to constrain 
the Wino and Higgsino productions, followed by
the decays into EW bosons,
e.g.~$pp \to \widetilde W \widetilde W, \widetilde H \widetilde H$
with $\widetilde W \widetilde W \to W Z(h)$
and
$\widetilde H \widetilde H \to W Z(h)$. 
To do this practically, we impose the limit on cross-section times
branching ratios for
$\tilde \eta \tilde \eta^\prime \to (h \tilde \xi) (W^\pm \tilde \xi^\prime)$ 
if $m_{\tilde \xi^{(\prime)}} - m_{\tilde \chi_1^0} \le 20$ GeV,
$m_{\tilde \eta^{(\prime)}} - m_{\tilde \eta} \le 20$ GeV
and 
$m_{\tilde \eta^{(\prime)}} - m_{\tilde \xi^{(\prime)}} \ge 125$ GeV,
where $\tilde \xi^{\prime} = (\tilde \chi^\pm_{1/2} \,\,{\rm or}\,\, \tilde \chi^0_{2/3})$
and $\tilde \eta^{\prime} = (\tilde \chi^\pm_1 \,\,{\rm or}\,\, \tilde \chi^0_{1/2})$.
As we will see below, we however do not find any region on our parameter planes where 
these simplified model limits give exclusion. 

ATLAS soft-$\ell$ \cite{ATLAS:2019lng} analysis
probed the compressed mass region ($1\,{\rm GeV} \lesssim \Delta m \lesssim 50\,{\rm GeV}$)
with soft leptons.
They interpreted their result 
for the
$\tilde \ell \tilde \ell \to (\ell \tilde \chi_1^0) (\ell \tilde \chi_1^0)$
and 
$\tilde \chi_2^0 \tilde \chi_1^\pm \to ( \ell^+ \ell^- \tilde \chi_1^0) (q \bar q \tilde \chi_1^0)$
with off-shell decays $Z^* \to \ell^+ \ell^-$ and $W^* \to q \bar q$ and published the signal cross-section limit 
on the 
($m_{\tilde \ell}$ vs $\Delta m(\tilde \ell, \tilde \chi_1^0)$)
and ($m_{\tilde \chi_2^0}$ vs $\Delta m(\tilde \chi_2^0, \tilde \chi_1^0)$)
assuming $m_{\tilde \chi_1^\pm} = m_{\tilde \chi_2^0}$, respectively. 
We apply these limits on our model points
in the same manner as described above for  
CMS $\ell^+ \ell^-$ analysis.

When $\chi_1^0$ is Wino-like,
the mass difference $\Delta m_{\pm, 0} = m_{\tilde \chi_1^\pm} - m_{\tilde \chi_1^0}$ becomes very small and the decay width of $\tilde \chi_1^\pm$ is phase-space suppressed.
In this region, the lifetime of $\tilde \chi_1^\pm$ can become as  large as $c \tau_{\tilde \chi_1^\pm} \sim {\cal O}(1)$\,cm.
Since $\tilde \chi_1^\pm$ decays
inside of the tracking system 
into unobservably soft particles, $X_{\rm soft}$, plus invisible $\tilde \chi_1^0$, it can give a disappearing track (DT) signature. 
ATLAS 
and CMS 
have been looking for this type of signature \cite{ATLAS:2022rme,CMS:2020atg}.
To constrain our parameter space 
with their experimental results,
we use the 95\% CL limit 
on the ($m_{\tilde \chi_1^\pm}, \tau_{\tilde \chi_1^\pm}$) plane 
provided in the ATLAS DT analysis \cite{ATLAS:2022rme}, which is based on the 13 TeV data with ${\cal L} = 136$ fb$^{-1}$.
The model point is excluded 
if the calculated $m_{\tilde \chi_1^\pm}$ and $\tau_{\tilde \chi_1^\pm}$ lie inside the excluded region in the ATLAS
($m_{\tilde \chi_1^\pm}, \tau_{\tilde \chi_1^\pm}$) plane.

\subsection{Results}
\label{sec:mssm_result}

\begin{figure}[t!]
\centering
      \includegraphics[width=0.44\textwidth]{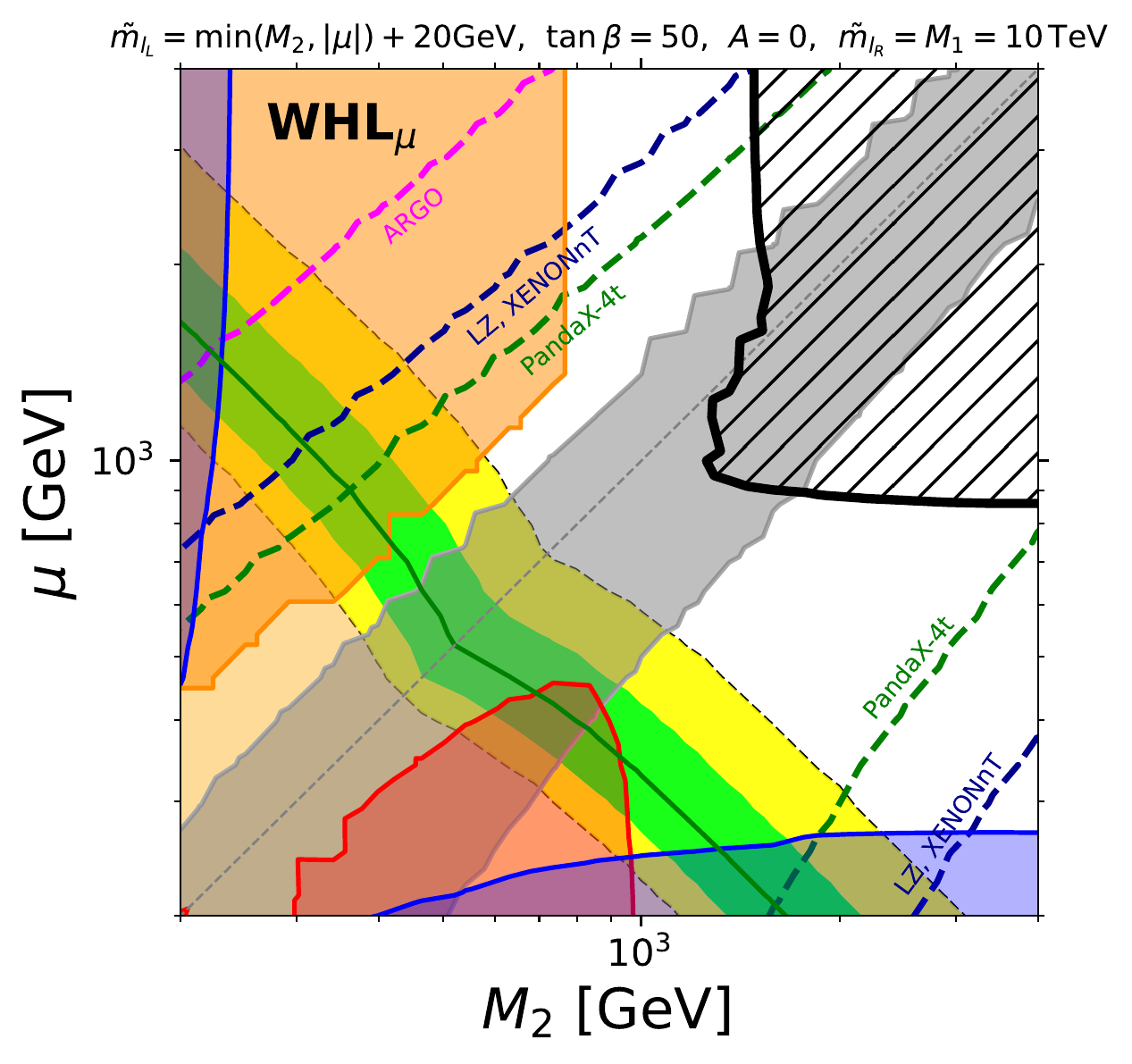}
      \includegraphics[width=0.45\textwidth]{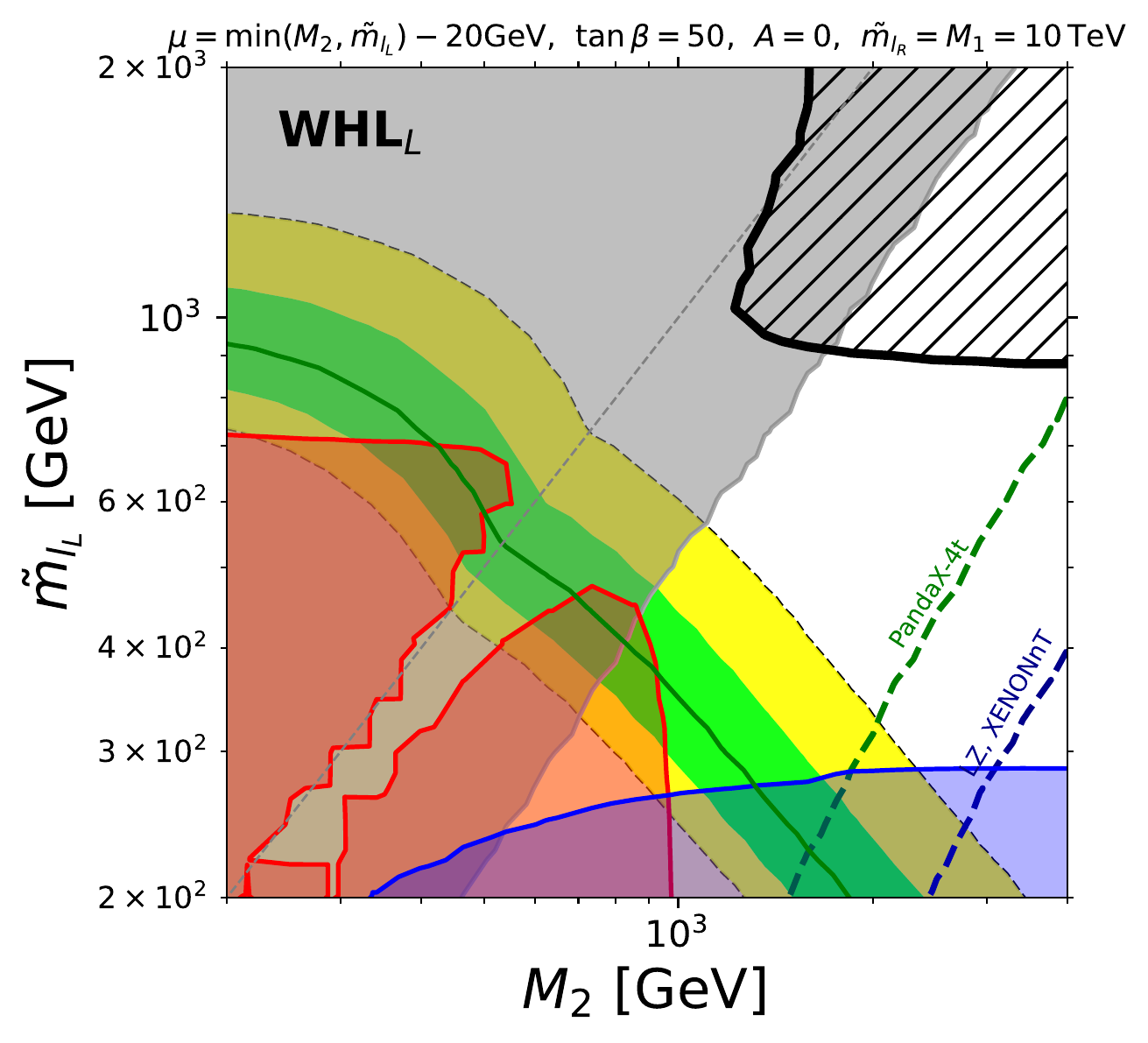}
\caption{\label{fig:MSSM_WHL} \small
Results for MSSM ${\bf WHL}_\mu$ (left) and ${\bf WHL}_L$ (right) planes. 
Region of parameter space allowed by the latest $a_\mu$ experimental results \cite{Muong-2:2021ojo} is depicted with green and yellow bands, corresponding to one and two sigma agreement respectively. The bottom left region shaded by the dull yellow colour corresponds to $a_\mu$ bigger than experimental value by more than 2 sigma.
Hatched region is excluded by dark matter abundance criterion. Blue and red shaded regions are excluded by LHC constraints. Light grey shaded area is excluded by Xenon1T experiment. Expected exclusion range by future DDMD experiments is also shown with appropriately labelled contours. 
}
\end{figure}

In this subsection 
we identify the parameter region
that can explain the $(g-2)_\mu$ anomaly
and confront it with 
relevant
phenomenological constraints.

We begin with 
the ${\bf WHL}_\mu$ (left)
and ${\bf WHL}_L$ (right) planes
shown in Fig.~\ref{fig:MSSM_WHL}.
Here, and also in the other plots below,
the area highlighted with green (yellow) 
corresponds to the region where
the predicted value of $a_\mu$
agrees with the measurement
within the 1 (2) $\sigma$ accuracy,
while in the bottom left region shaded by the dull yellow colour, the SUSY contribution to $a_\mu$ is too large and the model is disfavoured
by more than 2 $\sigma$.

The hatched region surrounded by the black curve indicates the region excluded by the neutralino overproduction, $\Omega_{\tilde \chi_1^0} \ge \Omega_{\rm DM}$.
At the boundary of this region (i.e.~on the black curve), the relic neutrino can explain the full amount of observed dark matter density.
One can see that in the WHL planes
the $(g-2)_\mu$ region 
is located outside of the 
black hatched region, implying 
that extra components, such as axions,
are needed to explain the dark matter.

The region shaded by light grey 
corresponds to the excluded 
area due to the upper bound on the spin-independent 
DM-nucleon scattering cross-section, $\sigma_{\rm SI}^{\rm UL}$,
set by the latest XENON1T experiment \cite{XENON:2018voc}. 
More precisely, the condition Eq.~\eqref{eq:SI_const} is violated 
in this area.
As discussed in the previous section,
$\sigma_{\rm SI}^{\tilde \chi_1^0}$
is enhanced when $\tilde \chi_1^0$
has both Higgsino and gaugino components.
In the WHL scenario, this happens 
when $\mu \sim M_2$.
In the ${\bf WHL}_\mu$ plane (Fig.~\ref{fig:MSSM_WHL} left),
we can explicitly see 
the region with $\mu \sim M_2$
is excluded by the XENON1T constraint.
In the ${\bf WHL}_L$ case (Fig.~\ref{fig:MSSM_WHL} right),
a half of the plane 
with $\tilde m_{l_L} \gtrsim M_2$
is excluded 
since $\mu \sim M_2$ is realised 
with the condition
$\mu = {\rm min}(M_2, \tilde m_{l_L}) - 20\,{\rm GeV}$.

The regions that can be probed by the future DMDD experiments are also shown in
Fig.~\ref{fig:MSSM_WHL}.
In the ${\bf WHL}_\mu$ plane,
the area between two green (blue) dashed lines are sensitive to  
PandaX-4t \cite{PandaX:2018wtu}
(LZ \cite{LUX-ZEPLIN:2018poe}
and XENONnT \cite{XENON:2020kmp}).
Also, the area below the magenta dashed line can be explored by ARGO \cite{argo}.
{In the ${\bf WHL}_L$ plane,
the region above 
the green and blue dashed lines 
can be probed by the PandaX-4t and LZ/XENONnT, respectively,
and the entire region in the plane will be explored by ARGO.}  
One can see that almost entire 
$(g-2)_\mu$ region in the
${\bf WHL}_L$ plane will be tested 
in the future DMDD experiments.

Constraints 
from direct BSM searches at the LHC
are also visible in Fig.~\ref{fig:MSSM_WHL}.
The red shaded area
corresponds to the 
95\,\% CL excluded region 
estimated using 
the slepton simplified model limit 
published in the CMS $\ell^+ \ell^-$ 
analysis \cite{CMS:2020bfa},
while the blue shaded area 
indicates the 95\,\% CL exclusion obtained 
from the published limit for 
a compressed slepton scenario 
given in the ATLAS soft-$\ell$
analysis \cite{ATLAS:2019lng}
(see Table~\ref{tb:stable}).
In the ${\bf WHL}_\mu$ plane
we also see the exclusion from 
the ATLAS DT search \cite{ATLAS:2022rme},
which is shaded by the orange colour.
We find that only these three constraints provide 95\,\% CL exclusions on the WHL parameter planes,
although other potentially relevant constraints listed in Table~\ref{tb:stable}
are also included in our analysis.

In the ${\bf WHL}_\mu$ plane
we see that 
the region with $\mu \lesssim M_2 \lesssim 1$\,TeV
is excluded by the CMS $\ell^+ \ell^-$ analysis.
In this region, the main SUSY processes,  
which contribute to the dilepton plus $\met$ signal region
targeted in this analysis,
are $pp \to \widetilde W^{+,0} \widetilde W^{-,0}$
followed by e.g.~$\widetilde W^{\pm} \to \ell^\pm \tilde \nu$ and
$\widetilde W^{0} \to \ell^\pm \tilde \ell^{\mp}$.
Note that since $\tilde m_{l_L} = \mu + 20$ GeV in this region,
the decay products of the slepton/sneutrino into the Higgsino-like LSP are too soft 
and may be neglected for this analysis.
On the other side of the diagonal dotted line, we do not see the corresponding excluded region with $M_2 \lesssim \mu$
for the Higgsino production,
$pp \to \widetilde H^{+,0} \widetilde H^{-,0}$.
This is because the Higgsino production cross-section is smaller than the Wino
and the main decay is given by
$\widetilde H^{\pm,0} \to h \widetilde W^{\pm,0}$, producing leptons much less often.

Exclusion from the disappearing track signature from long-lived Winos is visible in the top left corner of the ${\bf WHL}_{\mu}$ plane with $\mu \gg M_2$.  In this region, $\Delta m_{\pm, 0} = m_{\tilde \chi_1^\pm} - m_{\tilde \chi_1^0}$, becomes ${\cal O}(100)$ MeV and $c \tau_{\tilde \chi_1^\pm} \sim {\cal O}(1)$\,cm.
We see that the current ATLAS DT analysis 
\cite{ATLAS:2022rme} excludes 
the Wino-like LSP up to $M_2 \sim 760$ GeV.
Exclusion from the disappearing track signature does not show up in the ${\bf WHL}_L$ plane, 
since the $\tilde \chi_1^0$ is Higgsino-like across this plane.

The exclusion from 
the ATLAS soft-$\ell$ analysis
can be seen in the regions 
with (i) $\mu \lesssim 250$ GeV
and
(ii) $M_2 \lesssim 220$ GeV,
excluding the $M_2 \sim \mu$ region.
In these areas, the relevant SUSY process is 
$\tilde \xi \tilde \xi^\prime \to (\ell^+ \tilde \eta) (\ell^- \tilde \eta^\prime)$
where $\tilde \xi^{(\prime)} = (\tilde \ell \,\,{\rm or}\,\, \tilde \nu)$
is a left-handed slepton doublet 
and $\tilde \eta^{(\prime)} = (\tilde \chi^\pm_1 \,\,{\rm or}\,\, \tilde \chi^0_{1/2})$ 
is dominantly Higgsino- and Wino-like 
in regions (i) and (ii), respectively.
Although the typical mass difference between 
$\tilde \xi$ and $\tilde \eta$ is small
($\sim 20$ GeV), 
the ATLAS soft-$\ell$ analysis
can detect the signal by exploiting soft leptons.

Taking both the LHC and DM constraints into account, one can find two $(g-2)_\mu$ favoured regions that are phenomenologically allowed in 
the ${\bf WHL}_\mu$ plane;
(a) $250\,{\rm GeV} \lesssim M_2 \sim \tilde m_{l_L} \lesssim 500\,{\rm GeV}$,
$500\,{\rm GeV} \lesssim \mu \lesssim 800\,{\rm GeV}$
and
(b)
$1\,{\rm TeV} \lesssim M_2 \lesssim 2\,{\rm TeV}$,
$250\,{\rm GeV} \lesssim \mu \sim \tilde m_{l_L} \lesssim 500\,{\rm GeV}$.
Both of these regions will 
be probed in future by
the next generation DMDD experiments.

In the ${\bf WHL}_L$ plane
two regions are excluded by the 
CMS $\ell^+ \ell^-$ analysis.
One region with $M_2 > \tilde m_{l_L}$
is identical to the corresponding 
excluded region in 
the ${\bf WHL}_\mu$ plane,
which has already been discussed above.
The second excluded region has
$M_2 < \tilde m_{l_L} \lesssim 700\,{\rm GeV}$ and $\mu = M_2 - 20$\,GeV.
In this region, the main SUSY signature 
is
$\tilde \xi \tilde \xi^\prime \to (\ell^+ \tilde \eta) (\ell^- \tilde \eta^\prime)$
where $\tilde \xi^{(\prime)} = (\tilde \ell \,\,{\rm or}\,\, \tilde \nu)$
and $\tilde \eta^{(\prime)} = \widetilde W^{\pm,0}$.
The blue shaded region 
is excluded by the ATLAS soft-$\ell$,
which is identical to the 
corresponding region in the 
the ${\bf WHL}_\mu$ plane discussed above.
In the ${\bf WHL}_L$ plane
the allowed $(g-2)_\mu$ region 
appear around $M_2 \sim 1.5$\,TeV, $\mu \sim 350$\,GeV, which has been already found in the ${\bf WHL}_\mu$ plane,
and the entire $(g-2)_\mu$ region
will be probed in the early stage of next generation DMDD experiments.

\begin{figure}[t!]
\centering
      \includegraphics[width=0.45\textwidth]{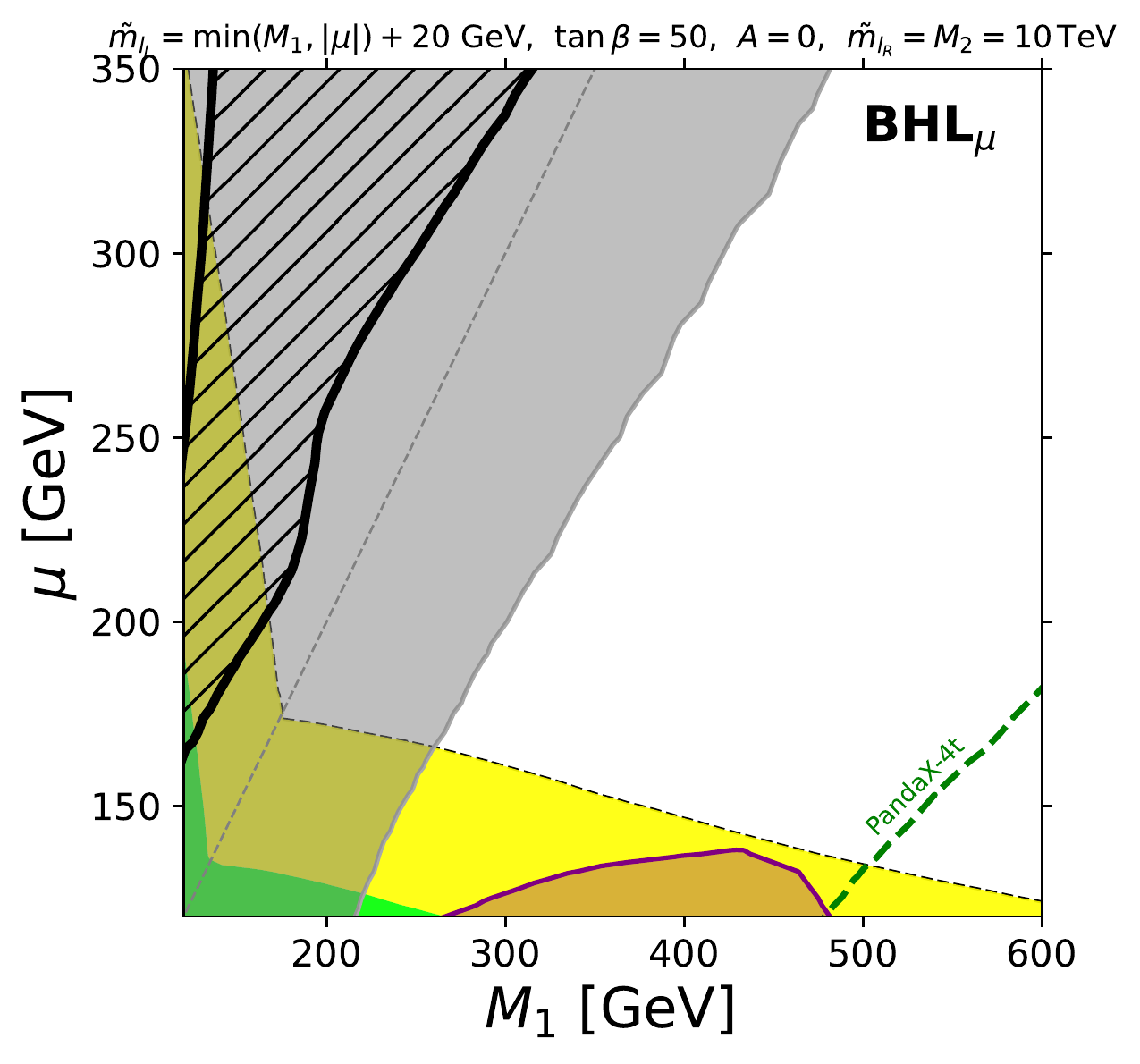}
      \includegraphics[width=0.45\textwidth]{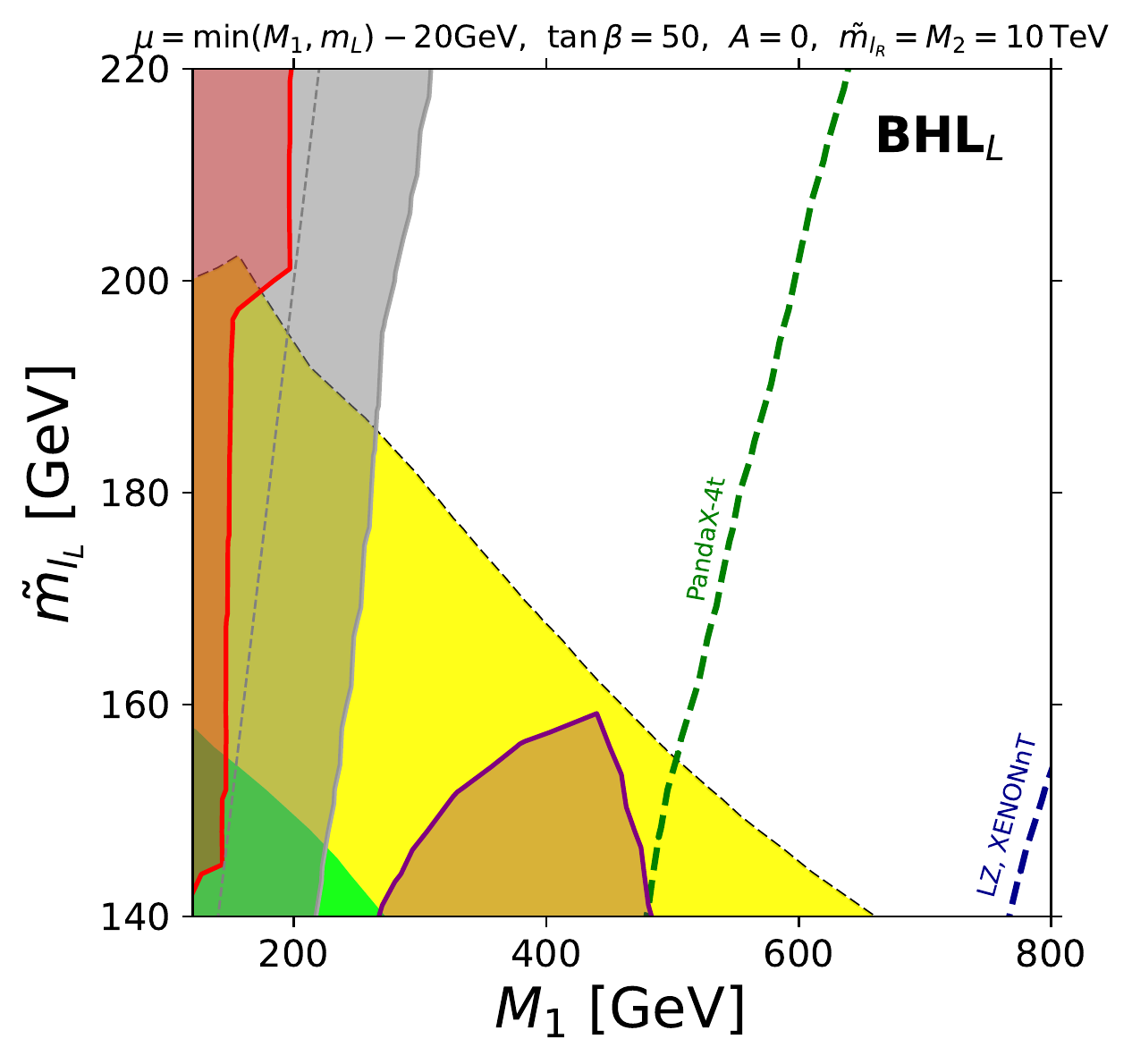}
      \includegraphics[width=0.45\textwidth]{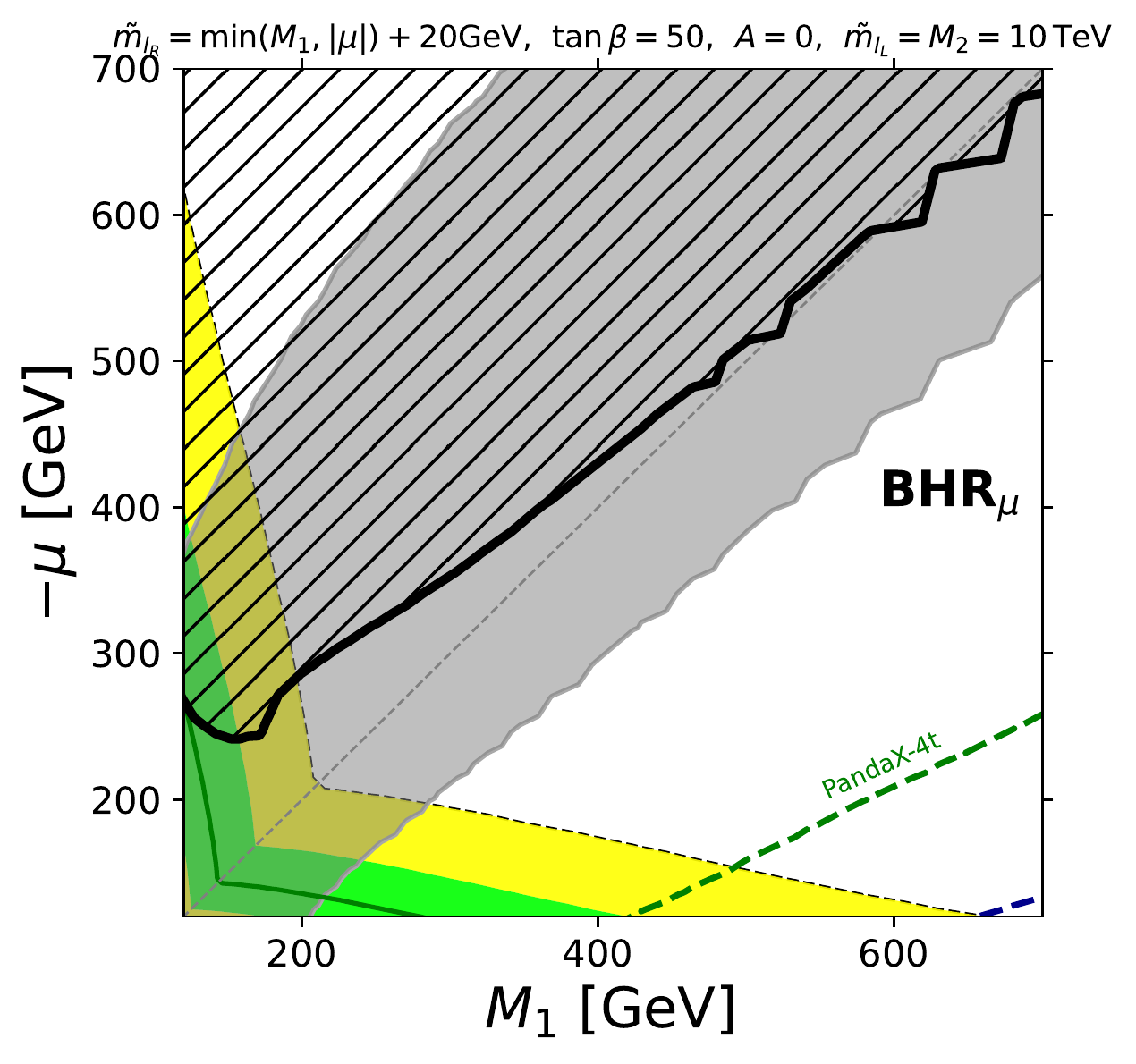}
      \includegraphics[width=0.45\textwidth]{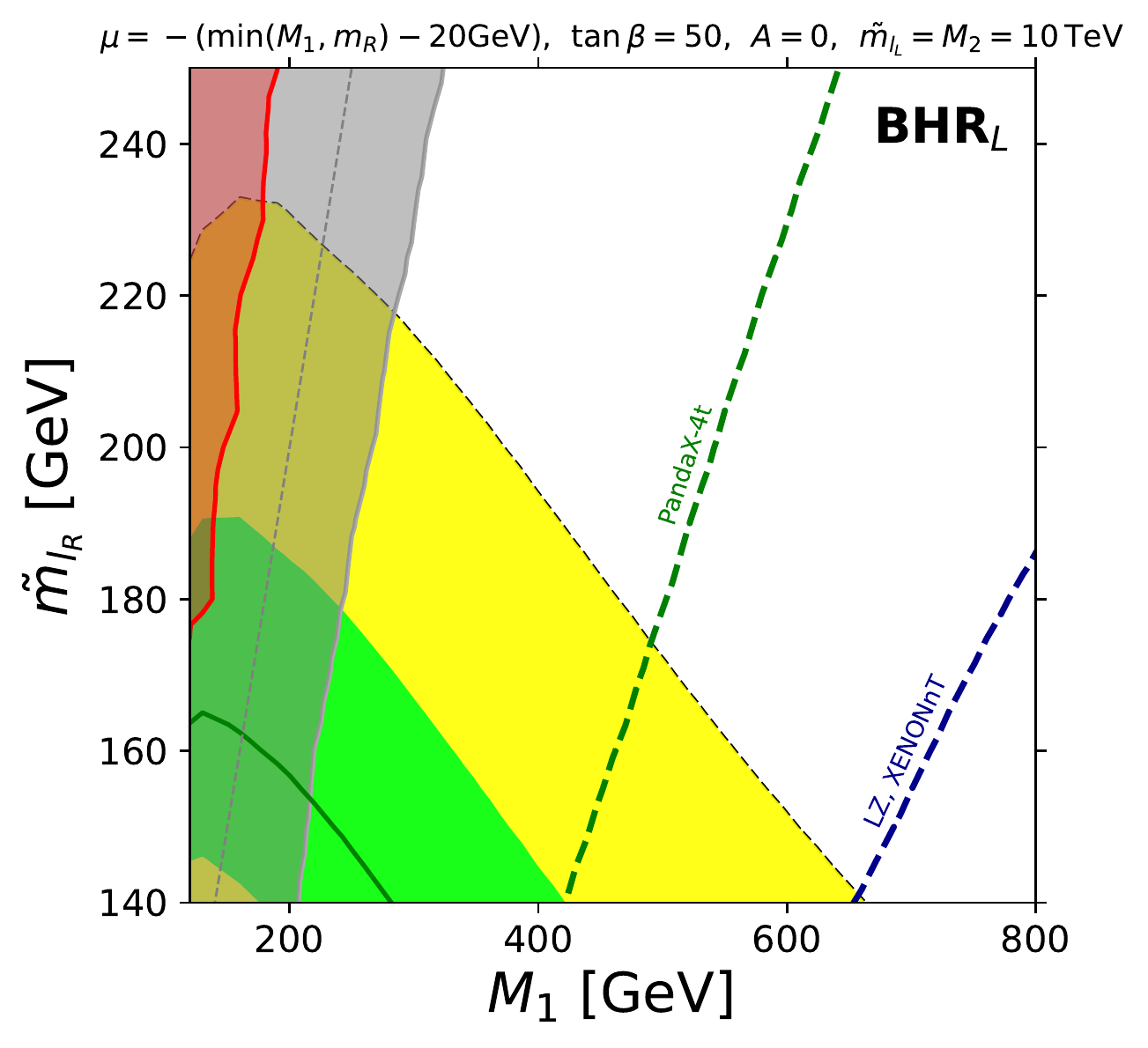}
\caption{\label{fig:MSSM_BHLR} \small
Results for MSSM ${\bf BHL}_\mu$ (upper left),  ${\bf BHL}_L$ (upper right),
${\bf BHR}_\mu$ (lower left) and {${\bf BHR}_L$ } (lower right) planes. 
Region of parameter space allowed by the latest $a_\mu$ experimental results \cite{Muong-2:2021ojo} is depicted with green and yellow bands, corresponding to one and two sigma agreement respectively.
Hatched region is excluded by dark matter abundance criterion. Purple and red shaded regions are excluded by LHC constraints. Light grey shaded area is excluded by Xenon1T experiment. Expected exclusion range by future DDMD experiments is also shown with appropriately labelled contours. 
}
\end{figure}

\medskip

We show the results of our analysis
for the BHL
and BHR scenarios
in the top and bottom panels 
of Fig.~\ref{fig:MSSM_BHLR}, respectively.
Note that we take the sign of $\mu$
to be negative for the BHR planes
in order to fit the $(g-2)_\mu$ anomaly
due to the minus sign in Eq.~\eqref{eq:aBHR}.
{First, we see in Fig.~\ref{fig:MSSM_BHLR} that
the SUSY masses required to fit 
the $(g-2)_\mu$ anomaly
are much smaller 
for the BHL and BHR scenarios 
than in the WHL case.
This is a reflection of the fact 
that the contributions from
the BHL and BHR diagrams are an order of magnitude smaller than that from the WHL diagram if all SUSY masses are taken to be equal. }
We also see in both ${\bf BHL}_\mu$ and 
${\bf BHR}_\mu$ planes
that
regions with $M_1 < |\mu|$
are severely constrained by the
neutralino overproduction.
This is because the $\tilde \chi_1^0$ 
is Bino-like in this region
and the relic abundance enhances 
due to a small annihilation cross-section for Binos.
In these two planes
one can also see that
regions
with $|\mu| \sim M_1$
are disallowed 
by the XENON1T constraint.
This can be understood since
the Higgsino-Bino mixing in $\chi_1^0$
becomes large if $|\mu| \sim M_1$,
which enhances 
the spin-independent neutralino-nucleon scattering cross-section, as discussed 
in the previous section.
XENON1T also excludes 
the $M_1 \sim \tilde m_{l_L}$ and 
$M_1 \sim \tilde m_{l_R}$ regions
in the ${\bf BHL}_L$ (top right)
and ${\bf BHR}_L$ (bottom right)
planes, respectively.
In these regions 
$|\mu| \sim M_1$ is realised 
due to the condition for the $\mu$-parameter.
We see that PandaX-4t (green dashed)
can explore the entire 1 $\sigma$
$(g-2)_\mu$ region in both the
BHL and BHR scenarios.
Also, the entire
2 $\sigma$ regions will be probed 
by the LZ and XENONnT (blue dashed) experiments.

Taking DM constraints into account,
only 
regions with $M_1 \gtrsim |\mu| + 100$\,GeV are allowed. 
In those regions the LHC constraints 
are rather weak.
This is because only Bino has a significant mass gap from the $\tilde \chi_1^0$, though the Bino cross-section at the LHC is minuscule when the squarks are decoupled.
The only LHC constraint that is significant in those areas 
comes from the limit imposed on the
compressed electroweakino simplified model provided by the ATLAS soft-$\ell$
analysis (see Table~\ref{tb:stable}).
The excluded region from this constraint 
only appear in 
the ${\bf BHL}_\mu$
and ${\bf BHL}_L$ planes,
which are shaded with purple.
The regions shown in these two plots are however identical.

\medskip

We now
show our result for
the ${\bf BLR}_{50}$ ($\tan\beta = 50$)
and ${\bf BLR}_{10}$ ($\tan\beta = 10$) planes
in the top and bottom panels,
respectively,
in Fig.~\ref{fig:MSSM_BLR}.
\begin{figure}[t!]
\centering
      \includegraphics[width=0.45\textwidth]{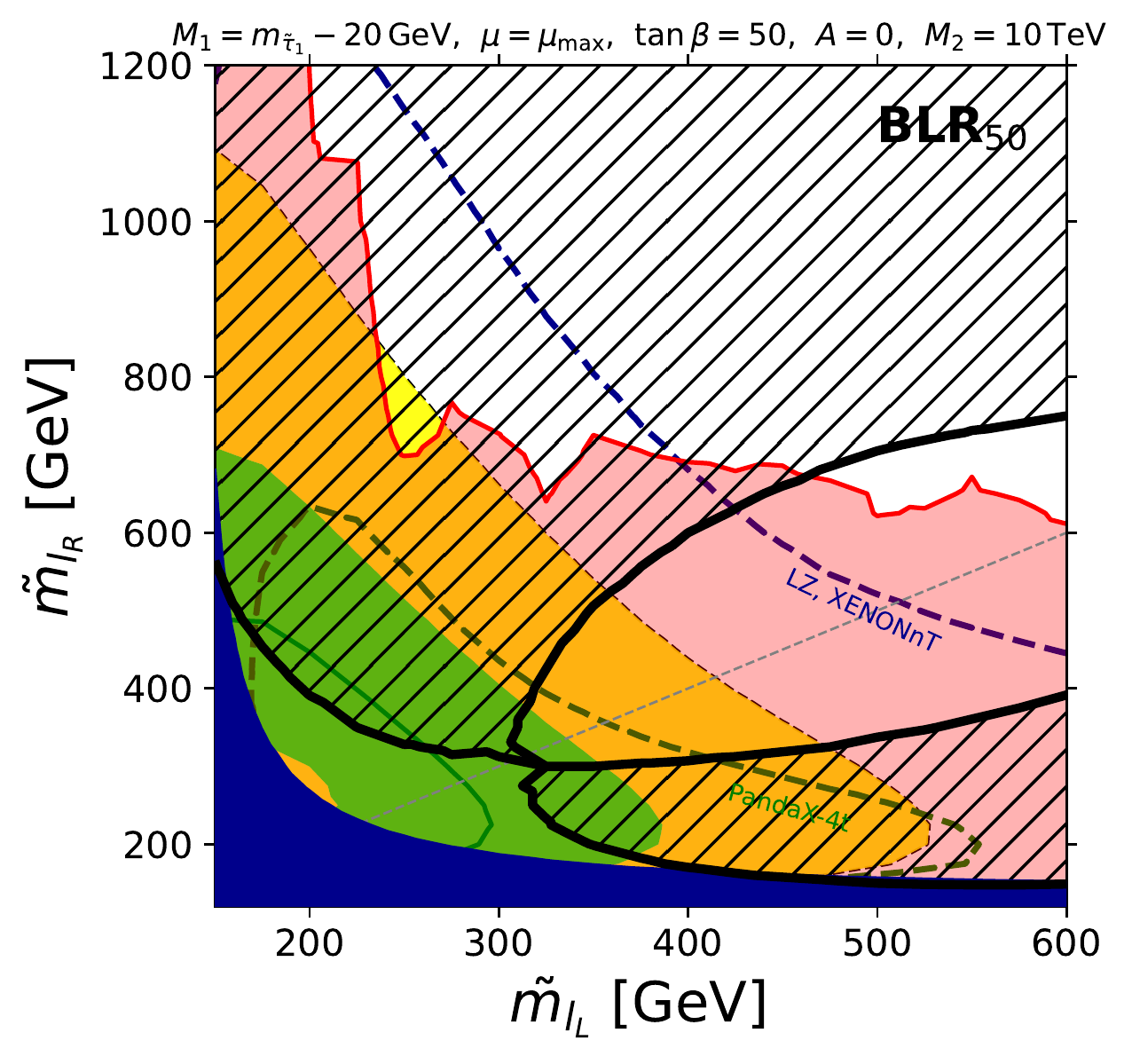}
      \includegraphics[width=0.45\textwidth]{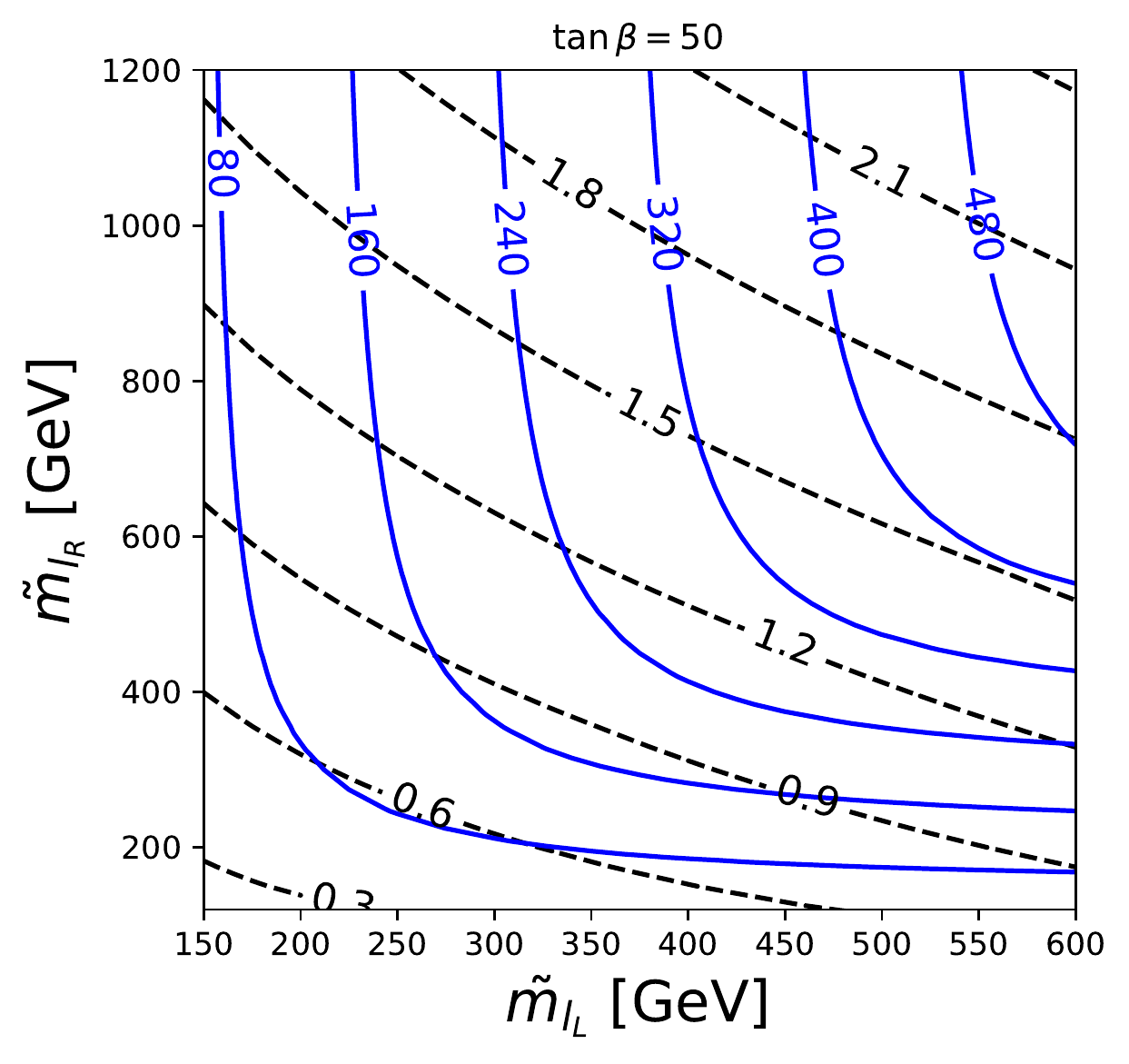}
      \includegraphics[width=0.45\textwidth]{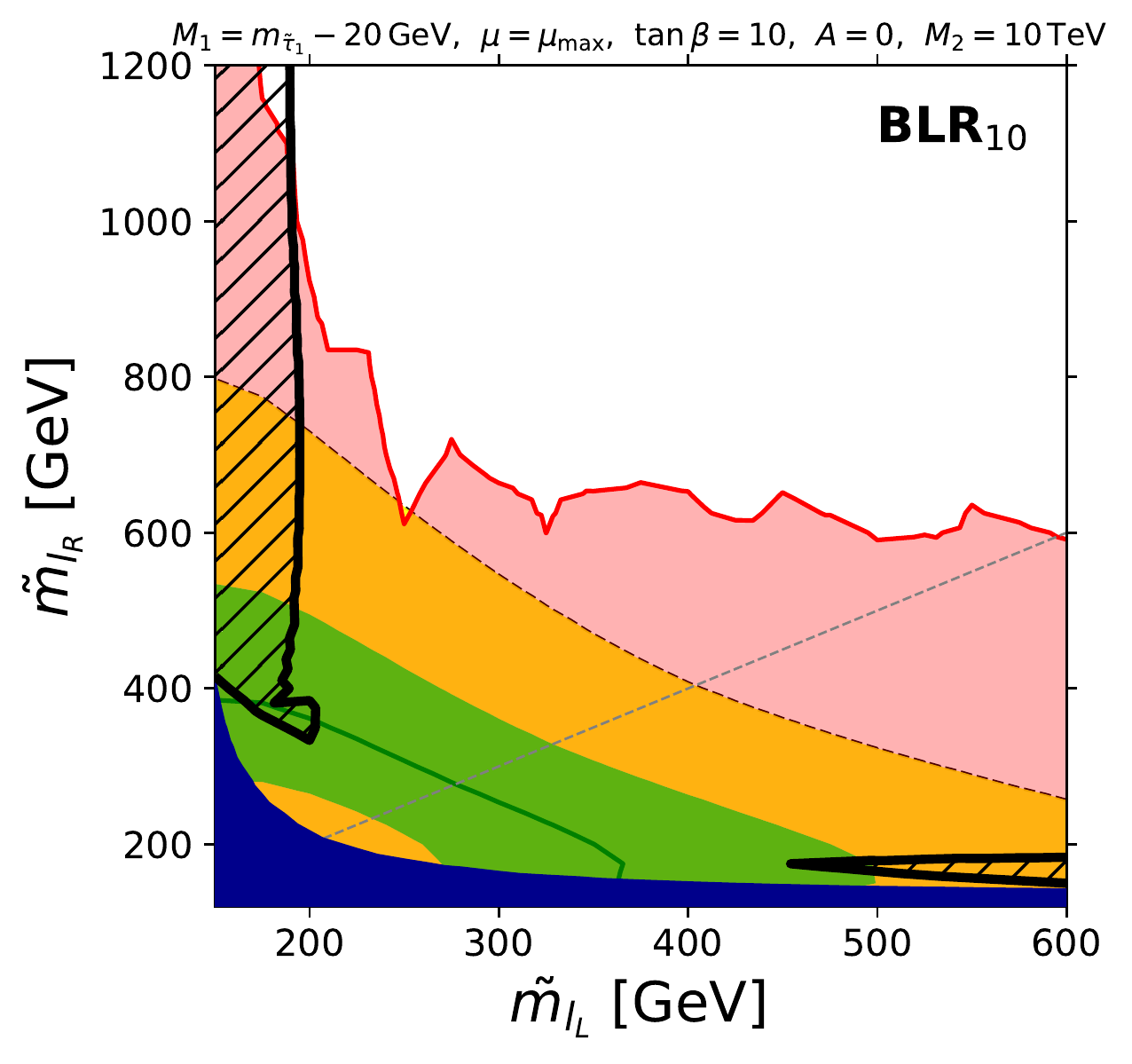}
      \includegraphics[width=0.45\textwidth]{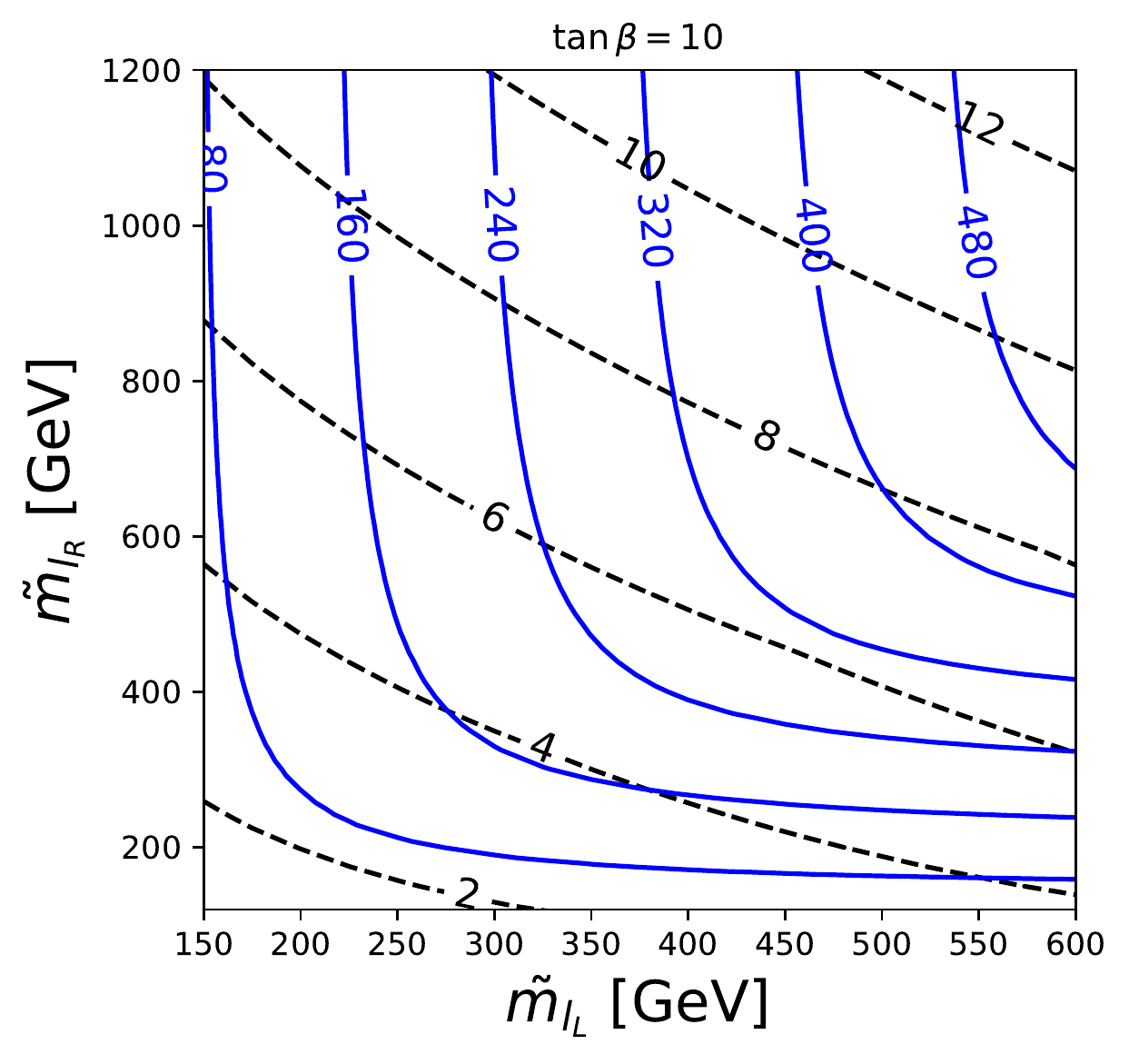}
\caption{\label{fig:MSSM_BLR} \small 
Results for MSSM ${\bf BLR}_{50}$ (upper row) and ${\bf BLR}_{10}$ (lower row) planes.
Plots on the left hand side show parameter regions and appropriate experimental constraints. 
Region of parameter space allowed by the latest $a_\mu$ experimental results \cite{Muong-2:2021ojo} is depicted with green and yellow bands, corresponding to one and two sigma agreement respectively.
Hatched region is excluded by dark matter abundance criterion. Red shaded region is excluded by LHC constraints. Dark blue shaded area is excluded by LEP stau mass bound \cite{Zyla:2020zbs}. Expected exclusion range by future DDMD experiments is also shown with appropriately labelled contours. 
Plots on the right hand side of the figure show Bino mass (blue) and $\mu$ parameter (dashed black) values, which are implicitly fixed by other parameters.
}
\end{figure}
As mentioned in subsection \ref{sec:planes},
in the BLR scenario the $\mu$-parameter is set to
the maximally allowed value 
from the vacuum stability constraint \cite{Kitahara:2013lfa,Endo:2013lva}.
Large $\mu$ is preferred
to ensure a large L-R smuon mixing necessary for the BLR scenario.
In this regime, however, 
one of the stau mass eigenstates
may become very light.
In order to avoid stau LSP, 
the Bino mass, $M_1$, is placed at 20 GeV below the lighter stau mass, $m_{\tilde \tau_1}$.
In the left panels,
the dark blue region
is excluded by the 
LEP stau mass bound 81.9 GeV
\cite{Zyla:2020zbs}.

In the previous section,
we have argued that
the relic neutralinos tend to be overproduced 
in the BLR scenario 
since 
the $\tilde \chi_1^0$
is Bino-like.
As expected, a large fraction of 
the parameter plane is excluded 
by this constraint as shown in the
hatched black regions.
In the region allowed by 
the $\Omega_{\tilde \chi_1^0}$
constraint, the stau-coannihilation
mechanism is operative.
We see that this constraint is milder for 
the ${\bf BLR}_{10}$ plane with $\tan \beta = 10$.
In this plane, $|\mu|$ is significantly larger than
that in ${\bf BLR}_{50}$, and the neutralino mass
is therefore slightly heavier.
Since we fix the stau-Bino mass difference, 
this results in smaller $\Delta m/m$ 
and more effective coannihilation.  

Since $\mu$ and $M_1$ are fixed implicitly at each point in the BLR planes, we show, in the right panels of Fig.~\ref{fig:MSSM_BLR},
contours of $\mu$ (black dashed in TeV) and $M_1$ (blue solid in GeV).
As can be seen, $M_1$, which is correlated with $m_{\tilde \tau_1}$,
is insensitive to $\tan \beta$.
This is because both $m_{\tilde \tau_1}$ and the vacuum tunnelling rate 
depend on the  product
$\mu \cdot \tan \beta$ (not on $\mu$ and $\tan \beta$ independently)
and they remain unchanged when
changing $\tan \beta$ and $\mu$ in such a way
that their product $\mu \tan \beta$ remains constant.

This also explains that
from $\tan \beta = 50$ to 10,
the value of $\mu$ is changed 
by a factor of 5,
as can be checked by comparing 
the top right and bottom right plots in Fig.~\ref{fig:MSSM_BLR}.

We finally comment on the LHC constraint on the BLR scenario.
As can be seen, the whole $(g-2)_\mu$ region is excluded 
by the limit coming from 
the slepton simplified model bounds in CMS $\ell^+ \ell^-$ analysis.
Although the Higgsino production can give a non-negligible contribution 
in the bottom left region of the ${\bf BLR}_{50}$ plane,
the main process contributing to the exclusion is
$pp \to \tilde \ell^+ \tilde \ell^-$ followed by
$\tilde \ell^{\pm} \to \ell^{\pm} \tilde \chi_1^0$,
which is identical to the process
targeted in CMS $\ell^+ \ell^-$ analysis.

{It is remarkable that the entire $(g-2)_\mu$ region is excluded in our BLR planes.
We note that this strong conclusion is based on our assumption of universal slepton masses.
In the BLR planes $\mu \tan \beta$ is taken to be at the maximum value allowed by the vacuum stability constraint.
For such a high value of $\mu \tan \beta$,
the lighter stau mass eigenvalue $m_{\tilde \tau_1}$ becomes significantly smaller than the masses of the other sleptons $\tilde \mu/\tilde e$ when the universal mass assumption is adopted.
This leads to a large mass gap between 
$\tilde \mu/\tilde e$ and Bino, since the Bino mass is set 20 GeV below $m_{\tilde \tau_1}$ to avoid stau being the LSP.
With such a large mass gap,
the slepton pair production leads to
the final state with two high $p_T$ leptons plus large $\met$
and the preferred $(g-2)_\mu$ region
with light sleptons are 
excluded by the CMS $\ell^+ \ell^-$ analysis.
In many popular scenarios 
(e.g.\ constrained MSSM, gauge mediation, anomaly mediation, mirage mediation, etc.) 
in which
the universal slepton masses are 
generated at a very high scale (e.g.~the GUT scale), some mass splitting is introduced at the electroweak scale via the renormalization group evolution \cite{Martin:1997ns}.\footnote{The contribution from the renormalization group is related to the Yukawa matrices and does not leads to flavour violation \cite{DAmbrosio:2002vsn}.} 
In this case, the SUSY breaking masses for staus are smaller than for the other sleptons, and the mass gap between the slepton and Bino must be taken even larger in the BLR planes.
We therefore expect the BLR contribution cannot explain the $(g-2)_\mu$ anomaly 
consistently with the LHC constraints also in those popular scenarios.

One way to avoid this situation is to relax the universality assumption on the slepton masses
and take the stau mass to be much heavier than the other sleptons.  
Generally, however, such a splitting leads to too large LFV and is not phenomenologically viable, unless the charged lepton Yukawa matrix and the L- and R-slepton mass matrices are tuned in such a way that they can be simultaneously diagonalised in the same basis to a very good precision.  
To see whether the $(g-2)_\mu$ anomaly can be explained by the BLR contribution in this case, we introduce two new planes, in which the parameters are defined 
similarly to {\bf BLR}$_{50/10}$. 
The difference from the previous scenarios is that
in the new planes,
$\tilde m_{l_{3}}$ and $\tilde m_{\tau_R}$ are fixed at 10 TeV independently from 
the first and second generation slepton masses 
(mass degeneracy is still assumed among the first two generations: 
$\tilde m_{l_L} \equiv \tilde m_{l_1} = \tilde m_{l_2}$
and 
$\tilde m_{l_R} \equiv \tilde m_{e_R} = \tilde m_{\mu_R}$).
Due to the large stau masses, the vacuum stability constraint is now relaxed and it is always possible to find a value for $\mu \tan \beta$ such that 
$(g-2)_\mu$ is fitted to the central value of the measurement as given in Eq.~\eqref{eq:gmin2}.   
We therefore fix $\mu \tan \beta$ at this value, instead of the maximum value allowed by the vacuum stability.
The Bino mass parameter, $M_1$, is taken at 20 GeV below the smuon mass, $m_{\tilde \mu_1}$ to avoid the appearence of a charged LSP and to relax the LHC constraints with compressed mass spectra.

Fig.~\ref{fig:MSSM_BLR_nostau} shows
the DM and LHC constraints on the new planes 
with the mass splitting assumption, $\tilde m_{\tilde \tau_R}, \tilde m_{\tilde l_3} \gg \tilde m_{l_L}, \tilde m_{l_R}$.
As mentioned above, the central value of the $(g-2)_\mu$ measurement is realised everywhere across the planes.
We see that the regions with $\tilde m_{l_L} \gtrsim 220$ GeV (for $\tan\beta = 50$)
and 
$\tilde m_{l_L} \gtrsim 300$ GeV (for $\tan\beta = 10$)
are excluded by the DM overproduction.
This bound may be relaxed if one adopts 
even more stringent compressed mass assumption, $m_{\tilde \mu_1} - M_1 < 20$ GeV, so that the coannihilation channel becomes more efficient.
We also see that the constraint
from the CMS $\ell^+ \ell^-$ analysis
(red shaded regions)
excludes $\tilde m_{l_L} \lesssim 300$ GeV and
$200 \lesssim \tilde m_{l_R}/{\rm GeV} \lesssim 600$.
This bound may also be relaxed 
by taking more compressed mass hierarchy, $m_{\tilde \mu_1} - M_1 < 20$ GeV.
We see that the central value of the $(g-2)_\mu$ measurement can be fitted  
in a large region of the parameter space 
(with $\tilde m_{l_R} \gtrsim 600$ GeV,
$\tilde m_{l_L} \gtrsim 220$ (300) GeV for $\tan \beta  = 50$ (10)) 
consistently with the other phenomenological constraints 
if the universal mass assumption is removed and the stau masses are taken to be much larger than the other slepton masses.

We also comment that for other parameter planes
adopting the mass splitting assumption with heavy staus
makes the LHC constraint even more stringent,
since the decay branching ratios of heavier electroweakinos 
into the eletron and muon final states relatively enhance
due to absence of staus in the low energy spectrum.
}

\begin{figure}[t!]
\centering
      \includegraphics[width=0.45\textwidth]{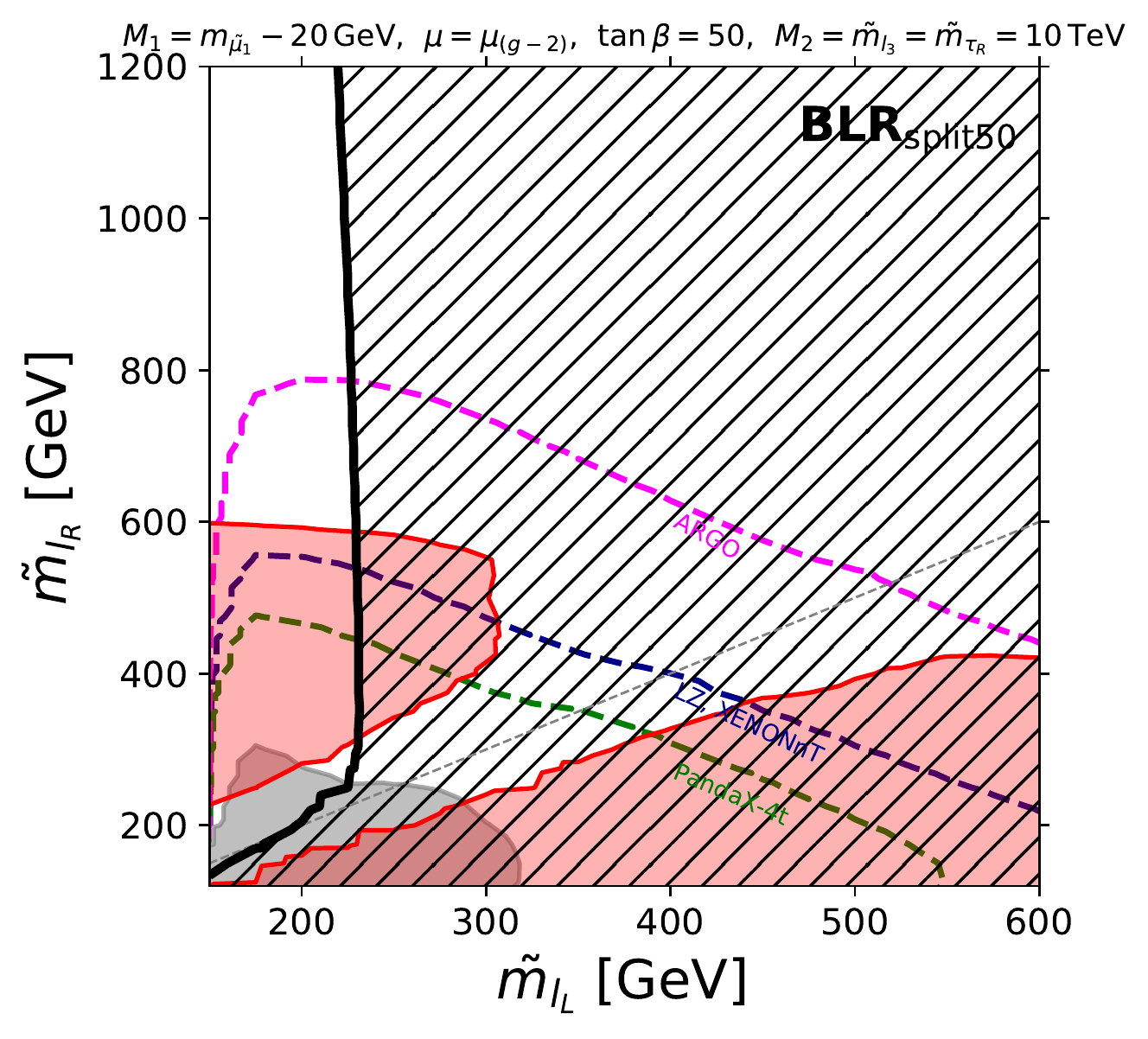}
    \includegraphics[width=0.45\textwidth]{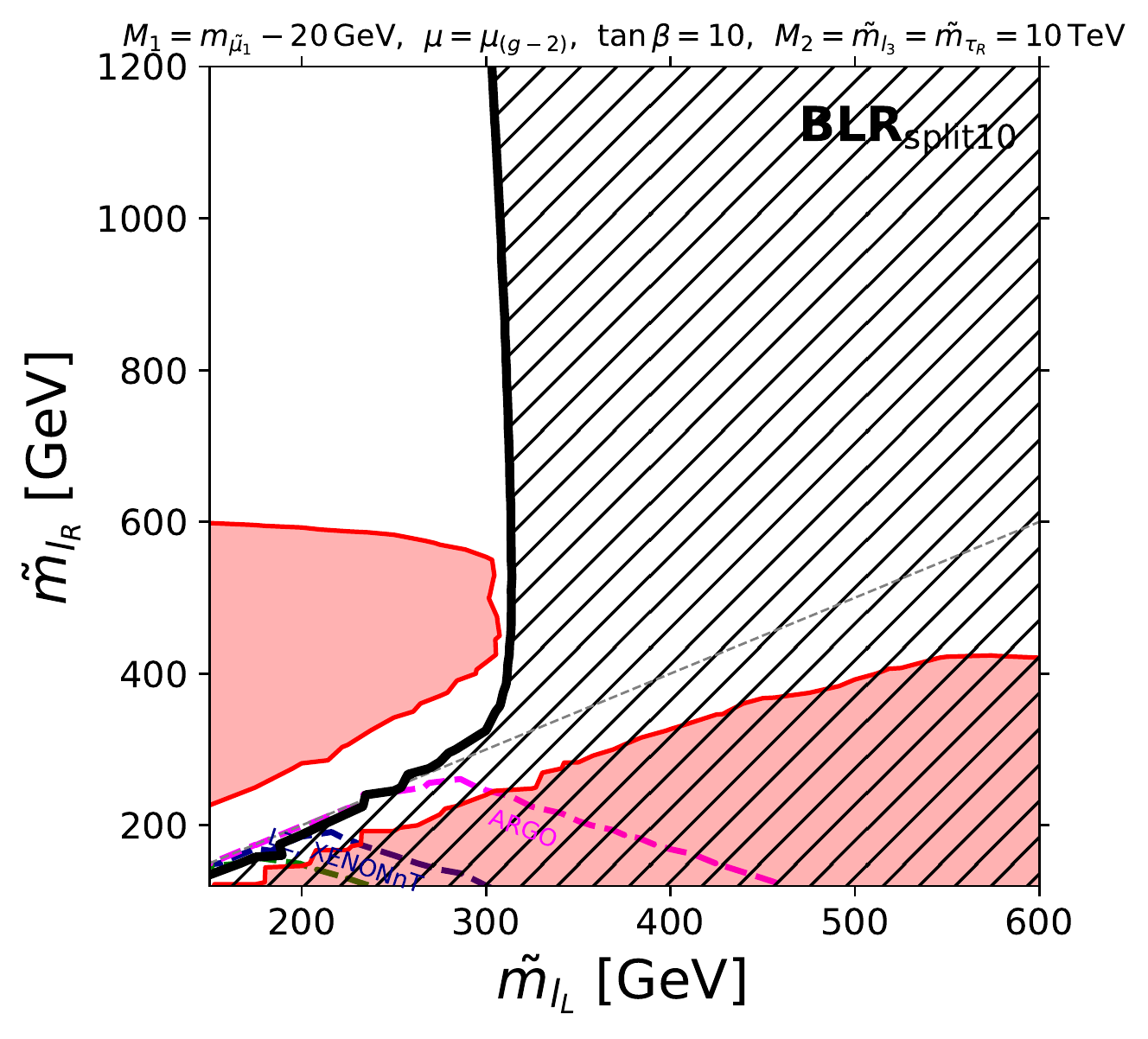}
\caption{\label{fig:MSSM_BLR_nostau} \small 
Similar planes to Fig.~\ref{fig:MSSM_BLR}
but $\tilde m_{l_3}$ and $\tilde m_{\tau_R}$ are taken to be 10 TeV
independently from the other slepton masses.
The value of $\tan \beta$ is taken to be 50 and 10 
in the left and right plots, respectively.
The $\mu$-parameter is fixed such that
it fits the central value of the $(g-2)_\mu$ measurement.
The hatched region and the grey shaded regions
are excluded by the dark matter overproduction
and the Xenon-1T limit, respectively.
The red shaded
region is excluded by the LHC constraint
from the CMS $\ell^+ \ell^-$ analysis.
The region below the dashed magenta, blue and green contours can be probed by the future ARGO, LZ/XENONnT and Panda-4t direct detection experiments, respectively.  
}
\end{figure}

\section{Baryonic R-parity violation}
\label{sec:rpv}

In the previous section, we have shown 
the SUSY $(g-2)_\mu$ regions are severely constrained 
by overproduction of relic neutralinos, 
DMDD experiments, and LHC searches in the large-$\met$ channel.
These constraints are a direct consequence of the assumption 
that $\tilde \chi_1^0$ is the LSP and stable.
If $\tilde \chi_1^0$ is the LSP but decays into visible SM particles,
the limits from $\tilde \chi_1^0$ overproduction
and DMDD experiments are entirely avoided
and the $\met$ signature at the LHC will not be
available.
On the other hand, the SUSY contribution to $(g-2)_\mu$
is basically unaffected unless the new particle or the operator introduced 
to permit the $\tilde \chi_1^0$ decay give 
large contribution to $(g-2)_\mu$.

One way to make $\tilde \chi_1^0$ unstable is to
introduce R-parity violation (RPV).
Without extending the MSSM particle contents,
the most general renormalisable superpotential 
admits four types of RPV operators (see, e.g., Ref.\cite{Barbier:2004ez} for a review).
Three of them break the lepton number ($L$) but not  
the baryon number ($B$) symmetry,
while it is the other way around for the last one.
In this section, we study the scenario with
the $B$-breaking RPV operator.
Namely, we extend the MSSM superpotential with
\begin{equation}
    W_{\slashed{R}} \,=\, \frac{1}{2} \lambda^{\prime \prime}_{ijk} U^c_i D^c_j D^c_k\,,
    \label{eq:WRPV}
\end{equation}
where $i,j,k$ are flavour indices
and $U^c$ ($D^c$) is the  
the right-handed anti-up(down)-quark superfield.
The $SU(3)_C$ invariance forces 
the colour indices to be
contracted in the totally antisymmetric manner, 
which results in
$\lambda^{\prime \prime}_{ijk} = - \lambda^{\prime \prime}_{ikj}$.

This scenario has three advantages. 
Firstly, the term in Eq.~\eqref{eq:WRPV} does not
generate the other three RPV operators radiatively,
since they are protected by the $L$-symmetry.
This justifies us to set the other $L$-violating RPV couplings to
zero at all scales.
Secondly, protons remain stable because, within the MSSM, proton decay requires both $B$- and $L$-violations.
Lastly, these operators
do not contribute to the $(g-2)_\mu$.
The one and two sigma $(g-2)_\mu$ regions 
found in the previous section 
are unchanged.

With the term in Eq.~\eqref{eq:WRPV}, the LSP neutralino can decay as $\tilde \chi_1^0 \to u_i d_j d_k$ and $\bar u_i \bar d_j \bar d_k$ via an off-shell squark.
If $i,j,k \neq 3$, a neutralino $\neut[1]$ decays into
three light-flavour quarks, and
the large-$\met$ signature of the SUSY events at the LHC
is replaced by
light-flavour jets.
Since jets are easily produced at the LHC,
discrimination of such a signature from 
the SM background is challenging.
Namely, this case provides a scenario 
that maximally opens up the allowed
SUSY $(g-2)_\mu$ parameter region. 
For simplicity, we switch off all the RPV couplings
but 
$\lambda^{\prime \prime}_{112}$ in our analysis.
However, the result may apply for 
non-zero $\lambda^{\prime \prime}_{ijk}$ couplings 
as far as $i,j,k \neq 3$.
The decay rate of the neutralino is
proportional to $|\lambda^{\prime \prime}_{112}/m^2_{\tilde q}|^2$.
{If all RPV couplings other than $\lambda^{\prime \prime}_{112}$ are switched off,
there is no strong phenomenological constraint on
$\lambda^{\prime \prime}_{112}$
\cite{Domingo:2018qfg}.
One can therefore safely assume 
that
$\lambda^{\prime \prime}_{112}$
is large enough so that 
the neutralino decay,
$\tilde \chi_1^0 \to u d s$ ($\bar u \bar d \bar s$), is prompt and does not yield displaced vertices.}



ATLAS and CMS searches targeting RPV scenarios 
are rather rare and simplified model limits 
are not widely available.
We therefore estimate the LHC constraints 
by a chain of Monte Carlo simulations
outlined in subsection \ref{sec:lhc}.
Although all 8 and 13 TeV analyses 
implemented in {\tt CheckMATE} (see Appendix \ref{app:checkmate} for the complete list) 
are included
in our analyses,
only a few provided 95\,\% CL exclusions.
Those analyses are listed in Table \ref{tb:rpv}.

\begin{table}[t!]
\begin{center}
\begin{tabular}{ |c|c|c|c|c| } 
\hline
Analysis & Ref & $E$/TeV & ${\cal L}$/fb$^{-1}$  & Colour  \\
\hline
ATLAS multijet+$\ell$ & \cite{ATLAS:2021fbt} & 13 & 139 & Red  \\
\hline
CMS multilepton & \cite{CMS:2017moi} & 13 & 35.9 & Blue  \\
\hline
ATLAS jets+$\met$ & \cite{ATLAS-CONF-2019-040} & 13 & 139 & Green  \\
\hline
\end{tabular}
\caption{\label{tb:rpv}
\small
Analyses in CheckMATE which are relevant for RPV scenario.}
\end{center}
\end{table}

\begin{figure}
\centering
\includegraphics[width=0.45\textwidth]{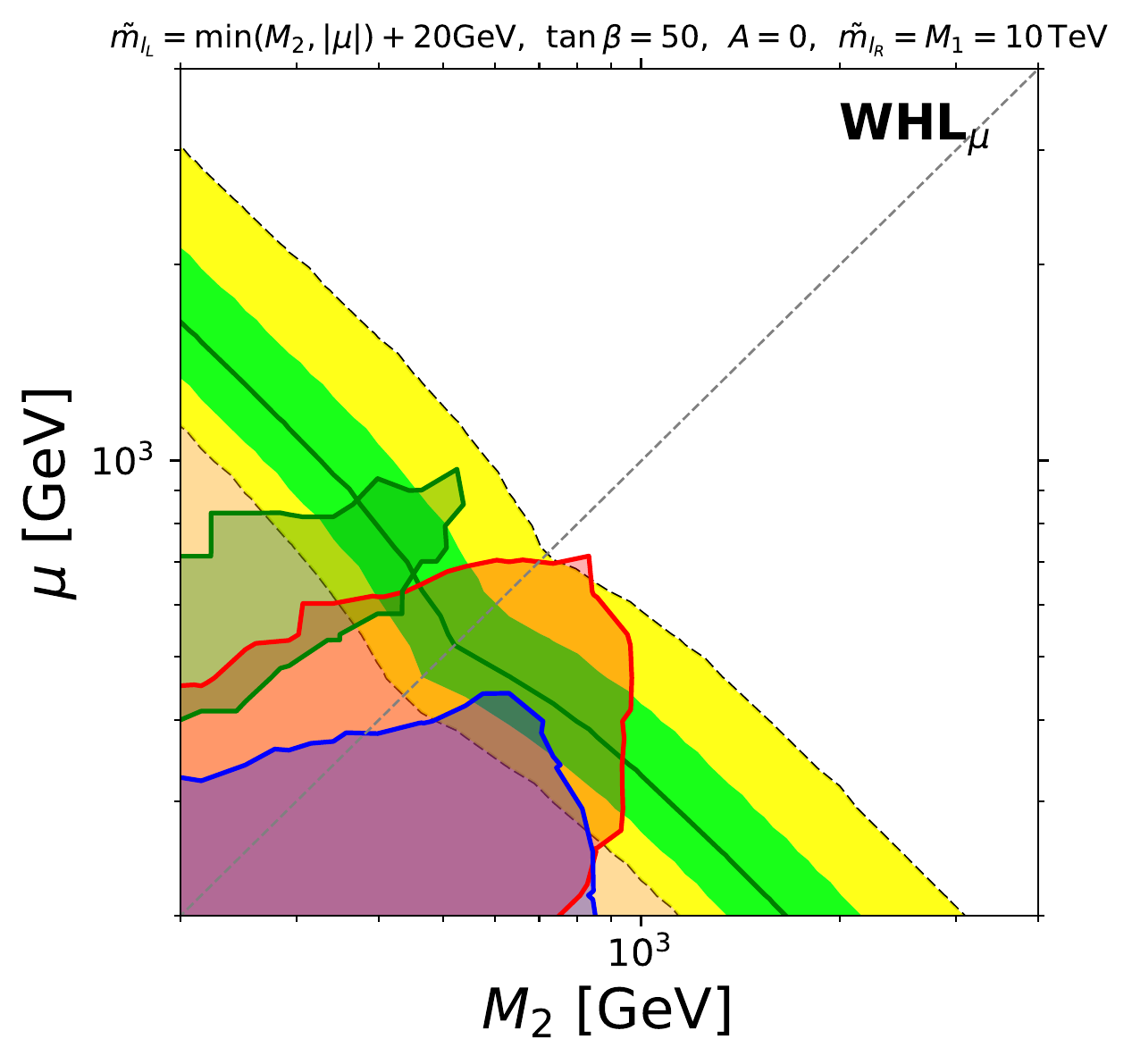}
\includegraphics[width=0.45\textwidth]{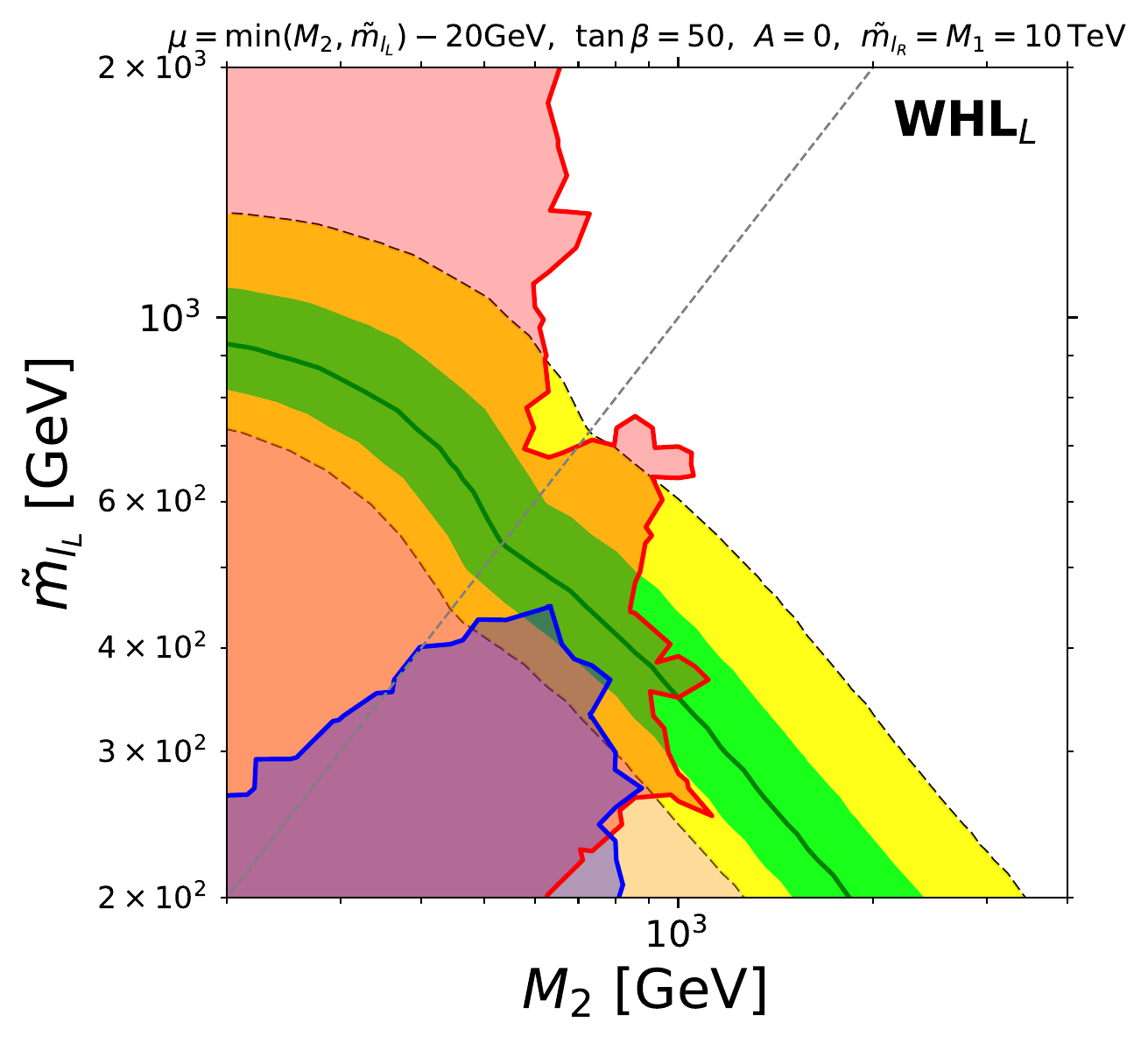}
\caption{\small 
Results for RPV ${\bf WHL}_\mu$ (left) and ${\bf WHL}_L$ (right) planes. 
Region of parameter space allowed by the latest $a_\mu$ experimental results \cite{Muong-2:2021ojo} is depicted with green and yellow bands, corresponding to one and two sigma agreement respectively. Orange shaded region corresponds to $a_\mu$ bigger than experimental value by more than 2 sigma.
 Red, green and blue shaded regions are excluded by LHC constraints.
}
\label{fig:RPV_1}
\end{figure}

\medskip

The left panel of Fig.~\ref{fig:RPV_1}
shows the ${\bf WHL}_\mu$ plane.
As in the previous plots,
the green and yellow regions 
correspond to 
the areas where the predicted $(g-2)_\mu$
agrees with the experimental value 
within the 1 and 2 $\sigma$ accuracy, respectively.
On the other hand, the red, blue and green shaded regions 
correspond to the 95 \% CL exclusions obtained from
ATLAS multijet+$\ell$ 
\cite{ATLAS:2021fbt},
CMS multilepton \cite{CMS:2017moi}
and
ATLAS jets+$\met$
\cite{ATLAS-CONF-2019-040}
analyses, respectively.
Due to the R-parity violating $UDD$ operator,
the LSP neutralino decays into three (anti-)quarks.
The constraints from the neutralino overproduction
and DM direct detection experiments are absent 
unlike the stable neutralino case.

For $M_2 > |\mu|$,
ATLAS multijet+$\ell$ (red)
and
CMS multilepton (blue) analyses
exclude 
Wino production followed by
$\tilde W^{\pm,0} \to \ell^{\pm} \tilde \eta$
($\tilde \eta = \tilde \ell, \tilde \nu$)
and $\tilde \eta \to \ell/\nu + {\rm jets}$.
In the opposite case, $|\mu| > M_2$,
Wino and Higgsino are significantly mixed,
and the production of heavier electroweakinos, $\tilde \chi_{\rm heavy}$,
significantly contributes to the exclusion.
The $\tilde \chi_{\rm heavy}$ can decay
to many different modes, e.g.,
into $\tilde \chi_{\rm light} + X$ with
$X = h,Z,W^{\pm}$
and
$l_3 \tilde l_3$ pairs with $l_3 = (\tau, \nu_\tau)$ and $\tilde l_3 = (\tilde \tau, \tilde \nu_\tau)$.
In the region excluded by
ATLAS jets+$\met$ (green),
Higgsino production dominantly contributes.
Higgsinos decay mainly into the Higgs boson and one of the
Wino states, leaving many jets in the final state.
Comparing the RPV ${\bf WHL}_\mu$ plane 
with the same plane in the stable neutralino 
(Fig.~\ref{fig:MSSM_WHL} left),
we see that the neutralino decay via the $UDD$ operator 
provides wider allowed region for the $(g-2)_\mu$.
In particular, 
the regions around (a) $M_2 \sim 250$ GeV, $\mu \sim 1.5$ TeV
and (b) $M_2 \sim 1.5$ TeV, $\mu \sim 250$ GeV
are only allowed for the RPV case.

{ In the ${\bf WHL}_L$ plane, shown in the right of Fig.~\ref{fig:RPV_1},
our results in the region with $M_2 > \tilde m_{l_L}$ (i.e., below the diagonal dotted line) are essentially identical to those of the 
$M_2 > \mu$ region
of the ${\bf WHL}_\mu$ plane, which is because $\tilde m_{l_L} = \mu + 20$\,GeV in the both regions and other parameters are fixed in the same way.
Meanwhile, in the other half ($\tilde m_{l_L} > M_2$), an wider region is excluded by the ATLAS multijet+$\ell$ (red).
We note that Wino and Higgsino are substantially mixed 
in the mass eigenstates in this region.}
The exclusion is mainly driven by the pair production of $\tilde \chi_{\rm heavy}$, which ends up into soft lepton(s) plus multijet final state,
where the leptons are originated from 
the $\tilde \chi_{\rm heavy}$ decay via an off-shell $W(Z)$ boson.

\medskip

\begin{figure}
\centering
\includegraphics[width=0.44\textwidth]{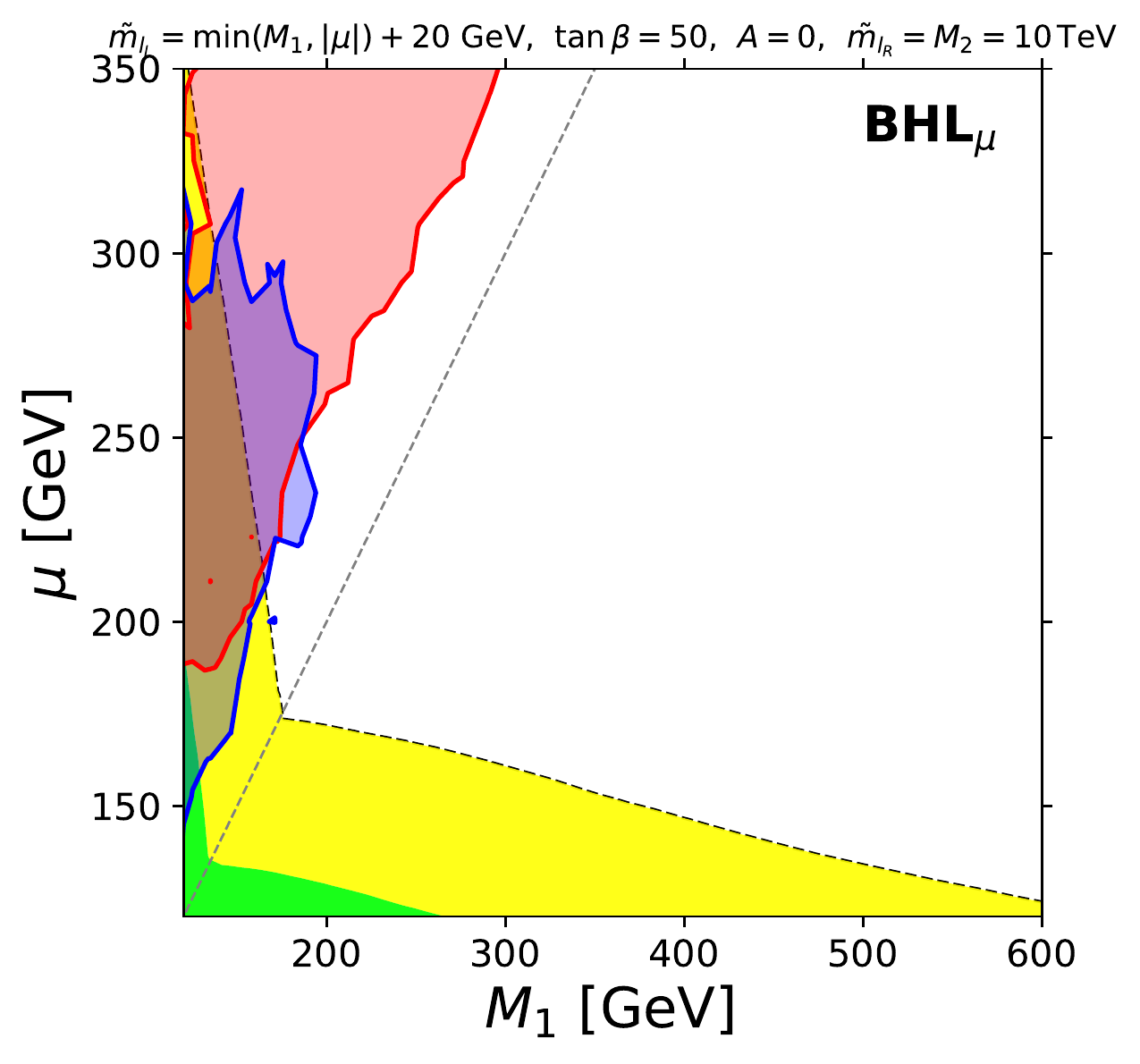}
\includegraphics[width=0.44\textwidth]{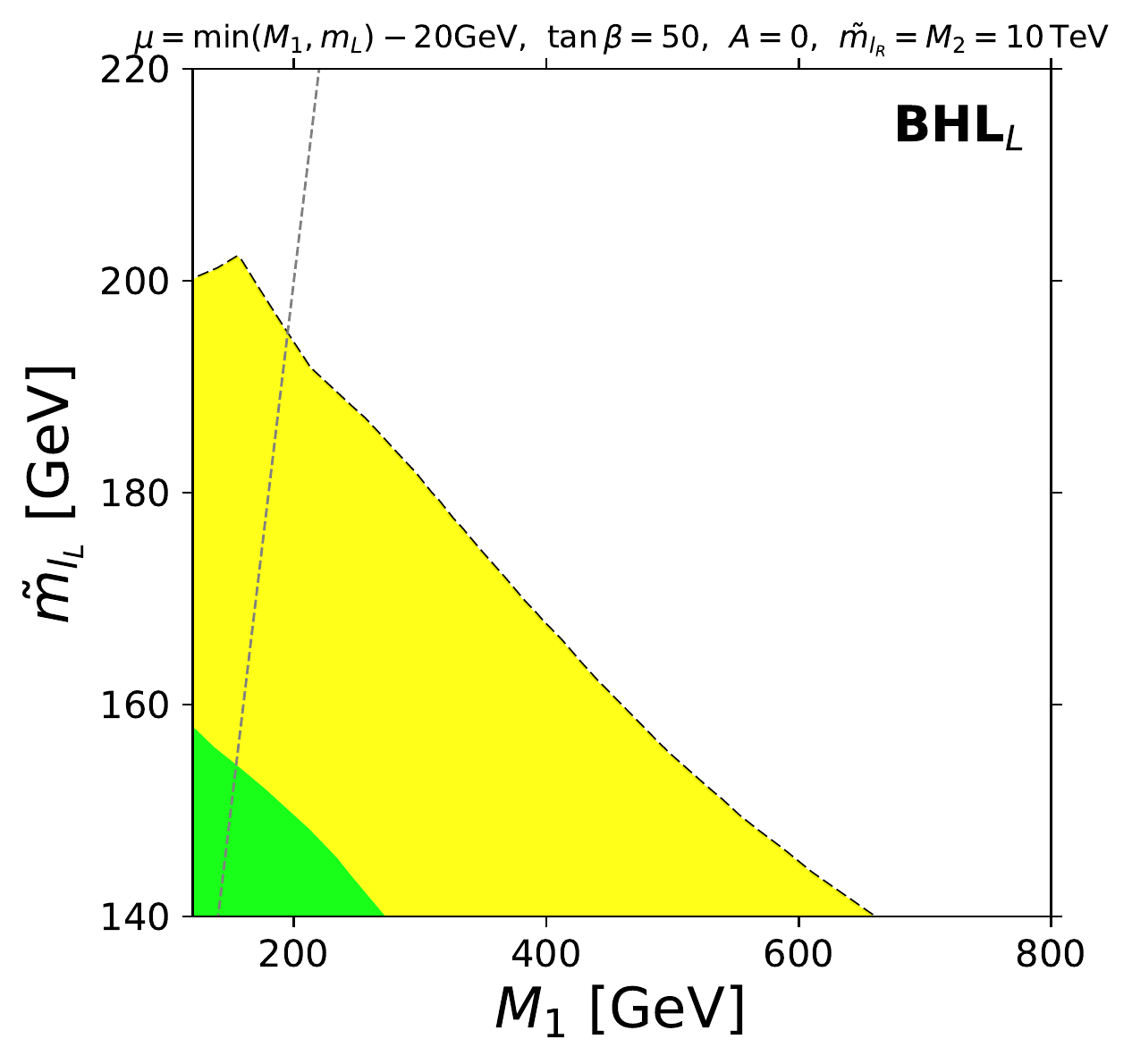}
\includegraphics[width=0.44\textwidth]{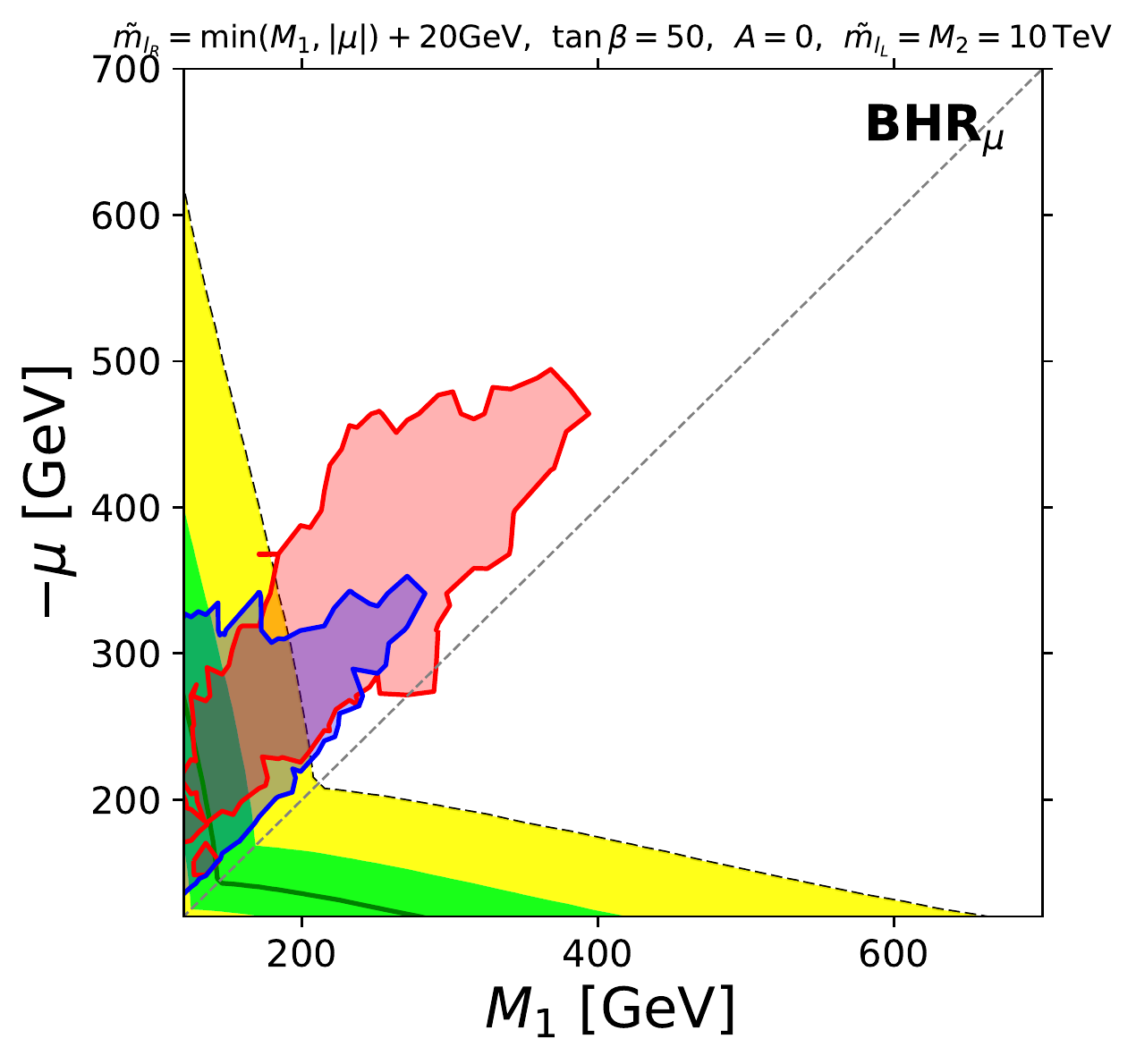}
\includegraphics[width=0.44\textwidth]{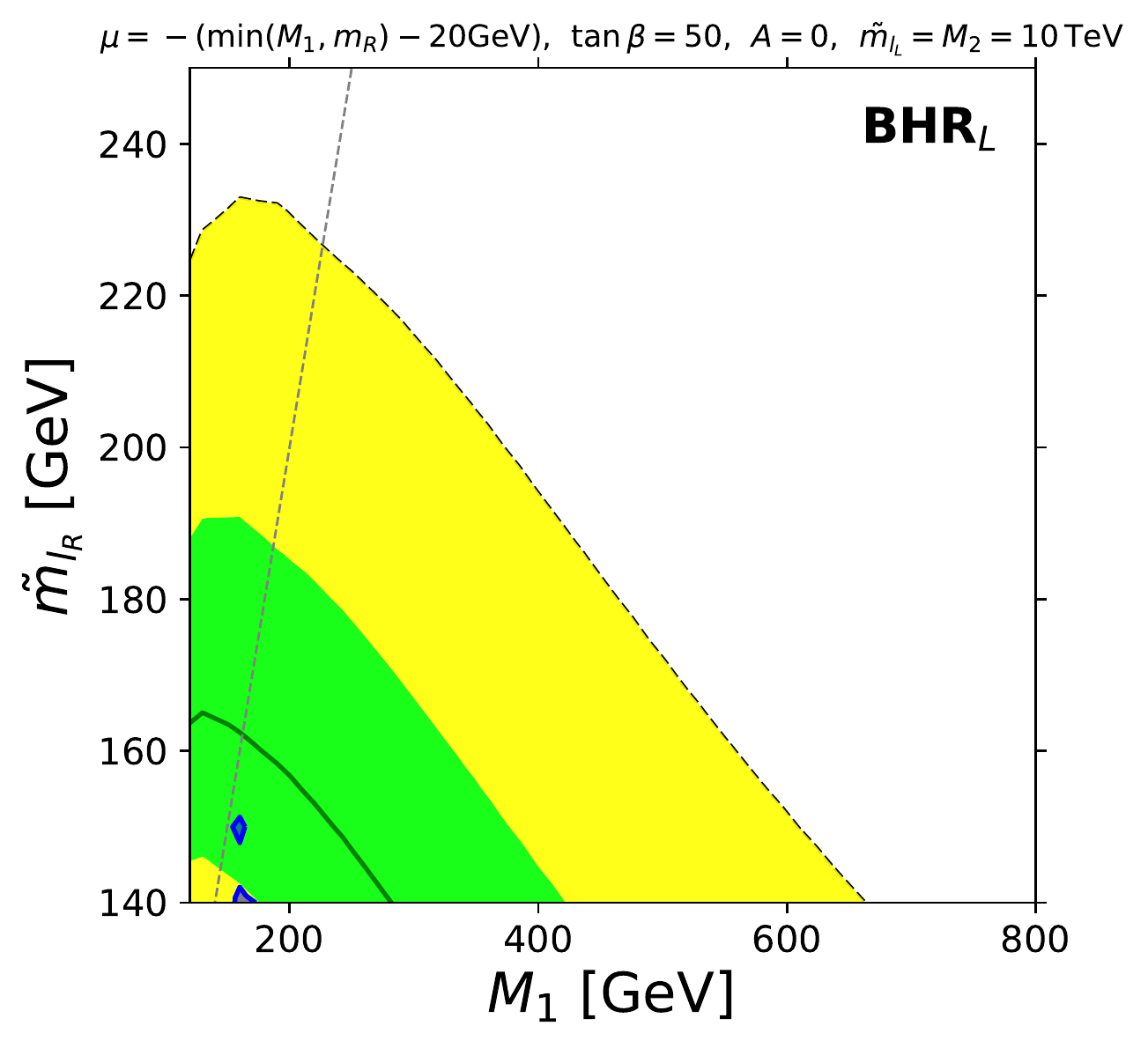}
\includegraphics[width=0.44\textwidth]{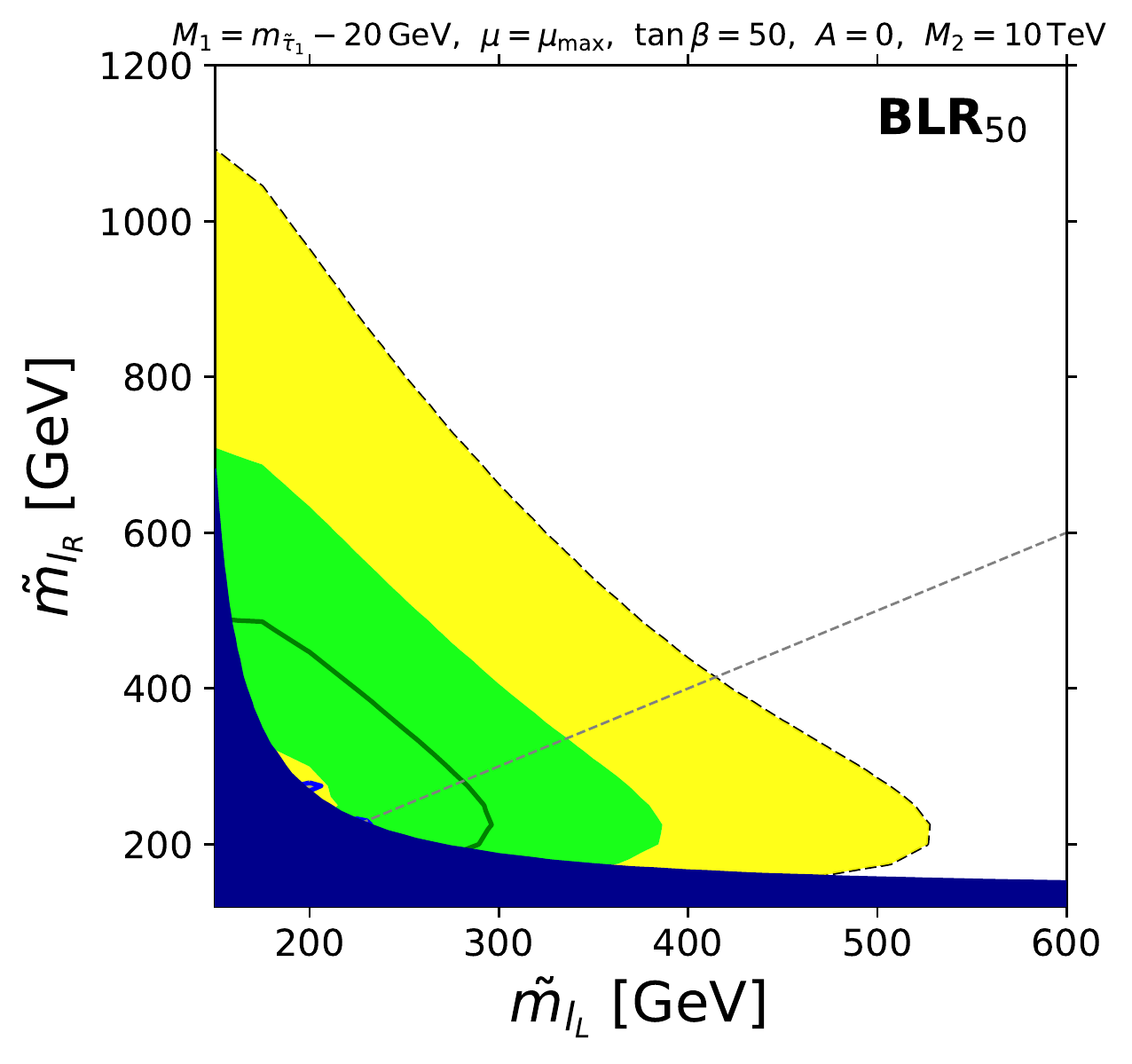}
\includegraphics[width=0.44\textwidth]{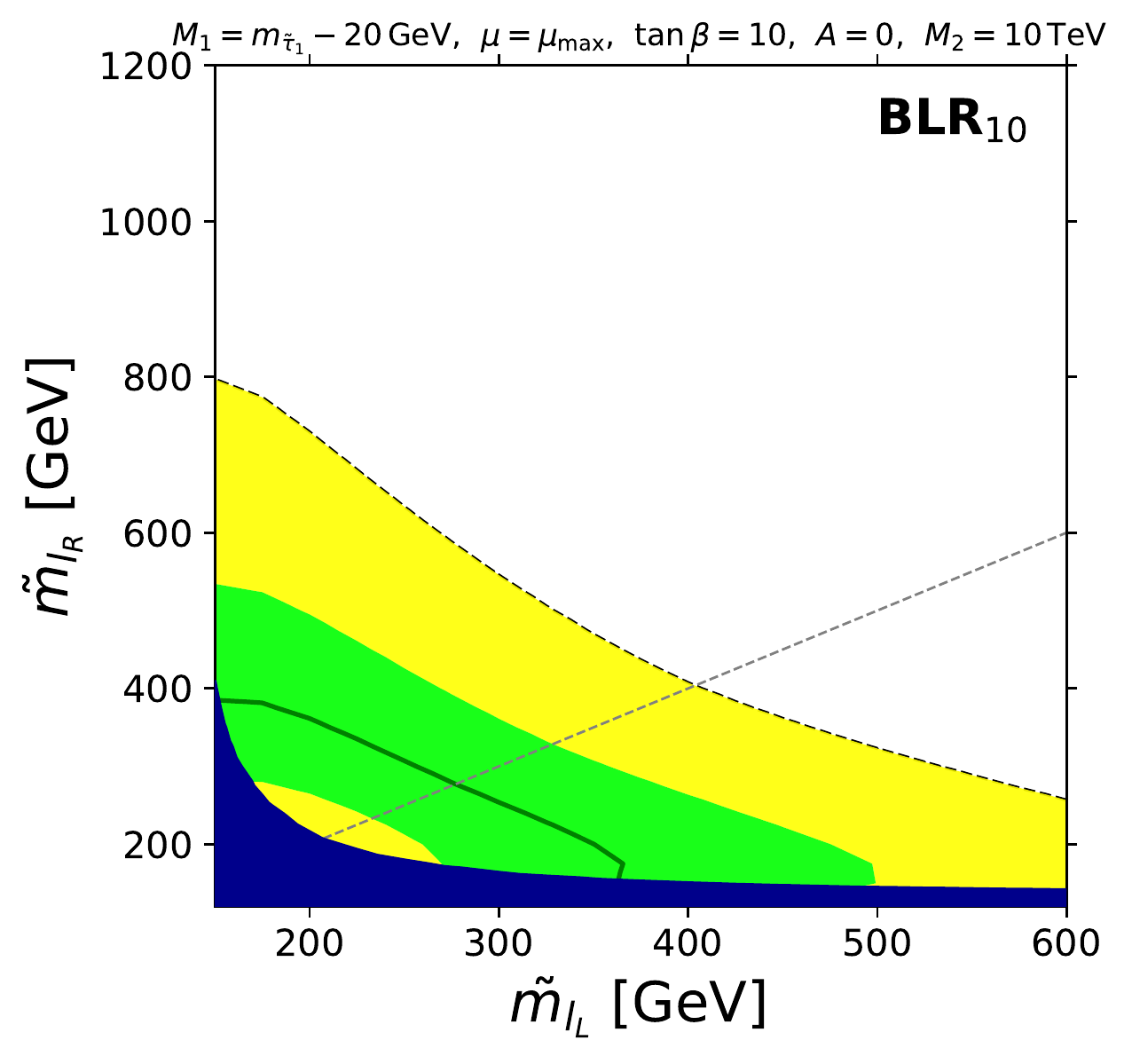}
\caption{\small
Results for RPV ${\bf BHL}_\mu$ (upper left),  ${\bf BHL}_L$ (upper right),
${\bf BHR}_\mu$ (middle left),  ${\bf BHR}_L$ (middle right),
${\bf BLR}_{50}$ (lower left) and ${\bf BLR}_{10}$ (lower right)
planes. 
Region of parameter space allowed by the latest $a_\mu$ experimental results \cite{Muong-2:2021ojo} is depicted with green and yellow bands, corresponding to one and two sigma agreement respectively.
Blue and red shaded regions are excluded by LHC constraints. 
Dark blue shaded area is excluded by LEP stau mass bound \cite{Zyla:2020zbs}.
}
\label{fig:RPV_2}
\end{figure}

The top and middle panels of Fig.~\ref{fig:RPV_2}
display the BHL and BHR planes, respectively. 
In the ${\bf BHL}_{\mu}$ (top left)
and ${\bf BHR}_{\mu}$ (middle left) planes with $|\mu| > M_1$, 
the excluded regions from 
ATLAS multijet+$\ell$ (red)
and
CMS multilepton (blue) analyses
are visible.
The main SUSY process contributing to the exclusion is
production of Higgsino-like states followed by
the decay into $l_3 \tilde l_3$ with $l_3 = (\tau, \nu_\tau)$
and $\tilde l_3 = (\tilde \tau, \tilde \nu_\tau)$.
The $\tilde l_3$ then decays into $l_3$ plus jets via an on-shell Bino.
The region with the opposite condition, $M_1 > |\mu|$,
is not constrained by the current LHC data,
since in this region, Higgsino states decay directly into
jets via the RPV operator and any leptons and neutrinos (as a source of $\met$) are unavailable.
In the ${\bf BHL}_{L}$ (top right) and ${\bf BHR}_{L}$ (middle right),
the regions on the right-hand side to the diagonal lines
are identical (in the sense that they share exactly the same parameters) to the
corresponding regions in the 
${\bf BHL}_{\mu}$ (top left) and ${\bf BHR}_{\mu}$ (middle left), respectively.
On the other hand, the regions on the left-hand side 
of the diagonal lines are new and unconstrained.
This is again because 
the Higgsino states decay directly into
jets in those regions.
Comparing with the BHL and BHR planes in the stable neutralino scenario (see Fig.~\ref{fig:MSSM_BHLR})
more $(g-2)_\mu$ region is opened up for the RPV case
due to the lack of DM direct detection constraints 
as well as the ATLAS soft-$\ell$ \cite{ATLAS:2019lng}
for $\tilde \chi_2^0 \tilde \chi_1^0 \to (Z^* \tilde \chi_1^0) (W^{\pm *} \tilde \chi_1^0)$.

\medskip

The bottom left and right plots in Fig.~\ref{fig:RPV_2}
show the ${\bf BLR}_{50}$ and ${\bf BLR}_{10}$ planes,
respectively.
We see that the entire $(g-2)_\mu$ region is unconstrained 
for both planes.
This provides a dramatic contrast 
with the stable neutralino case, 
where the whole $(g-2)_\mu$ region
is excluded by the 
CMS $\ell^+ \ell^-$ \cite{CMS:2020bfa}
analysis as well as the neutralino overproduction.
{In the RPV BLR planes, the dominant SUSY process 
is the slepton pair production,
$pp \to \tilde \ell^+ \tilde \ell^-$.}
However, small slepton cross-sections
cannot lead to statistically distinguishable 
signature when the missing transverse energy is replaced by hadronic jets.

\section{ Gravitino LSP }
\label{sec:grav}

{The gravitino LSP scenario, often realised in  gauge mediated SUSY breaking (GMSB) \cite{Dine:1993yw,Dine:1994vc,Dine:1995ag},}
is another well-known case in which $\neut[1]$ becomes unstable even if it is the lightest among the MSSM particles.
In GMSB,
SUSY is spontaneously broken by the $F$-term of a singlet field $\Phi$ 
in a hidden sector.
The hidden sector also contains messenger fields ($\Psi, \overline{\Psi}$) with non-trivial representations under the SM gauge group.
Assuming a direct coupling between the SUSY breaking field
and messenger fields in the superpotential, $W_{\rm hid} \ni \kappa \Phi \overline{\Psi} \Psi$,
the scalar and fermion masses in the messenger multiplets
split and the SUSY breaking is mediated to
the MSSM sector radiatively via gauge interaction.
In the GMSB, the ratio between the gravitino mass, $m_{\widetilde G}$, 
and the MSSM soft SUSY breaking mass scale, $\widetilde m$, is roughly given by
\begin{equation}
\frac{m_{\widetilde G}}{\widetilde m} \,\sim\, \frac{1}{\epsilon} \frac{M_{\rm mess}}{M_P},
\end{equation}
where $\epsilon \sim 0.1 - 0.01$ is a loop factor,
$M_P$ is the Planck scale and $M_{\rm mess}$ 
is the mass scale of the messenger fields.
Since we often take $M_{\rm mess} \ll M_{P}$,
the gravitino LSP is realised.

In fact, very light gravitinos have several theoretical motivations. 
One of them is to relax apparent fine-tuning in the Higgs sector.
In GMSB, the radiative correction to the SUSY breaking Higgs mass parameter 
depends logarithmically on the messenger mass, $\delta m_{H_u}^2 \propto \ln(M_{\rm mess}/\widetilde m)$, for a fixed $\widetilde m$, 
and the fine-tuning may be ameliorated by taking $M_{\rm mess} \sim \widetilde m$.
For example, taking $M_{\rm mess} \sim (10-100)$ TeV and $\widetilde m \sim 1$ TeV leads to $m_{\widetilde G} \sim (1-10)$ eV.

Another motivation is in cosmology.
Gravitinos tend to be overproduced at the reheating era 
unless the reheating temperature $T_R$ is very low \cite{Moroi:1993mb}.
However, such a low $T_R$ is inconsistent with 
the thermal leptogenesis scenario \cite{Fukugita:1986hr, Buchmuller:2005eh}, which typically requires $T_R > 10^9$ GeV.
This is known as the cosmological gravitino problem.
Very light gravitino with $m_{\widetilde G} \lesssim 100$ eV provides an attractive solution.
Such light gravitinos are thermalised in the early Universe 
and the relic density becomes independent of $T_R$
and can be smaller than the observed DM density 
if $m_{\widetilde G} \lesssim 100$ eV.\footnote{Very 
light gravitinos are however not free from other cosmological constraints
\cite{Viel:2005qj, Ichikawa:2009ir, Osato:2016ixc}.
For example, their large free-streaming length 
may suppress density fluctuations of (sub-)galactic length scales and the effect can be imprinted 
in the cosmic microwave background (CMB).
{The authors of Ref.~\cite{Osato:2016ixc} have} derived the upper bound, 
$m_{\widetilde G} < 4.7$ eV, using the CMB lensing 
and the cosmic shear measured by Plack and 
the Canada France Hawaii Lensing Survey, respectively, combined with analyses of the primary CMB anisotropies and the baryon acoustic oscillations in galaxy distributions.}

Backed by these observations, we study LHC constraints 
on the SUSY $(g-2)_\mu$ solution assuming approximately massless 
gravitinos.\footnote{See also \cite{Kim:2019vcp} for LHC constraints on the gravitino LSP scenario.}
Although in the minimal GMSB model
sfermion and gaugino masses are uniquely determined by the messenger scale, they are less constrained in non-minimal models \cite{Meade:2008wd, Buican:2008ws, Grajek:2013ola}.
Since the aim of this study is to understand how the LHC constraints 
change from the stable neutralino case to the gravitino LSP
scenario, we take the bottom-up approach 
and treat the MSSM parameters as free.
In the first part of this section 
we keep the same parameter planes 
defined in section \ref{sec:planes}
and put the approximately massless gravitino at the bottom of the spectrum.
We then introduce small modifications 
to the definition of our planes 
to accommodate non-neutralino NLSP cases 
in the second part of the section.
Since the gravitino contribution to $(g-2)_\mu$
is negligible, the preferred $(g-2)_\mu$ regions
on the parameter planes are essentially unchanged 
from the previous cases.
We also assume that the decay of the 
next-to-the-lightest SUSY particle (NLSP)
is prompt and does not yield displaced vertices.
As we will see below this assumption is consistent 
with very light gravitinos with $m_{\widetilde G} \lesssim 10$ eV.

\subsection{ Neutralino NLSP }
\label{sec:grav_neu}

In our analysis the NLSP $\tilde \chi_1^0$ decays into 
the (almost) massless gravitino together with a neutral boson ($\gamma$, $Z$ or $h$). 
The partial decay rates are given by~\cite{Ambrosanio:1996jn, Meade:2009qv}:
\begin{eqnarray}
\Gamma(\tilde \chi_1^0 \to \tilde G  \gamma) &=& \big| N_{11} c_W + N_{12} s_W \big|^2 {\cal A} \,,
\nonumber \\
\Gamma(\tilde \chi_1^0 \to \tilde G  Z) &=& \Big( \big| N_{12} c_W - N_{11} s_W \big|^2 
+ \frac{1}{2} \big| N_{13} c_\beta - N_{14} s_\beta \big|^2 \Big)
\Big( 1 - \frac{m_Z^2}{m^2_{\tilde \chi_1^0}} \Big)^4
{\cal A} \,,
\nonumber \\
\Gamma(\tilde \chi_1^0 \to \tilde G  h) &=&  \frac{1}{2} \big| N_{13} c_\beta + N_{14} s_\beta \big|^2
\Big( 1 - \frac{m_h^2}{m^2_{\tilde \chi_1^0}} \Big)^4
{\cal A} \,,
\label{eq:br_neut}
\end{eqnarray}
where $N_{ij}$ is the element of the neutralino mixing matrix,
$(c_W, s_W) = (\cos \theta_W, \sin \theta_W)$,
$(c_\beta, s_\beta) = (\cos \beta, \sin \beta)$
and
\begin{equation}
{\cal A} ~=~ \frac{m^5_{\tilde \chi_1^0}}{16 \pi m^2_{\widetilde G} M^2_{\rm pl}}
\,\sim \,
\frac{1}{0.3\,\rm mm}
\Big( \frac{m_{\tilde \chi_1^0}}{100\,\rm GeV} \Big)^5
\Big( \frac{m_{\widetilde G}}{10\,\rm eV} \Big)^{-2} \,.
\end{equation}
We see from this expression the neutralino decay is prompt, $c \tau_{\tilde \chi_1^0} \lesssim 1$ mm,
for $m_{\tilde \chi_1^0} \gtrsim 100$ GeV
and $m_{\widetilde G} \lesssim 10$ eV.

\begin{table}[t!]
\begin{center}
\begin{tabular}{ |c||c|c|c| } 
\hline
$\tilde \chi_1^0$ & ${\rm Br}(\tilde \chi_1^0 \to \gamma \widetilde G)$ & ${\rm Br}(\tilde \chi_1^0 \to Z \widetilde G)$ & ${\rm Br}(\tilde \chi_1^0 \to h \widetilde G)$ \\ 
\hline \hline
$\widetilde B$ &  $\frac{c_W^2}{c_W^2 + s_W^2 x_Z^4}$  &  $\frac{s_W^2 x_Z^4}{c_W^2 + s_W^2 x_Z^4}$ & $0$ \\ 
\hline
$\widetilde W$ &  $\frac{s_W^2}{s_W^2 + c_W^2 x_Z^4}$  &  $\frac{c_W^2 x_Z^4}{s_W^2 + c_W^2 x_Z^4}$ & $0$ \\ 
\hline
$\widetilde H$ &  $0$  &  $\frac{x_Z^4}{x_Z^4 + x_h^4}$ & $\frac{x_h^4}{x_Z^4 + x_h^4}$ \\ 
\hline
\end{tabular}
\caption{\label{tb:br}
\small
Branching ratios for the pure neutralino states 
at the large $\tan \beta$ regime.
$x_I \equiv (1 - m^2_I / m^2_{\tilde \chi_1^0})$
with $I = Z, h$ are used. 
In the heavy $\tilde \chi_1^0$ limit ($x_I \to 1$),
we find ${\rm Br}(\widetilde B \to \gamma \widetilde G) = c_W^2 \simeq 0.77$,
${\rm Br}(\widetilde W \to Z \widetilde G) = c_W^2 \simeq 0.77$
and 
${\rm Br}(\widetilde H \to Z \widetilde G) \sim {\rm Br}(\widetilde H \to h \widetilde G) \sim 0.5$.
}
\end{center}
\end{table}

Branching ratios for the pure neutralino states can readily be computed 
by taking $N_{1i} = \delta_{1i}$ for pure-Bino, $N_{1i} = \delta_{2i}$ for pure-Wino,
and $N_{13} = - N_{14} = 1/\sqrt{2}$, $N_{11} = N_{22} = 0$
for pure-Higgsino.
{The branching ratios 
of those pure-ino states}
in the large $\tan \beta$ regime
are listed in Table \ref{tb:br}.
In the heavy $\tilde \chi_1^0$ limit
we have ${\rm Br}(\widetilde B \to \gamma \widetilde G) = {\rm Br}(\widetilde W \to Z \widetilde G) = c_W^2 \simeq 0.77$
and 
${\rm Br}(\widetilde H \to Z \widetilde G) \sim {\rm Br}(\widetilde H \to h \widetilde G) \sim 0.5$.


In Ref.~\cite{CMS:2018szt}, CMS analysed 
 13 TeV, $35.9$ fb$^{-1}$ data
in the multilepton channel 
and excluded NLSP Higgsinos {with $\mu \lesssim 650$ GeV}
together with a massless gravitino for all values
of ${\rm Br}(\widetilde H \to Z \widetilde G)$.
Similar conclusion has also been obtained 
in more recent CMS $\ell^+ \ell^-$ analysis ($137$ fb$^{-1}$) \cite{CMS:2020bfa},
where the pure-Higgsino was excluded 
up to 650 GeV assuming 
${\rm Br}(\widetilde H \to Z \widetilde G) = {\rm Br}(\widetilde H \to h \widetilde G) = 0.5$.
In the same CMS paper
the $[Z \widetilde G] [Z \widetilde G]$
final state is analysed 
and the limit $\sigma < 3$ fb is derived for $m_{\tilde \chi_1^0} \gtrsim 500$ GeV.
For the pure-Wino this limit translates to $m_{\widetilde W} > 780$ GeV.

\medskip

At this point, it is already clear that
the preferred $(g-2)_\mu$ region 
in the WHL planes
is entirely excluded,
since this region requires either $\mu \lesssim 650$ GeV
or $M_2 \lesssim 750$ GeV (see Fig.~\ref{fig:MSSM_WHL} or \ref{fig:RPV_1}).
Note also that $\mu \sim 650$ GeV implies 
$M_2 \sim 750$ GeV in the $(g-2)_\mu$ region,
and the exclusion around this region 
is even stronger,
since both Higgsino- and Wino-like states are produced
and contribute to signal regions.

\medskip 

The same observation can be made for 
the BHL and BHR planes.
In the half of those planes ($|\mu| < M_1$)
$\tilde \chi_1^0$ is Higgsino-like
and 
a good $(g-2)_\mu$ fit requires $m_{\tilde \chi_1^0} \lesssim 200$ GeV, which violates the aforementioned CMS bounds, $\gtrsim 650$ GeV.

In order to constrain the Bino-like neutralino region,
we use the information provided in CMS $\gamma$+$\met$ analysis \cite{CMS:2017brl}
(13 TeV, 35.9 fb$^{-1}$).
In this analysis, the final state with an energetic photon
and large $\met$ is selected.
The signal region is defined by the following requirements:
at least one photon with $\pT > 180$ GeV,
jets are separated from photons and the missing transverse energy as
$\Delta R(\gamma, {\rm jet}) > 0.5$
and $\Delta \phi({\bf p}_T^{\rm miss}, {\bf p}_T^{\rm jet}) > 0.3$
for jets with $\pT > 100$ GeV,
$\met > 300$ GeV,
$m_T \equiv 2 \pT^{\gamma_1} \met (1 - \cos \Delta \phi({\bf p}_T^{\gamma_1}, {\bf p}_T^{\rm miss}) ) > 300$ GeV and
$S_T^\gamma \equiv \met + \sum_i \pT^{\gamma_i} > 600$ GeV, where $\gamma_i$ ($i = 1,2,\cdots$)
are sorted in the descending order of $\pT$.
We note that the event selection is rather inclusive 
and neither jets nor leptons are required.
The result of this analysis is interpreted for
the simplified model of electroweakinos with
${\rm Br}(\tilde \chi_1^0 \to \gamma \widetilde G) = 0.5$ and
${\rm Br}(\tilde \chi_1^0 \to Z \widetilde G) = {\rm Br}(\tilde \chi_1^0 \to h \widetilde G) = 0.25$,
and the cross-section limit is derived as a function of 
$m_{\tilde \chi_1^0}$.
In particular, $\sigma < 100$ fb is derived for $m_{\tilde \chi_1^0} \sim 300$ GeV.
The same result is also interpreted for 
the squark simplified model, $\tilde q \tilde q \to [q \tilde \chi_1^0] [q \tilde \chi_1^0]$ followed by $\tilde \chi_1^0 \to \gamma \widetilde G$, which results in the $[\gamma \widetilde G] [\gamma \widetilde G] + {\rm jets}$ final state.
For $m_{\tilde \chi_1^0} \sim (100 - 300)$ GeV 
the cross-section limit, $\sigma \lesssim 10$ fb,
has been derived.

In the Bino-like neutralino region ($M_1 < |\mu|$)
of the BHL and BHR planes,
a good $(g-2)_\mu$ fit
requires $m_{\tilde \chi_1^0} \lesssim 200$ GeV and
$|\mu| \lesssim 500$ GeV.
In this region $\tilde \chi_1^0$ decays almost exclusively 
to $\gamma \widetilde G$, since the $Z \widetilde G$ mode
is phase-space suppressed.
In this region Higgsino productions (with $\mu \lesssim 500$ GeV) have cross-section $\gtrsim 33$ fb, and
they contribute to the $[\gamma \widetilde G] [\gamma \widetilde G] + {\rm jets}$ final state via
$\widetilde H \to h \widetilde B$, $\widetilde B \to \gamma \widetilde G$.
This exceeds the aforementioned bound, $\sigma \lesssim 10$ fb, for
$[\gamma \widetilde G] [\gamma \widetilde G] + {\rm jets}$,
and we conclude the SUSY $(g-2)_\mu$ solution in the BHL and BHR
planes is incompatible with LHC constraints.

\medskip

In the BLR planes the NLSP neutralino is almost pure-Bino.
We see in the right panels of Fig.~\ref{fig:MSSM_BLR} 
that a good $(g-2)_\mu$ fit requires $M_1 \simeq m_{\tilde \chi_1^0} \lesssim 250$ GeV.
In this region 
the cross-section bound, $\sigma \lesssim 10$ fb, 
may apply to the inclusive SUSY processes 
giving rise to 
the $[\gamma \widetilde G] [\gamma \widetilde G] + {\rm jets}$
final state.
The dominant SUSY processes in the BLR planes
are found to be
the left-handed slepton/sneutrino 
and the right-handed slepton productions,
and
the bound $\sigma \gtrsim 10$ fb translates to
$\tilde m_{l_L} \gtrsim 400$ GeV 
and  
$\tilde m_{l_R} \gtrsim 200$ GeV. 
We see that these mass bounds exclude the whole 1 $\sigma$ 
$(g-2)_\mu$ region in the ${\bf BLR}_{50}$ plane.
At the vicinity of the LEP stau mass bound,
a small part of the 2 $\sigma$ region in the ${\bf BLR}_{50}$ and ${\bf BLR}_{10}$ planes as well as 
of the 1 $\sigma$ region in the ${\bf BLR}_{10}$ plane 
is still allowed.
More detailed analysis for the NLSP neutralino scenario 
is provided in Appendix \ref{app:n1_nlsp}.

\subsection{Slepton/Stau NLSP}
\label{sec:stau}

In the previous subsection, we showed that
the SUSY $(g-2)_\mu$ solution is incompatible
with the LHC constraints in the gravitino LSP scenario 
when the NLSP is $\tilde \chi_1^0$.
In this case the neutralino decays into the massless gravitino and a neutral gauge boson, $\tilde \chi_1^0 \to \widetilde G + \gamma/Z/h$, and the final states must contain 
two energetic neutral bosons plus large $\met$.
Such final states are much easier to be detected 
compared to the stable neutralino case studied in section \ref{sec:stable}.\footnote{We designed the parameter planes such that the two smallest mass parameters are 20 GeV apart from each other.  
When lighter particles are pair produced and decay into the LSP neutralinos, the momentum direction of $\tilde \chi_1^0$ tends not to alter significantly from the original direction of the mother particle due to the small mass difference.
Consequently the two neutralinos in such an event are almost back-to-back
in the transverse plane and cancel the net missing transverse energy.
This cancellation is, on the other hand, lost in the gravitino LSP scenario, since the mass difference between $\tilde \chi_1^0$
and the massless $\widetilde G$ cannot be small and
the decay $\tilde \chi_1^0 \to \widetilde G + \gamma/Z/h$
is energetic.
}

{In this subsection, we study the gravitino LSP scenario 
where the NLSP is given by the slepton, sneutrino or stau.}
To do this systematically, we take the
${\bf WHL}_\mu$,
${\bf BHL}_\mu$ and
${\bf BHR}_\mu$ planes
and simply put the slepton soft mass parameter 
20 GeV {\it below} the smallest of the gaugino and Higgsino masses.  Namely we set
\begin{itemize}
\item 
${\bf WHL}_\mu$\,;~~~$\tilde m_{l_L} = \min(M_2, \mu) - 20$ GeV, ~~~NLSP:~$\tilde \nu$
\item 
${\bf BHL}_\mu$\,;~~~~$\tilde m_{l_L} = \min(M_1, \mu) - 20$ GeV, ~~~NLSP:~$\tilde \nu$  
\item 
${\bf BHR}_\mu$\,;~~~\,$\tilde m_{l_R} = \min(M_1, |\mu|) - 20$ GeV, ~~~NLSP:~$\tilde e_R$, $\tilde \mu_R$, $\tilde \tau_R$ 
\end{itemize}
For the BLR planes 
we set $M_1$ at 20 GeV {\it above} the stau mass;
\begin{itemize}
\item
${\bf BLR}_{50/10}$\,;~~~~$M_1 = m_{\tilde \tau_1} + 20$ GeV,~~~NLSP:~$\tilde \tau_1$.
\end{itemize}
In this setup, the NLSP turns out to be
sneutrinos
in ${\bf WHL}_\mu$ and ${\bf BHL}_\mu$,
since the electroweak symmetry breaking
introduces a small mass splitting in the left-handed slepton 
doublet in such a way that the sneutrino becomes lighter
than the charged slepton.
In ${\bf BHR}_\mu$ the NLSP is given by
three mass-degenerate right-handed charged sleptons,
$\tilde e_{R}, \tilde \mu_{R}$ and $\tilde \tau_R$.
In ${\bf BLR}_{50/10}$, on the other hand,
the NLSP becomes the lighter stau, $\tilde \tau_1$,
in which $\tilde \tau_R$ and $\tilde \tau_L$ are largely mixed.

\begin{table}[t!]
\begin{center}
\begin{tabular}{ |c|c|c|c|c| } 
\hline
Analysis & Ref & $E$/TeV & ${\cal L}$/fb$^{-1}$  & Colour  \\
\hline
CMS multilepton & \cite{CMS:2017moi} & 13 & 35.9 & Blue  \\
\hline
CMS soft $\ell^+ \ell^-$ & \cite{CMS:2018kag} & 13 & 35.9 & Orange  \\
\hline
ATLAS soft-$\ell$ & \cite{ATLAS:2017vat} & 13 & 36.1 & Purple  
\\
\hline
ATLAS $\tau^+ \tau^-$ & \cite{ATLAS:2019gti} & 13 & 139 & Magenta  \\
\hline
\end{tabular}
\caption{\label{tb:gmsb_stau}
\small
Analyses in CheckMATE which are relevant for GMSB scenario with NLSP different from neutralino.}
\end{center}
\end{table}

Since the LHC signature of these models 
is unconventional and simplified model limits 
are not available,
we estimate the LHC constraints 
using the MC chain outlined in subsection \ref{sec:lhc}.
Although we include in our analysis all of 8 and 13 TeV analyses
listed in the tables in Appendix \ref{app:checkmate},
only a few give the 95 \% CL exclusion.
We list those relevant ATLAS and CMS analyses in Table \ref{tb:gmsb_stau}.

\begin{figure}[t!]
\centering
\includegraphics[width=0.42\textwidth]{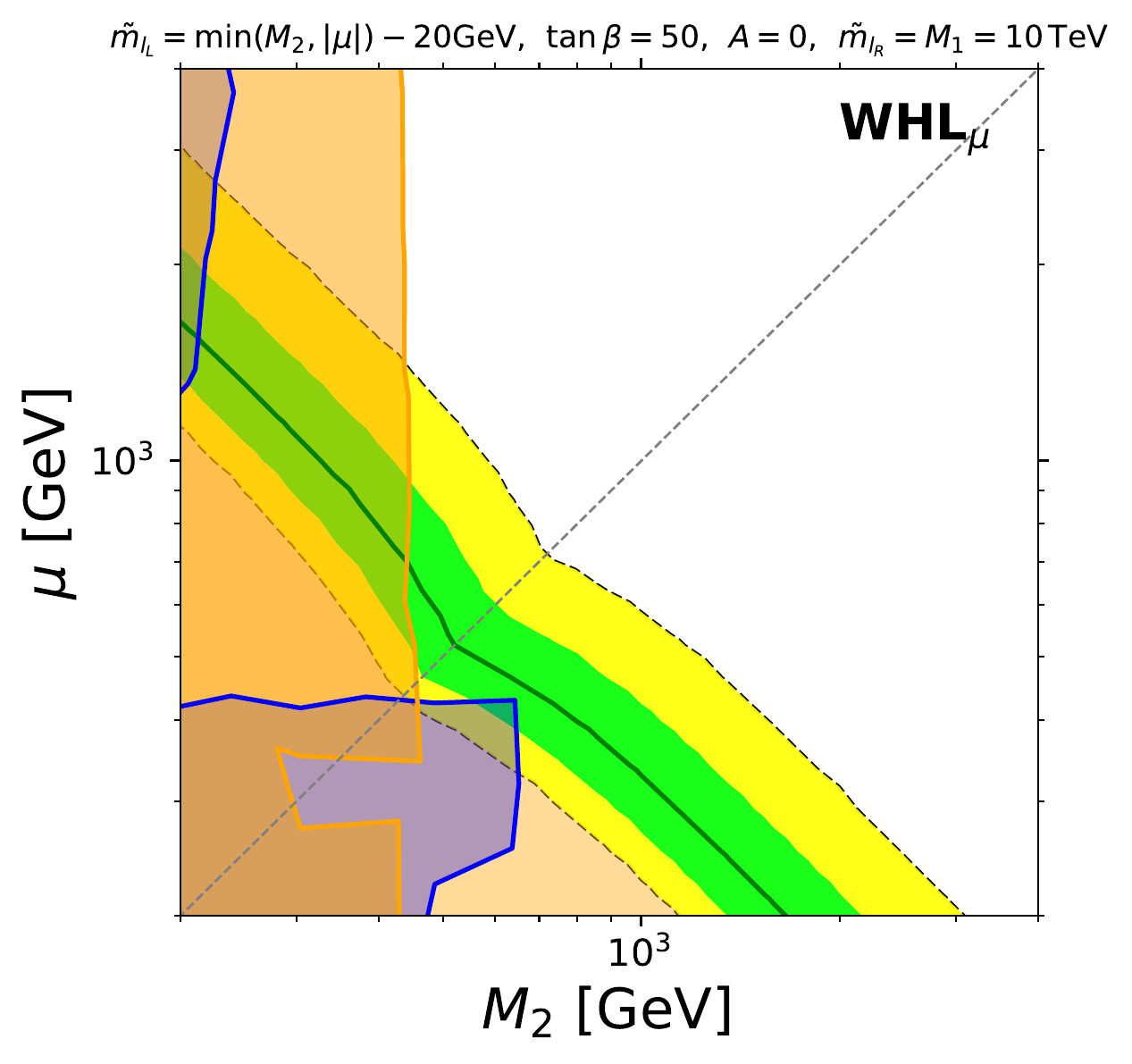}
\includegraphics[width=0.42\textwidth]{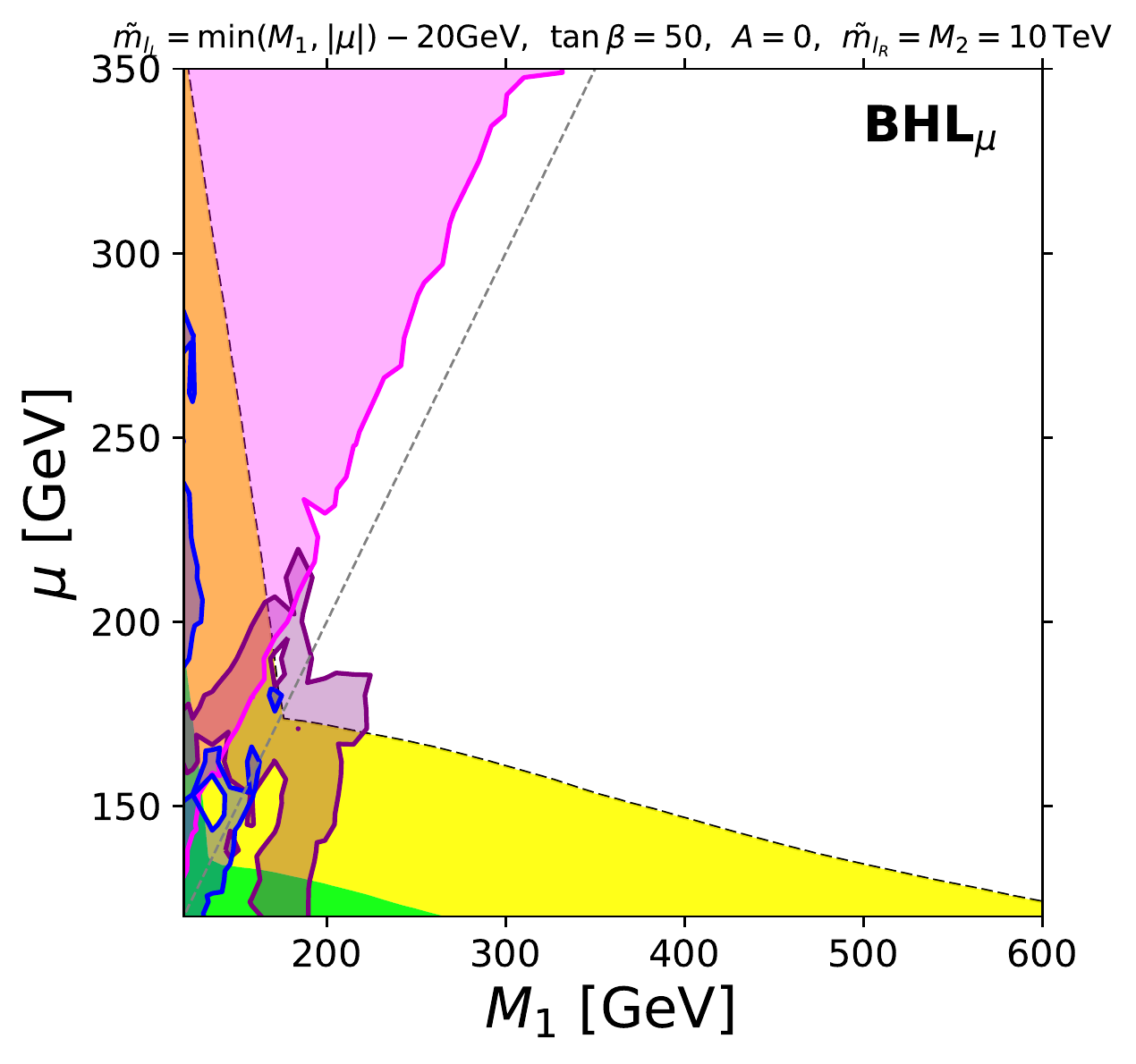}
\includegraphics[width=0.42\textwidth]{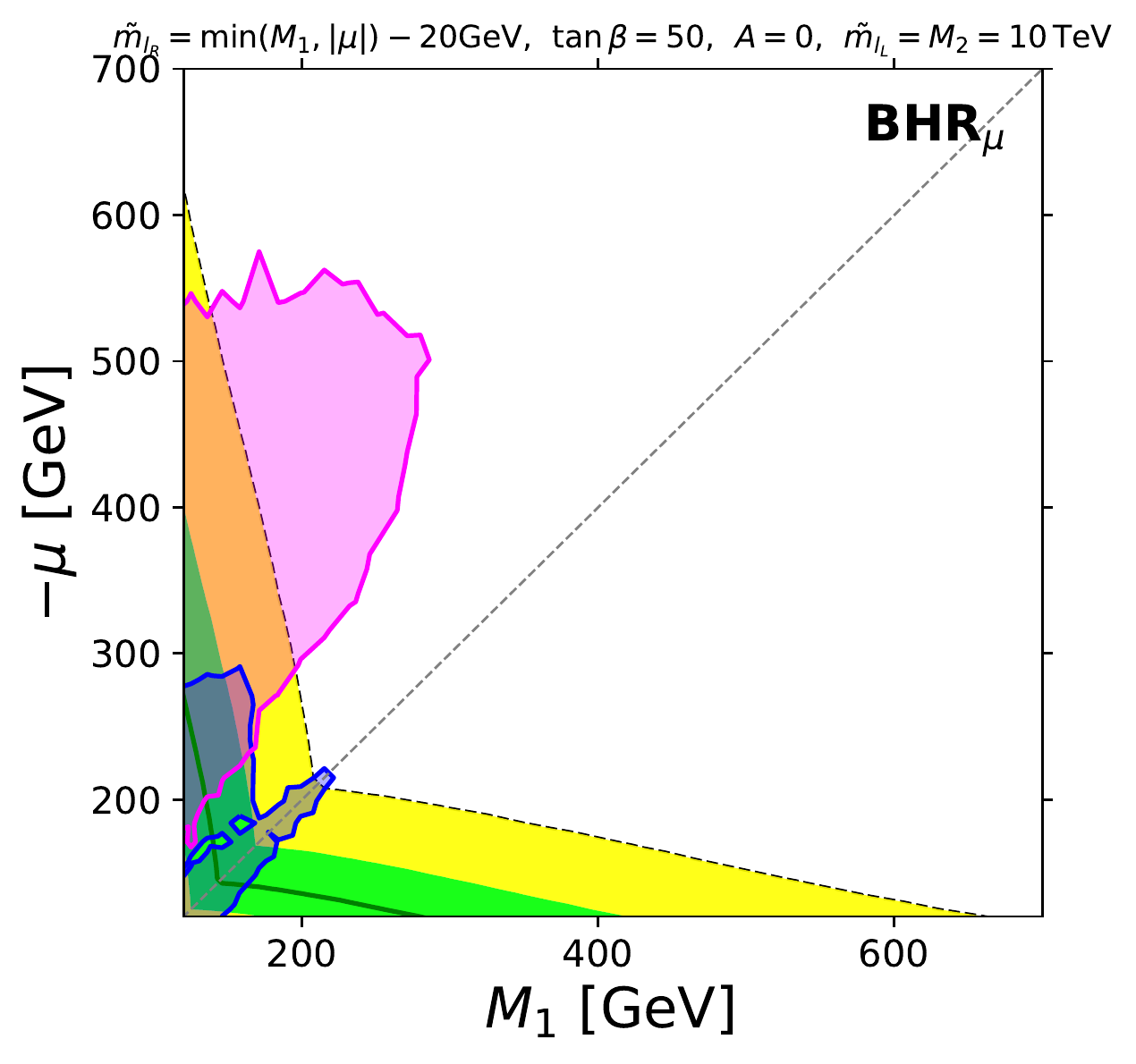}
\\
\includegraphics[width=0.42\textwidth]{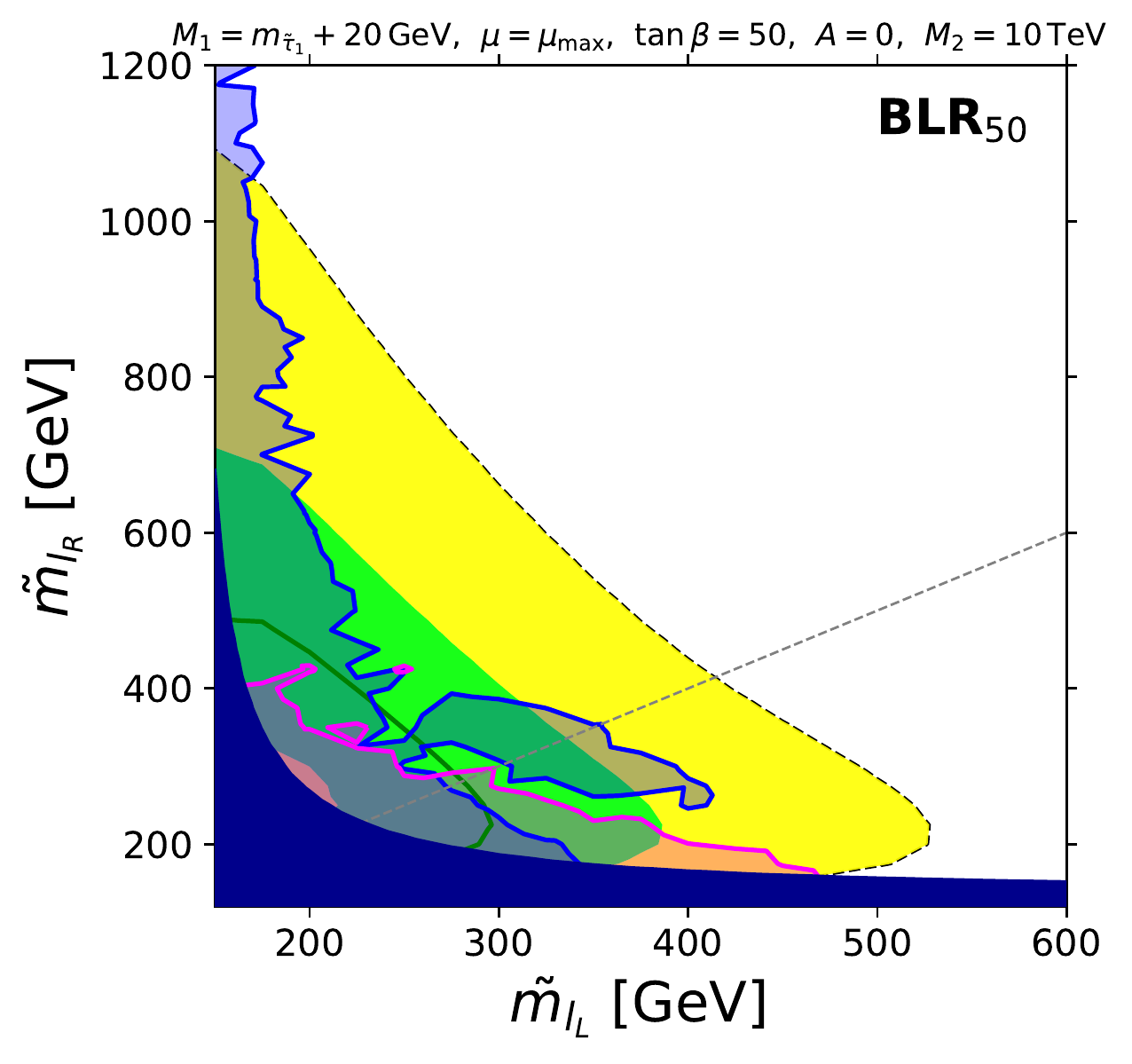}
\includegraphics[width=0.42\textwidth]{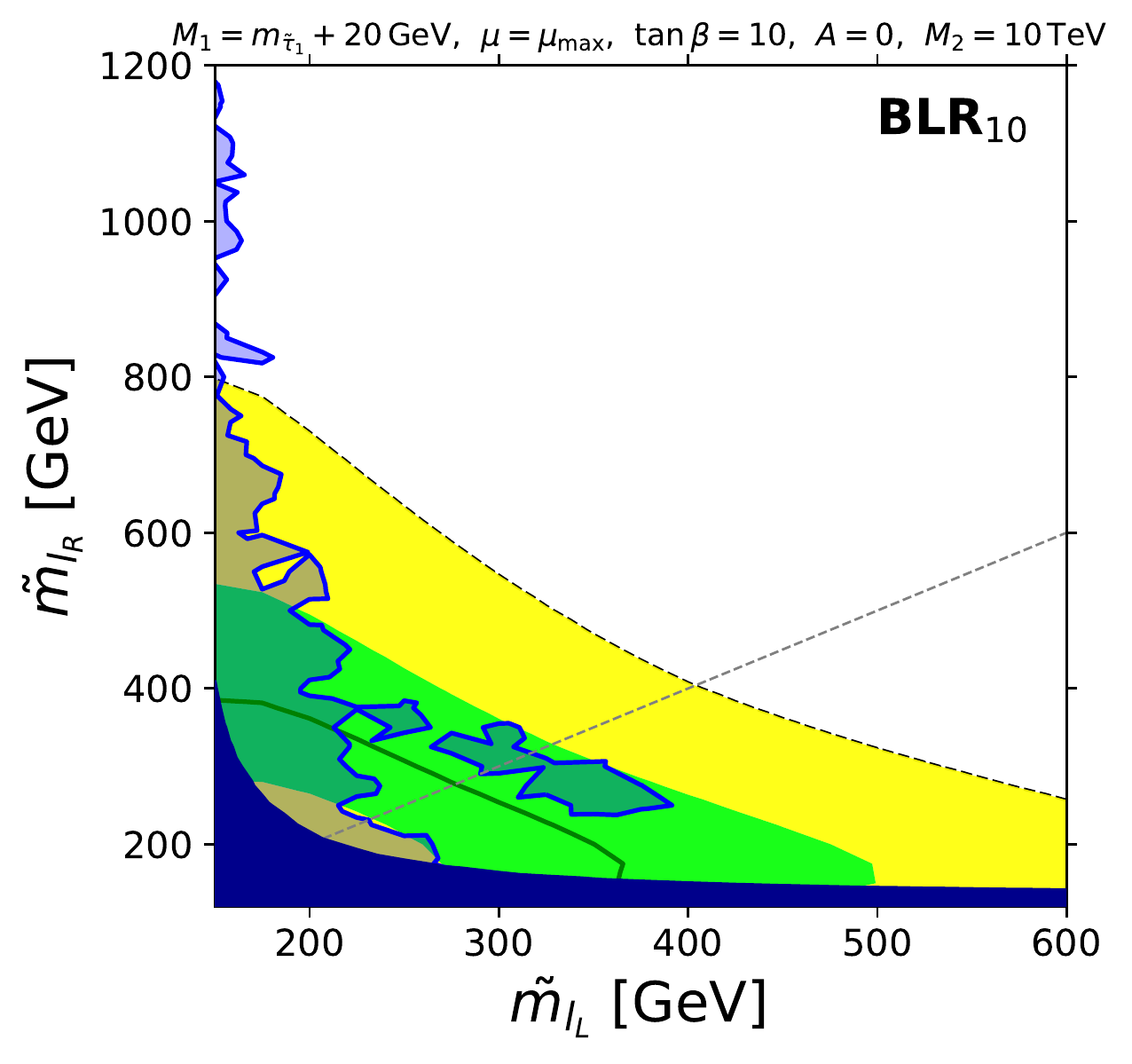}
\caption{\small
Results for GMSB with NLSP different from neutralino: 
${\bf WHL}_\mu$ (upper left),  ${\bf BHL}_\mu$ (upper right),
${\bf BHR}_\mu$ (middle),
${\bf BLR}_{50}$ (lower left) and  ${\bf BLR}_{10}$ (lower right)
planes. 
Region of parameter space allowed by the latest $a_\mu$ experimental results \cite{Muong-2:2021ojo} is depicted with green and yellow bands, corresponding to one and two sigma agreement respectively.
Blue, orange, purple and magenta shaded regions are excluded by LHC constraints. 
Dark blue shaded area is excluded by LEP stau mass bound \cite{Zyla:2020zbs}.
}
\label{fig:gmsb_stau}
\end{figure}


\medskip

We first look at the ${\bf WHL}_\mu$ plane shown 
in the top left panel of Fig.~\ref{fig:gmsb_stau}.
In the plot the blue and orange shaded regions
are excluded at 95 \% CL 
by the CMS multilepton \cite{CMS:2017moi} 
and
CMS soft $\ell^+ \ell^-$ \cite{CMS:2018kag} 
analyses, respectively.
The CMS multilepton analysis requires 
the signal electrons (muons) to have
$\pT > 25 (20)$ GeV.
This analysis is therefore sensitive to
production of heavier electroweakino states, $\tilde \chi_{\rm heavy}$, since relatively hard leptons may be produced 
from the decay of $\tilde \chi_{\rm heavy}$.
On the other hand, CMS soft $\ell^+ \ell^-$
exploits soft leptons with $5 < \pT/{\rm GeV} < 30$.
Therefore, the analysis is sensitive even for 
production of lighter electroweakino states.
In the orange shaded region in the plot
the production of Wino pairs followed by 
$\widetilde W \to l \tilde l^{(\prime)}$
($l^{(\prime)} = (e, \mu, \nu_e, \nu_\mu)$)
mainly contributes to the exclusion.
We find that the charged sleptons then decay
into the sneutrino and a $f \bar f$ pair via an off-shell $W^{\pm}$.
The decay of NLSP sneutrino 
is totally invisible, $\tilde \nu \to \nu \widetilde G$,
and the NLSP sneutrinos may effectively be treated as the invisible LSP.
In the $\mu < M_2$ region, 
Higgsino production cannot 
give a large contribution to the 
CMS soft $\ell^+ \ell^-$ analysis,
since the Higgsinos predominantly decay into
the third generation sleptons; $\widetilde H \to l_3 \tilde l^{(\prime)}_3$ ($l_3^{(\prime)} = (\tau, \nu_\tau)$).
Due to this almost entire 
1 $\sigma$ $(g-2)_\mu$ region
remains unconstrained 
in the $\mu \lesssim M_2$ region.

\medskip

The top right panel of Fig.~\ref{fig:gmsb_stau} 
displays the ${\bf BHL}_\mu$ plane.
We see that the region with
$\tilde m_{l_L} \lesssim M_1 < \mu$
is excluded by
the ATLAS $\tau^+ \tau^-$ (magenta).
The excluded signature comes from
the Higgsino pair production followed by
$\widetilde H^0 \to \tau \tilde \tau_L \to \tau \tau \widetilde G$
and 
$\widetilde H^\pm \to \tau \tilde \nu_L$.
In the bottom left region
with $\tilde m_{l_L} \lesssim M_1 < \mu$,
we also see the exclusion from
the ATLAS soft-$\ell$ {(purple)}
and the CMS multilepton (blue) analyses.
The direct slepton/sneutrino production, $pp \to \tilde l \tilde l$, followed by $\tilde l \to l \widetilde G$ (with $l = \ell$ or $\nu$)
is responsible for these exclusions.
We also see the ATLAS soft-$\ell$ exclusion
is slightly extended to 
the $\mu \lesssim M_1$ region.
{There, the NLSP is given by $\tilde \chi_1^0$
due to large mixing between Higgsino and Bino
and
sleptons decay into 
$\tilde l \to l + \tilde \chi_1^0$, producing soft leptons.}
However, in general, the region with $\mu < M_1$
is rather unconstrained since
all SUSY particles with large production cross-sections (Higgsinos and sleptons/sneutrinos) are mass degenerate.

\medskip

We turn to the ${\bf BHR}_\mu$ plane
shown in the middle panel of 
Fig.~\ref{fig:gmsb_stau}.
Two excluded regions are visible;
ATLAS $\tau^+ \tau^-$ (magenta)
and 
CMS multilepton (blue).
The ATLAS $\tau^+ \tau^-$ exclusion
appears for the same reason as in
the ${\bf BHL}_\mu$ plane.
It excludes the production of Higgsino pair 
followed by $\widetilde H \to \tau \tilde \tau_R$
up to $|\mu| \sim 520$ GeV
provided a large mass difference 
between $\widetilde H$ and $\tilde \tau_R$. 
In this plane energetic taus can also be produced
from the NLSP stau decay, $\tilde \tau_L \to \tau + \widetilde G$.
We also see that 
CMS multilepton 
analysis
excludes 
some region with $300\,{\rm GeV} \gtrsim |\mu| \gtrsim M_1$.
In this region, Bino and Higgsino are largely mixed and leptons are produced from
the decay of EW bosons originated from
heavier electroweakino decays,
$\tilde \chi_{\rm heavy} \to \tilde \chi_{\rm light} + X$ ($X = W^{\pm},Z,h$).
As can be seen in the plot,
the 1 $\sigma$ $(g-2)_\mu$ region
is allowed for $|\mu| < M_1$.

\medskip

Finally, we show the ${\bf BLR}_{50}$ (left)
and ${\bf BLR}_{10}$ (right) planes
in the bottom panels of Fig.~\ref{fig:gmsb_stau}.
In the BLR scenario,
$\tilde \tau_1$ becomes significantly lighter 
than the other sleptons due to the large 
L-R mixing.
Since $M_1$ is fixed at 20 GeV above $m_{\tilde \tau_1}$, there is a mass gap between slepton and Bino.
In the region excluded by 
the CMS multilepton (blue),
the slepton production followed by 
$\tilde \ell^{\pm} \to \ell^{\pm} \widetilde B$
contributes,
where the Bino further decays into
the NLP $\tilde \tau_1$ and a soft-$\tau$, then
$\tilde \tau_1 \to \tau + \widetilde G$.

In ${\bf BLR}_{50}$, the region excluded 
by ATLAS $\tau^+ \tau^-$ (magenta) is also visible, where
the Higgsino production followed by
$\widetilde H \to l_3 \tilde l_3$
and $\tilde \tau_1 \to \tau + \widetilde G$
contributes.
Together with the CMS multilepton constraint, 
almost all 1 $\sigma$ $(g-2)_\mu$ region is excluded

In the ${\bf BLR}_{10}$ plane, on the other hand,
the exclusion from ATLAS $\tau^+ \tau^-$
is absent.
This is because the Higgsinos in this plane are about five times heavier 
than those in ${\bf BLR}_{50}$ as discussed in Section 
\ref{sec:stable} (see also the right panels of Fig.~\ref{fig:MSSM_BLR}).
Because of this and
the fact that the 1 $\sigma$ $(g-2)_\mu$ region is more extended to larger $\tilde m_{l_L}$,
the 1 $\sigma$ $(g-2)_\mu$ region
is allowed for $\tilde m_{l_R} < \tilde m_{l_L}$
in the ${\bf BLR}_{10}$ plane.

{
We have seen that several regions 
(i.e.\ the magenta regions in ${\bf BHL}_\mu$,
${\bf BHR}_\mu$ and ${\bf BLR}_{50}$ in Fig.~\ref{fig:gmsb_stau})
in the slepton NLSP scenario studied in this section
are excluded by the ATLAS $\tau^+ \tau^-$ analysis.
We briefly comment what we expect for these regions if staus are artificially decoupled as we considered for the stable neutralino case at the end of subsection \ref{sec:mssm_result}.
In the ${\bf BHL}_\mu$ plane,
the magenta region was excluded due to  
the Higgsino pair production followed by
$\widetilde H^{0(\pm)} \to \tau \tilde \tau_L (\tilde \nu_\tau)$.
If $\tilde \tau_L$ and $\tilde \nu_\tau$ are decoupled, the Higgsinos predominantly decay into
$V + \tilde \chi_1^0$ ($V = W, Z, h$) if these modes are kinematically allowed.
Otherwise, the Higgsinos decay into $l \tilde l^{(\prime)}$ ($l^{(\prime)} = e,\mu,\nu_e, \nu_\mu$).
The latter case is severely constrained
by the analyses targeting the leptonic final states
\cite{CMS:2017moi, CMS:2018kag, ATLAS:2017vat}.
In the former case, the constraint is milder 
and the region will be allowed as long as the Higgsino-Bino mass splitting is smaller than $\sim$150 GeV \cite{CMS:2020bfa}.
In the ${\bf BHR}_\mu$ plane, the NLSPs will be $\tilde e_R$ and $\tilde \mu_R$ after decoupling $\tilde \tau_R$.
Hard leptons are produced when they decay into
the LSP gravitino and even larger region will be excluded if $\tilde \tau_R$ is decoupled.
In the ${\bf BLR}_{50}$ plane, if staus are decoupled, the vacuum stability constraints can easily be satisfied even for very large $\mu$
and $\mu$ can be fixed such that the SUSY $(g-2)_\mu$ fits the experimental central value.
However, in this case, the NLSPs are given by light flavour sleptons.
Their production and decay into the massless gravitino contributes to the
final state with hard leptons plus large $E_T^{\rm miss}$, which is strongly constrained by the CMS $\ell^+ \ell^-$ analysis \cite{CMS:2020bfa}. 
Overall, we do not expect in general more regions to open up for SUSY $(g-2)_\mu$ by decoupling staus 
in the slepton NLSP scenario discussed in this subsection.
}

\section{Conclusion}
\label{sec:concl}

We have investigated in this article 
supersymmetric solutions to 
the muon $(g-2)$ anomaly
with and without the stable neutralino.
We have organised our analysis
based on the four types of 1-loop 
diagrams generating the SUSY
$(g-2)_\mu$ contribution,
denoted by WHL, BHL, BHR and BLR,
as shown in Fig.~\ref{fig:diagrams}.

For significant contribution
the first three types require 
large gaugino-Higgsino mixing
in the neutralino state.
However, such a neutralino is severely 
constrained by the DM direct detection experiments.
On the other hand, when the BLR contribution
is dominant, the neutralino is almost 
pure Bino, which suffers from the problem 
of neutralino overproduction.
In order to fit the observed 
$(g-2)_\mu$ anomaly, 
some of the electroweakly interacting sparticles 
must be very light, $\lesssim {\cal O}(500)$ GeV.
Those light particles would be produced at the LHC and give rise to the signature with large $\met$.

In section \ref{sec:stable}
we have demonstrated these points numerically.
We showed that the preferred $(g-2)_\mu$ region
is in fact severely constrained by the combination of constraints from the neutralino overproduction, DM direct detection and direct BSM searches at the LHC, assuming the neutralino is stable.
We nevertheless found still-allowed 
1 $\sigma$ $(g-2)_\mu$ regions 
in the WHL planes with $(M_2, \mu, \widetilde m_{l_L}) \sim (1500,\, 300,\, 320)$ GeV 
and $(350,\, 600,\, 370$) GeV
(Fig.~\ref{fig:MSSM_WHL})
and the BHR planes with $(M_1, \mu, \widetilde m_{l_R}) \sim (300,\, -120,\, 150)$ GeV
(Fig.~\ref{fig:MSSM_BHLR}),
though these regions will be explored in the next generation DM direct detection experiments.
Interestingly the whole $(g-2)_\mu$ region
is excluded in the BLR planes by
the LHC constraint in the dilepton plus large $\met$ channel (Fig.~\ref{fig:MSSM_BLR}).

Moving to supersymmetric scenarios without  stable neutralino, the DM constraints can be trivially avoided.  
The LHC constraints may also be relaxed 
though the details of the LHC signature
and constraints depend on the exact scenarios.

In section \ref{sec:rpv} we studied 
the scenario where the neutralino decays 
into three (anti-)quarks via the R-parity violating $UDD$ operator.
We have shown in Figs.~\ref{fig:RPV_1} and \ref{fig:RPV_2} that
although the large ${\met}$ signature 
is unavailable in this scenario, some regions 
are excluded by the LHC searches 
exploiting the multijet and multilepton channels.
Nevertheless, the allowed $(g-2)_\mu$ region
is significantly extended compared to the stable neutralino case.
In particular, the BLR solution, which is excluded in the stable neutralino case, is entirely allowed in the RPV scenario.

Another well-known example where
the neutralino is unstable is 
the scenario with gravitino LSP, 
which we studied in section \ref{sec:grav}.
Adding the approximately massless gravitino
in the spectrum, the NLSP neutralino
decays into the gravitino in association with
one of the neutral boson, $\gamma/Z/h$.
There are several LHC analyses targeting these final states.  Using the cross-section limits 
derived in these analyses, we showed that
the SUSY $(g-2)_\mu$ solution in our 2D parameter planes is incompatible with the LHC constraints.
A more detailed analysis is provided in Appendix \ref{app:n1_nlsp}.
This is an interesting example showing 
that making the neutralino unstable does not always relax the phenomenological constraints for $(g-2)_\mu$.

In subsection \ref{sec:grav_neu}
we studied a gravitino LSP scenario
where the NLSP is provided by $\tilde \ell/\tilde \nu/\tilde \tau_1$.
In order to study these cases systematically we 
simply set the soft breaking slepton mass 
20 GeV below the minimum of the gaugino and the Higgsino mass in the ${\bf WHL}_\mu$, ${\bf BHL}_\mu$ and ${\bf BHR}_\mu$ planes.
For the ${\bf BLR}_{50/10}$ planes 
we set $M_1$ 20 GeV above the lighter stau mass, $m_{\tilde \tau_1}$.
In this setup, the NSLP is found to be
sneutrino in 
${\bf WHL}_\mu$ and ${\bf BHL}_\mu$,
right-handed charged sleptons 
in ${\bf BHR}_\mu$
and the lighter stau in ${\bf BLR}_{50/10}$.
As shown in Fig.~\ref{fig:gmsb_stau}
some regions are constrained 
by the LHC searches exploiting 
the multilepton, soft-$\ell$ and di-$\tau$
channels.
Unlike the neutralino NLSP case,
we found several 1 $\sigma$ $(g-2)_\mu$ regions
particularly in the ${\bf WHL}_\mu$,
${\bf BHR}_\mu$ and ${\bf BLR}_{10}$ planes.

For unstable neutralinos,
the region favoured by the $(g-2)_\mu$ anomaly 
can only be probed by collider experiments.
If the observed $(g-2)_\mu$ anomaly
is indeed a sign of supersymmetry,
it is very important to understand 
the collider signature and develop 
a methodology to find it 
for both stable and unstable neutralino 
scenarios.
We however leave this task as an interesting future work.

\section*{Acknowledgments}

The work of R.M.\
is partially supported by the National Science Centre, Poland,
under research grant 2017/26/E/ST2/00135
and the Beethoven grant DEC-2016/23/G/ST2/04301.
The work of
K.S.\ is partially supported 
by the National Science Centre, Poland, under research grant\\ 2017/26/E/ST2/00135 and 
the Norwegian Financial Mechanism for years 2014-2021, grant DEC-2019/34/H/ST2/00707.

\newpage

\appendix

\begin{appendices}
\section{CheckMATE analyses}
\label{app:checkmate}

In the tables below, 
we list the ATLAS and CMS analyses used in our numerical analyses with {\tt CheckMATE}.
Tables 
\ref{tab:CM_ATLA13TeV},
\ref{tab:CM_CMS13TeV},
\ref{tab:CM_ATLAS8TeV},
and \ref{tab:CM_CMS8TeV}
list 
the 13 TeV ATLAS, 
13 TeV CMS,
8 TeV ATLAS and 
8 TeV CMS analyses,
respectively.

\begin{center}
\begin{table}[!h]
\begin{tabular}{ |c|c|c|c| } 
 \hline
 Name & $E$/TeV & ${\cal L}$/fb$^{-1}$ & Description \\ 
 \hline \hline
 atlas\_1604\_01306 & 13 & 3.2 & Monophoton \\ 
 \hline
 atlas\_1605\_09318 & 13 & 3.3 & 3 b-jets + 0-1 lepton + MET \\ 
 \hline
 atlas\_1609\_01599 & 13 & 36 & Monophoton \\ 
 \hline
 atlas\_1704\_03848 & 13 & 36 & Monophoton \\ 
 \hline
 atlas\_conf\_2015\_082 & 13 & 3.2 & 2 leptons (Z) + jets + MET \\ 
 \hline
 atlas\_conf\_2016\_013 & 13 & 3.2 & 1 lepton + jets (4 tops, VVL quarks) \\ 
 \hline
 atlas\_conf\_2016\_050 & 13 & 13.3 & 1 lepton + (b) jets + MET \\ 
 \hline
 atlas\_conf\_2016\_054 & 13 & 13.3 & 1 lepton + (b) jets + MET \\ 
 \hline
 atlas\_conf\_2016\_076 & 13 & 13.3 & 2 lepton + jets + MET \\ 
 \hline
 atlas\_conf\_2016\_096 & 13 & 13.3 & Multi-lepton + MET \\ 
 \hline
 atlas\_conf\_2017\_060 & 13 & 36 & Monojet \\ 
 \hline
 atlas\_conf\_2016\_066 & 13 & 13.3 & Photons, jets and MET \\ 
 \hline
 atlas\_1712\_08119 & 13 & 36 & soft leptons (compressed EWKinos) \\ 
 \hline
 atlas\_1712\_02332 & 13 & 36 & squarks and gluinos, 0 lepton, 2-6 jets \\ 
 \hline
 atlas\_1709\_04183  & 13 & 36 & Jets + MET (stops) \\ 
 \hline
 atlas\_1802\_03158  & 13 & 36 & search for GMSB with photons \\ 
 \hline
 atlas\_1708\_07875   & 13 & 36 & EWKino search with taus and MET \\ 
 \hline
 atlas\_1706\_03731 & 13 & 36 & Multilepton + Jets + MET (RPC and RPV) \\ 
 \hline
 atlas\_1908\_08215  & 13 & 36 & 2 leptons + MET (EWKinos)  \\ 
 \hline
 atlas\_1909\_08457 & 13 & 139 & SS lepton + MET (squark, gluino) \\ 
 \hline
 atlas\_conf\_2019\_040 & 13 & 139 & Jets + MET (squark, gluino) \\ 
 \hline
 atlas\_conf\_2019\_020 & 13 & 139 & 3 leptons (EWKino) \\ 
 \hline
 atlas\_1803\_02762 & 13 & 36 & 2 or 3 leptons (EWKino) \\ 
\hline
 atlas\_conf\_2018\_041 & 13 & 80 & Multi-$b$-jets (stops, sbottoms) \\ 
 \hline
 atlas\_2101\_01629 & 13 & 139 & 1 lepton + jets + MET \\ 
 \hline
 atlas\_conf\_2020\_048 & 13 & 139 & Monojet \\ 
 \hline
 atlas\_2004\_14060  & 13 & 139 & $t \bar t$ + MET \\ 
 \hline
 atlas\_1908\_03122 & 13 & 139 & Higgs bosons + $b$-jets + MET \\ 
 \hline
 atlas\_2103\_11684  & 13 & 139 & 4 or more leptons (RPV, GMSB) \\ 
 \hline
 atlas\_2106\_09609 & 13 & 139 & Multijets + leptons (RPV) \\ 
 \hline
 atlas\_1911\_06660 & 13 & 139 & Search for Direct Stau Production\\ 
 \hline
\end{tabular}
\caption{\label{tab:CM_ATLA13TeV}List of 13 TeV ATLAS analyses available in CheckMATE and which have been validated against published experimental results.}
\end{table}
\end{center}

\begin{center}
\begin{table}
\begin{tabular}{ |c|c|c|c| } 
 \hline
 Name & $E$/TeV & ${\cal L}$/fb$^{-1}$ & Description \\ 
 \hline \hline
 cms\_pas\_sus\_15\_011 & 13 & 2.2 & 2 leptons + jets + MET\\
 \hline
 cms\_sus\_16\_039 & 13 & 35.9 & electroweakinos in multilepton final state\\
 \hline
 cms\_sus\_16\_025 & 13 & 12.9 & electroweakino and stop compressed spectra\\
 \hline
 cms\_sus\_16\_048 & 13 & 35.9 & two soft opposite sign leptons\\
 \hline
\end{tabular}
\caption{\label{tab:CM_CMS13TeV}List of 13 TeV CMS analyses available in CheckMATE and which have been validated against published experimental results.}
\end{table}
\end{center}

\begin{center}
\begin{table}
\begin{tabular}{ |c|c|c|c| } 
 \hline
 Name & $E$/TeV & ${\cal L}$/fb$^{-1}$ & Description \\ 
 \hline \hline
 atlas\_1308\_1841 & 8 & 20.3 &  0 lepton + $\geq$ 7 jets + MET \\
 \hline
 atlas\_1308\_2631 &  8 & 20.1 &0 leptons + 2 b-jets + MET\\
 \hline
 atlas\_1402\_7029 & 8 & 20.3 & 3 leptons + MET (chargino+neutralino)\\
 \hline
atlas\_1403\_4853 & 8 & 20.3 & 2 leptons + MET (direct stop)\\
\hline
atlas\_1403\_5222 & 8 & 20.3 & stop production with Z boson and b-jets\\
\hline
atlas\_1404\_2500 & 8 & 20.3 & Same sign dilepton or 3 lepton\\
\hline
atlas\_1405\_7875 & 8 & 20.3 & 0 lepton + 2-6 jets + MET\\
\hline
atlas\_1407\_0583 & 8 & 20.3 & ATLAS, 1 lepton + (b-)jets + MET (stop)\\
\hline
atlas\_1407\_0608 & 8 & 20.3 & Monojet or charm jet (stop)\\
\hline
atlas\_1411\_1559 & 8 & 20.3 & monophoton plus MET\\
\hline
atlas\_1501\_07110 & 8 & 20.3 & 1 lepton + 125GeV Higgs + MET\\
\hline
atlas\_1502\_01518 & 8 & 20.3 & Monojet + MET \\
\hline
atlas\_1503\_03290 & 8 & 20.3 & 2 leptons + jets + MET\\
\hline
atlas\_1506\_08616 & 8 & 20.3 & di-lepton and 2b-jets + lepton\\
\hline
atlas\_1507\_05493 & 8 & 20.3 & photonic signatures of gauge-mediated SUSY\\
\hline
atlas\_conf\_2012\_104 & 8 & 20.3 & 1 lepton + $\geq$ 4 jets + MET\\
\hline
atlas\_conf\_2013\_024 & 8 & 20.3 & 0 leptons + 6 (2 b-)jets + MET\\
\hline
atlas\_conf\_2013\_049 & 8 & 20.3 & 2 leptons + MET\\
\hline
atlas\_conf\_2013\_061 & 8 & 20.3 & 0-1 leptons + $\geq$ 3 b-jets + MET\\
\hline
atlas\_conf\_2013\_089 & 8 & 20.3 & 2 leptons (razor)\\
\hline
atlas\_conf\_2015\_004 & 8 & 20.3 & invisible Higgs decay in VBF\\
\hline
atlas\_1403\_5294 & 8 & 20.3 & 2 leptons + MET, (SUSY electroweak)\\
\hline
atlas\_higg\_2013\_03 & 8 & 20.3 & 2 leptons + MET, (invisible Higgs)\\
\hline
atlas\_1502\_05686 & 8 & 20.3 & search for massive sparticles decaying to many jets\\
\hline
\end{tabular}
\caption{\label{tab:CM_ATLAS8TeV}List of 8 TeV ATLAS analyses available in CheckMATE and which have been validated against published experimental results.}
\end{table}
\end{center}

\begin{center}
\begin{table}[!h]
\begin{tabular}{ |c|c|c|c| } 
 \hline
 Name & $E$/TeV & ${\cal L}$/fb$^{-1}$ & Description \\ 
 \hline \hline
 cms\_1303\_2985 & 8 & 11.7 & $\alpha_T$ + b-jets \\
 \hline
 cms\_1408\_3583 & 8 & 19.7 & monojet + MET\\
 \hline
 cms\_1502\_06031 & 8 & 19.4 & 2 leptons, jets, MET (only on-Z)\\
 \hline
 cms\_1504\_03198 & 8 & 19.7 & 1 lepton, $\geq$ 3 jets, $\geq$ 1 b-jet, MET (DM + 2 top)\\
 \hline
 cms\_sus\_13\_016 & 8 & 19.5 & OS lepton 3+ b-tags\\
 \hline
 \end{tabular}
 \caption{\label{tab:CM_CMS8TeV}List of 8 TeV CMS analyses available in CheckMATE and which have been validated against published experimental results.}
\end{table}
\end{center}


\vskip -2cm
\FloatBarrier
\section{LHC constraints on the neutralino NLSP scenario}
\label{app:n1_nlsp}
\FloatBarrier

We show in Fig.~\ref{fig:GMSB_neut}
the preferred $(g-2)_\mu$ regions
(green; 1 $\sigma$ and yellow; 2 $\sigma$)
and the excluded regions by
the 
$[Z \widetilde G][Z \widetilde G]$ (blue)
$[Z \widetilde G][h \widetilde G]$ (blue)
$[\gamma \widetilde G][Z(h) \widetilde G]$ (red)
and
$[\gamma \widetilde G][\gamma \widetilde G] + {\rm jets}$ (Magenta) final states.
The information of the analyses used here are listed in Table \ref{tab:GMSB_neut}.
As discussed in section \ref{sec:grav_neu}
almost all preferred $(g-2)_\mu$ regions
are excluded in the gravitino LSP scenario 
with the neutralino NLSP.

\begin{table}[h!]
\begin{center}
\begin{tabular}{ |c|c|c|c|c| } 
\hline
Analysis & $E$/TeV & ${\cal L}$/fb$^{-1}$ & Final State & Colour \\
\hline
\multirow{2}{7em}{CMS $\ell^+ \ell^-$ \cite{CMS:2020bfa}} & \multirow{2}{1em}{13} & \multirow{2}{2em}{137} &
$[Z \widetilde G] [Z \widetilde G]$ & Blue \\ 
\cline{4-5}
& & & $[Z \widetilde G] [h \widetilde G]$ & Green \\ 
\hline
\multirow{2}{10em}{CMS $\gamma$+$\met$ \cite{CMS:2017brl}} & \multirow{2}{1em}{13} & \multirow{2}{2em}{35.9} &
$[ \gamma \widetilde G ] [ Z(h) \widetilde G ]$ & Red \\ 
\cline{4-5}
& & & $[ \gamma \widetilde G] [\gamma \widetilde G] + {\rm jets}$ & Magenta \\ 
\hline
\end{tabular}
\caption{\label{tab:GMSB_neut}
\small
Analyses in CheckMATE which are relevant for GMSB scenario with neutralino NLSP.}
\end{center}
\end{table}

\begin{figure}[h!]

\includegraphics[width=0.31\textwidth]{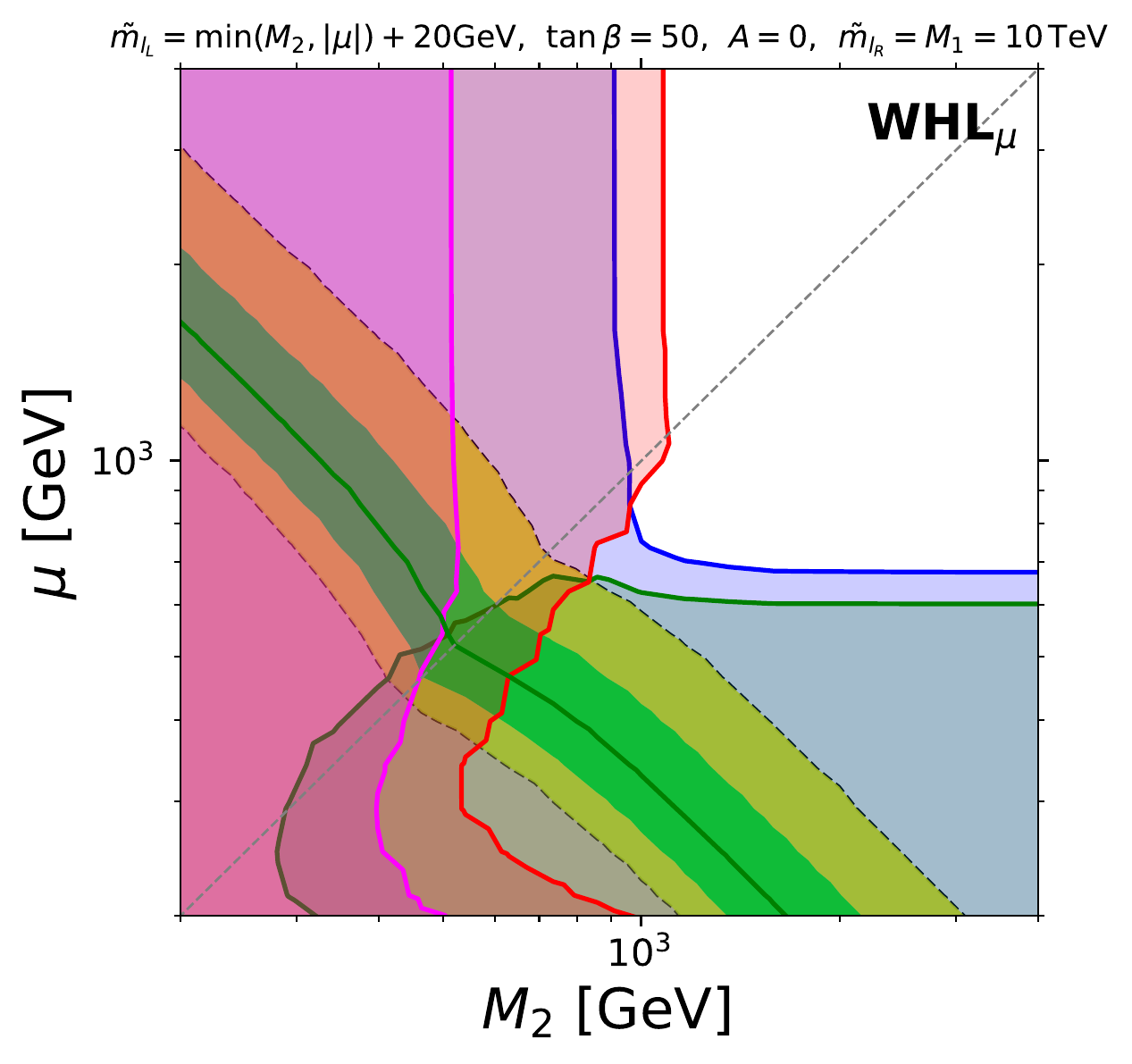}
\includegraphics[width=0.31\textwidth]{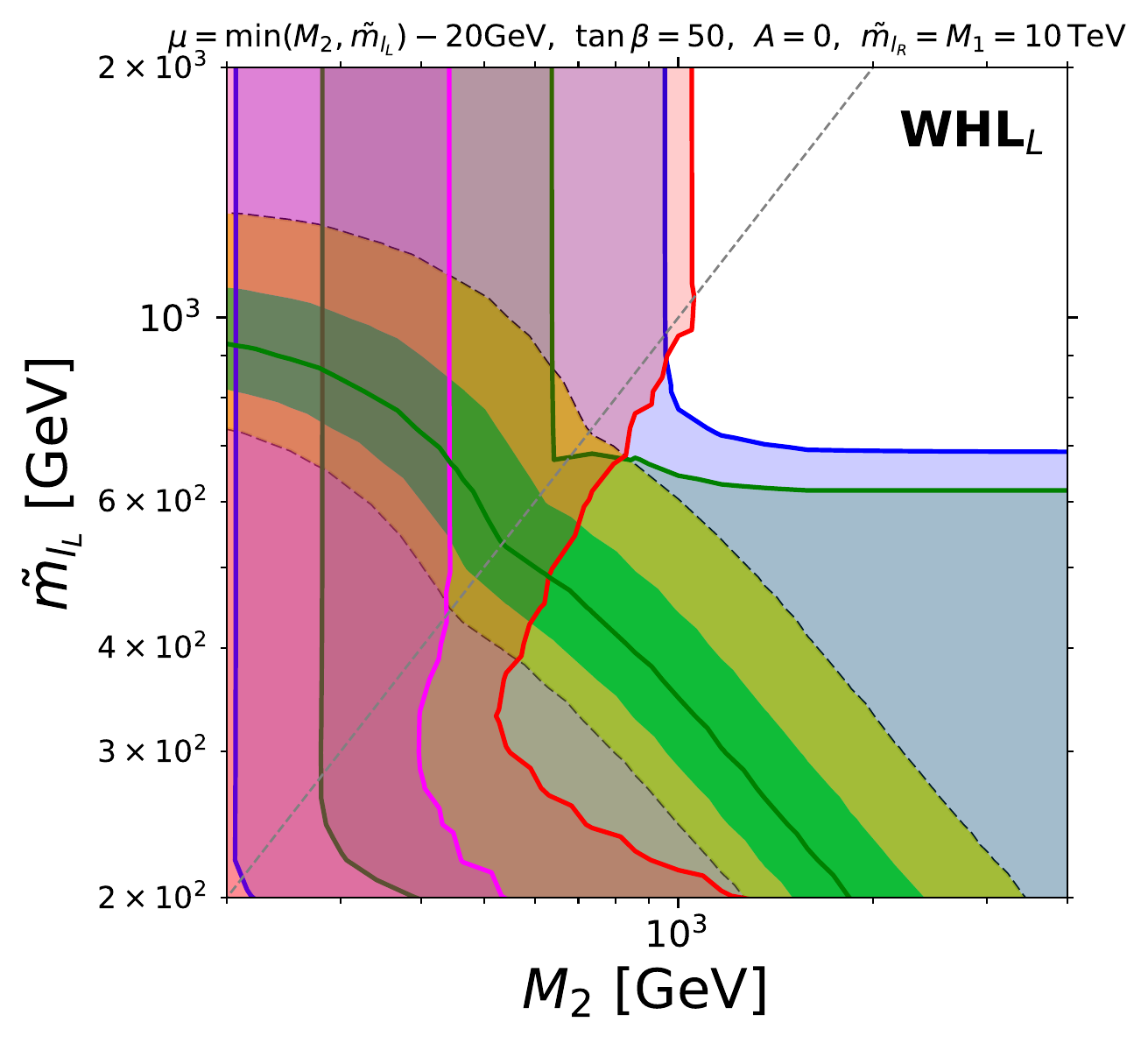}
\includegraphics[width=0.31\textwidth]{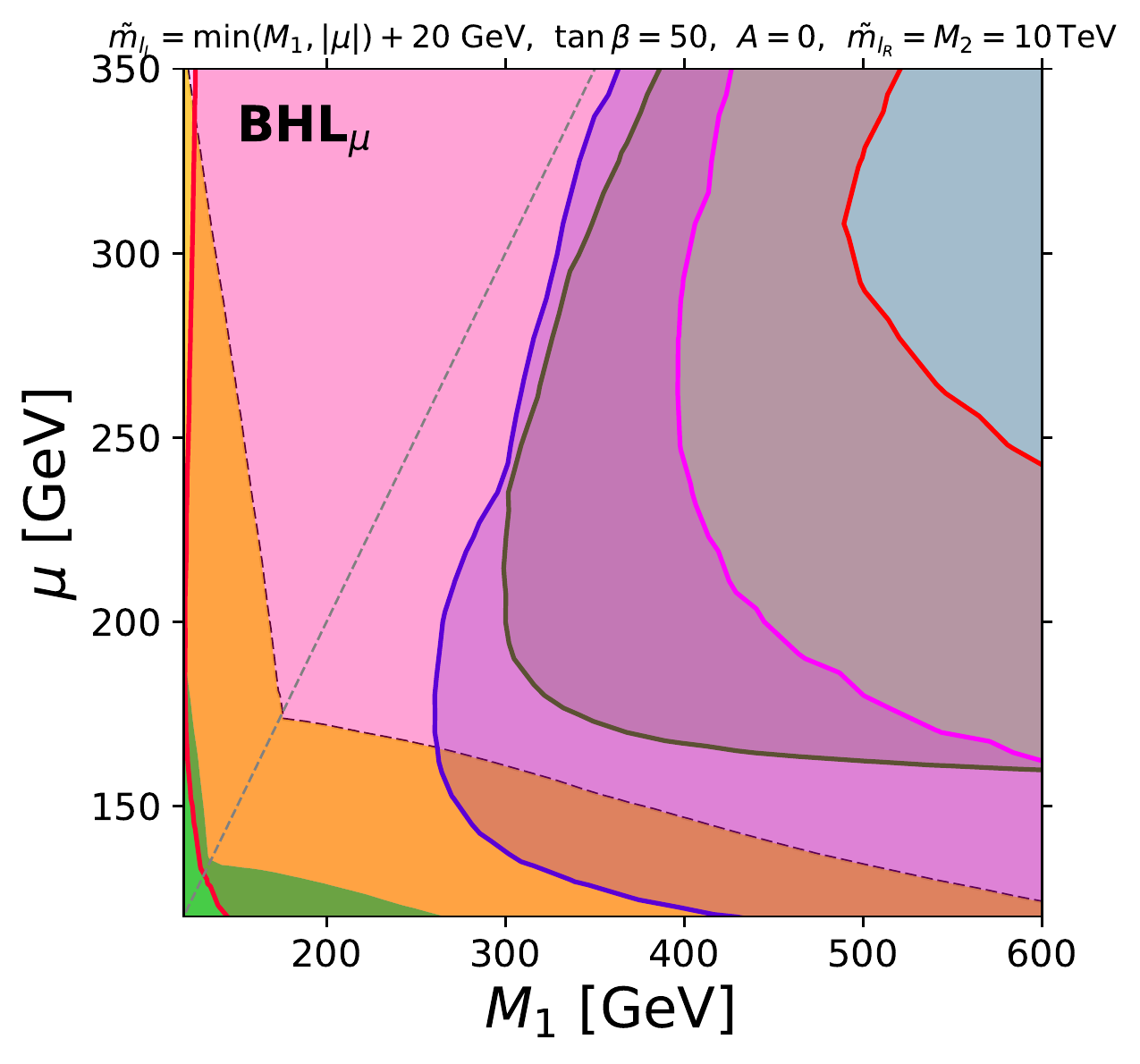}
\includegraphics[width=0.31\textwidth]{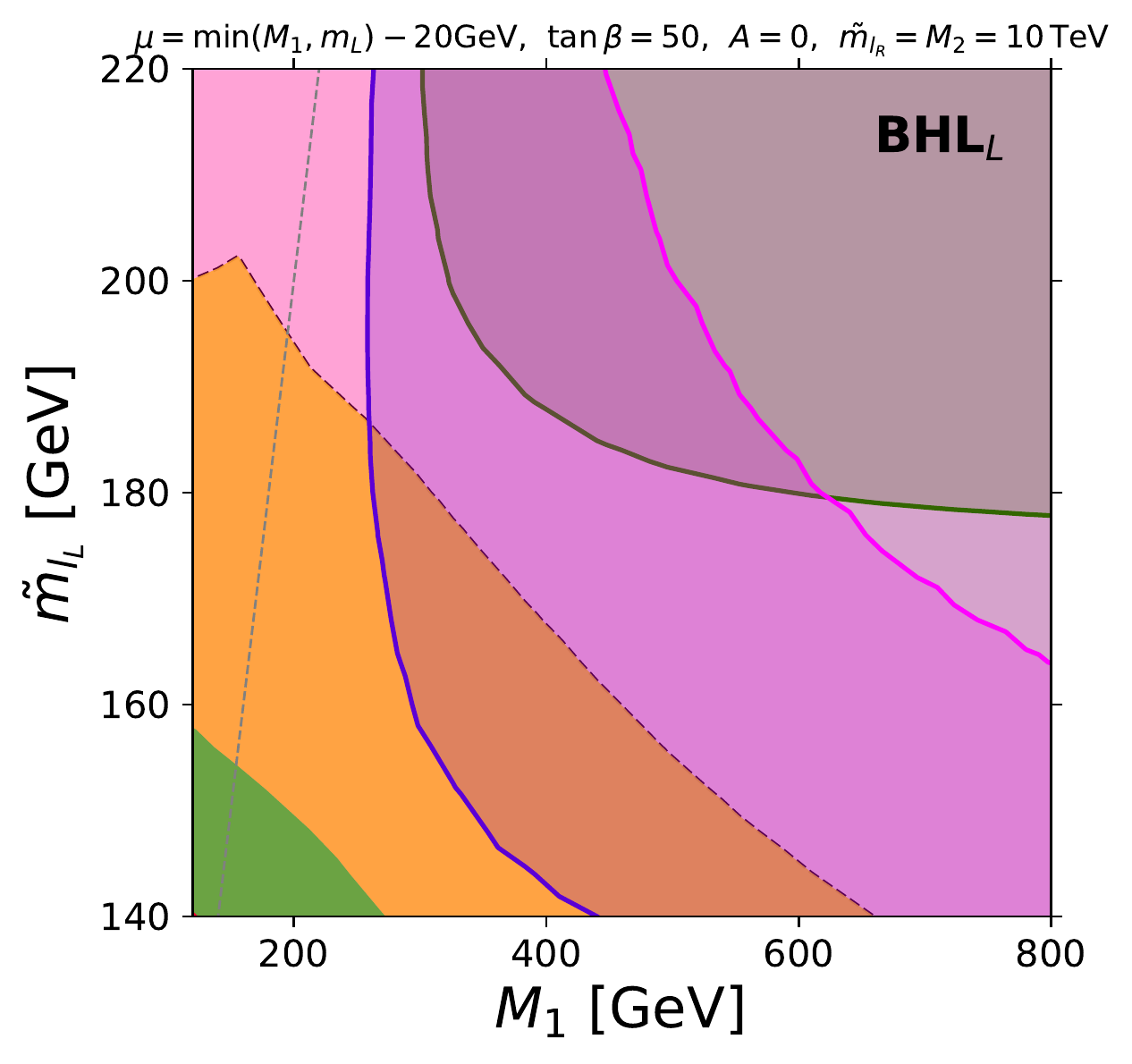}
\includegraphics[width=0.31\textwidth]{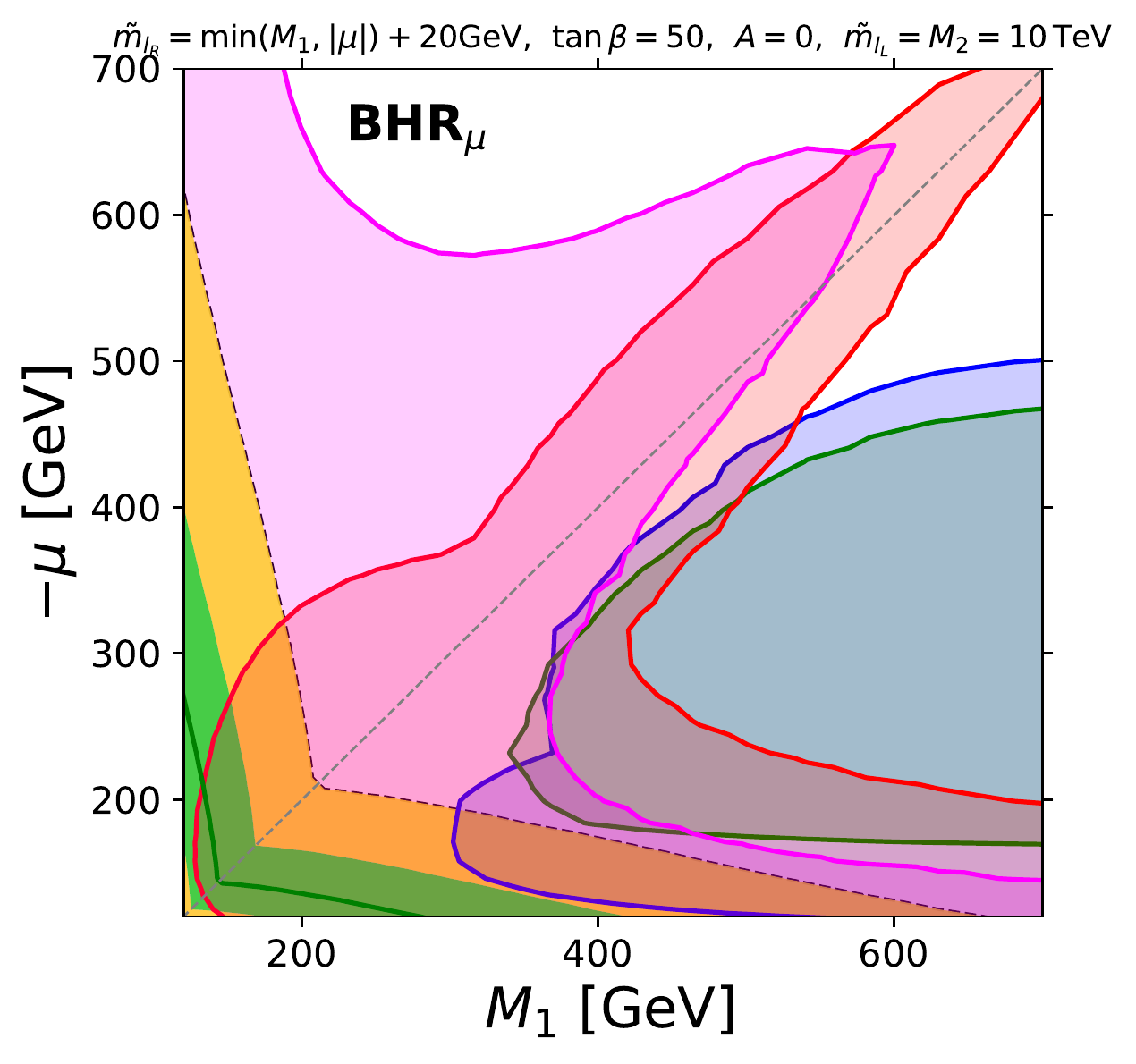}
\includegraphics[width=0.31\textwidth]{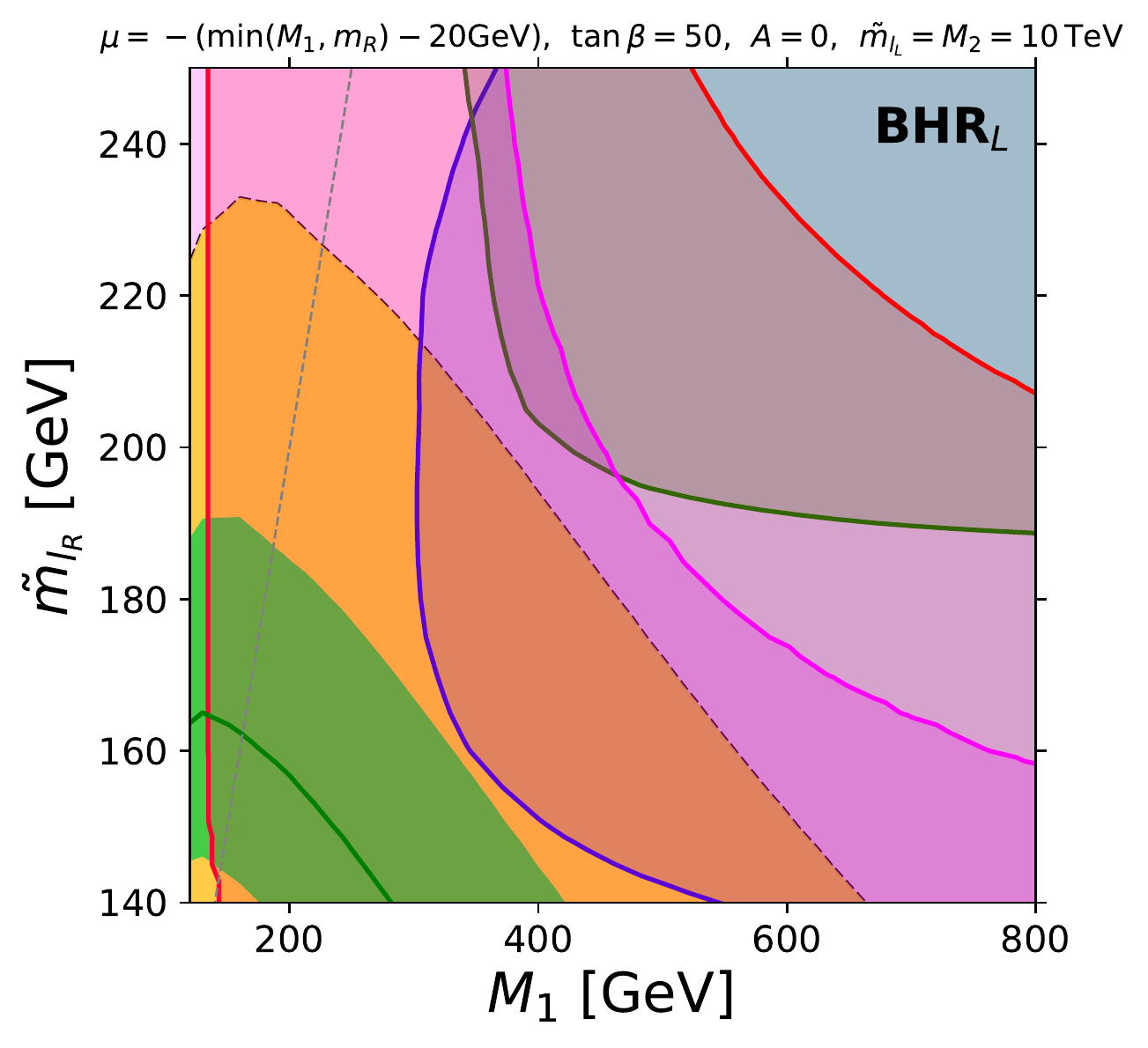}
\includegraphics[width=0.31\textwidth]{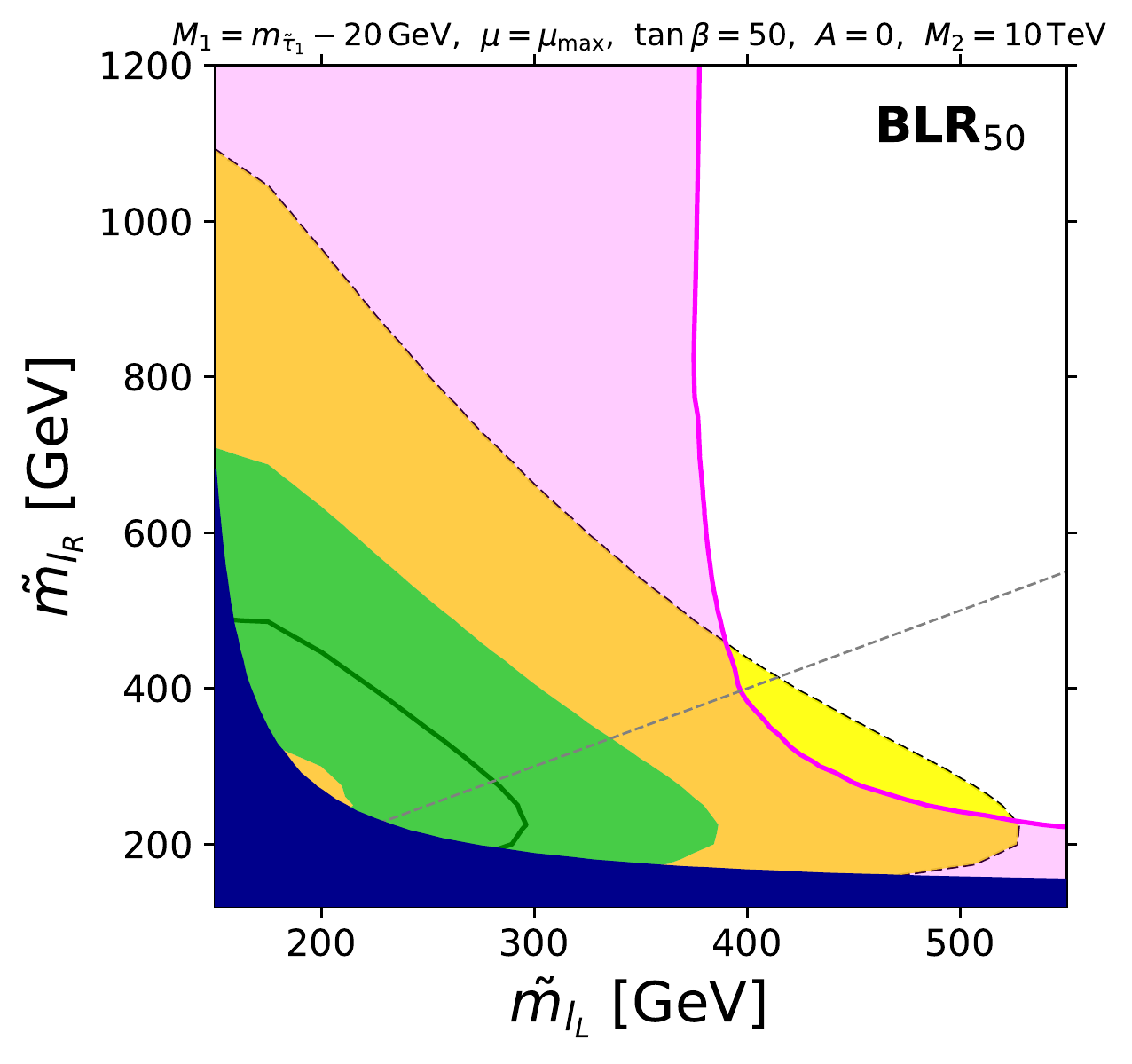}
\includegraphics[width=0.31\textwidth]{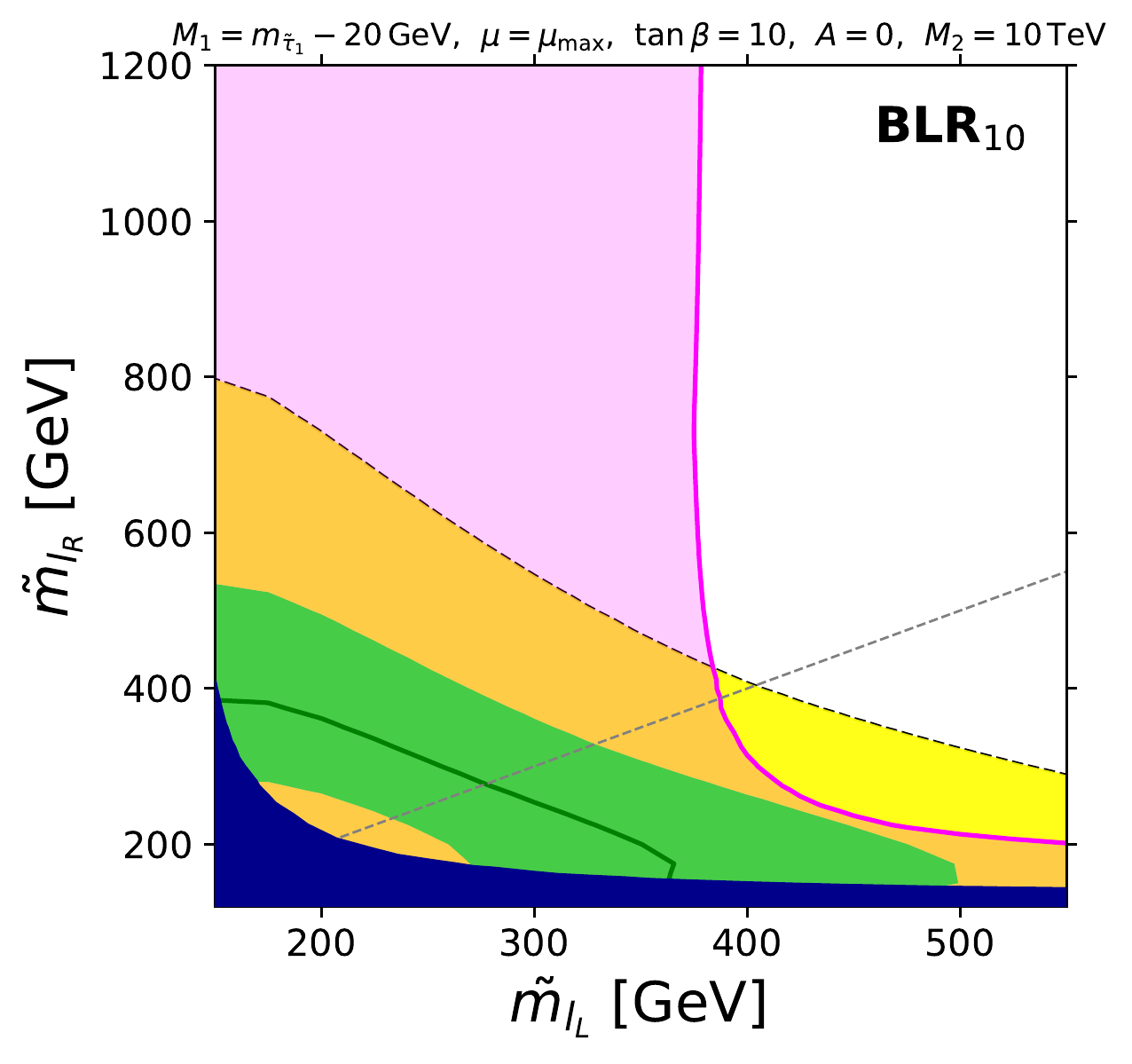}
\caption{\small
All considered planes for GMSB scenario with neutralino NLSP. 
Region of parameter space allowed by the latest $a_\mu$ experimental results \cite{Muong-2:2021ojo} is depicted with green and yellow bands, corresponding to one and two sigma agreement respectively.
Blue, green, red and magenta shaded regions are excluded by LHC constraints. 
Dark blue shaded area is excluded by LEP stau mass bound \cite{Zyla:2020zbs}.
}
\label{fig:GMSB_neut}
\end{figure}


\end{appendices}
\newpage

\bibliography{ref}
\bibliographystyle{utphys28mod}

\end{document}